\documentclass[11pt]{article}
\pdfoutput=1
\usepackage{amsmath,amssymb,graphicx,color,cancel,cite}
\usepackage{placeins}
\usepackage{feynmp,multirow,array}

\def\be{\begin{equation}}
\def\ee{\end{equation}}
\def\bea{\begin{eqnarray}}
\def\eea{\end{eqnarray}}

\def\bead{\begin{aligned}}
\def\eead{\end{aligned}}
\def\beal#1\eeal{\begin{align}#1\end{align}}
\def\beald#1\eeald{\begin{align}\aligned#1\endaligned\end{align}}

\def\Hc{\text{H.c.}}

\def\nSUSY{{\cancel{\text{SUSY}}}}
\def\nSUSYEWSB{{{\cancel{\text{SUSY}}}_\text{EWSB}}}
\def\nSUSYEWS{{{\cancel{\text{SUSY}}}_\text{EWS}}}
\def\Msoft{m_\text{soft}}

\def\TeV{\text{TeV}}
\def\GeV{\text{GeV}}
\def\eV{\text{eV}}

\def\OP{\text{OP}}
\def\OPnu{\OP_\nu}
\def\SOP{\widehat{\OP}}
\def\SOPnu{\SOP_\nu}

\def\pext{p_\text{ext}}

\newcommand{\figref}[1]{Fig.~\ref{#1}}
\newcommand{\tbref}[1]{Tab.~\ref{#1}}
\newcommand{\secref}[1]{Sec.~\ref{#1}}
\newcommand{\appref}[1]{Appendix~\ref{#1}}
\renewcommand{\eqref}[1]{Eq.\,(\ref{#1})}

\newcommand{\Dbar}[1]{\bar D_{\dot #1}}
\newcommand{\D}[1]{D_{#1}}

\def\vspUfig{0.0in}
\def\vspDfig{-0.2in}

\title{Neutrino masses from SUSY breaking in radiative seesaw models} 
\date{}
\author{Ant\'onio J.\ R.\ Figueiredo}
\begin{document}

	\vspace*{-2cm}
	\begin{flushright}
	CFTP 14-009
	
	\vspace*{2mm}
	\today
	\end{flushright}
	\begin{center}
	\vspace*{10mm}
	{\Large\bf Neutrino masses from SUSY breaking in radiative seesaw models}\\
	\vspace{1cm}
	{\bf Ant\'onio J.\ R.\ Figueiredo}
	
	 \vspace*{.5cm} 
	Centro de F\'isica Te\'orica de Part\'iculas (CFTP), \\
	Instituto Superior T\'ecnico - University of Lisbon, \\ Av.\ Rovisco Pais 1, 
	1049-001 Lisboa, Portugal. \\
	ajrf@cftp.ist.utl.pt
	
	\end{center}
	
	\vspace*{5mm}
	\begin{abstract} 
		Radiatively generated neutrino masses ($m_\nu$) are proportional to supersymmetry (SUSY) 
		breaking, as a result of the SUSY non-renormalisation theorem. In this work, we 
		investigate the space of SUSY radiative seesaw models with regard to their dependence 
		on SUSY breaking ($\nSUSY$). In addition to contributions from sources of $\nSUSY$ that 
		are involved in electroweak symmetry breaking ($\nSUSYEWSB$ contributions), and which are 
		manifest from $\langle F^\dagger_H \rangle = \mu \langle \bar H \rangle \neq 0$ and 
		$\langle D \rangle = g \sum_H \langle H^\dagger \otimes_H H \rangle \neq 0$, 
		radiatively generated $m_\nu$ can also receive contributions from $\nSUSY$ sources 
		that are unrelated to EWSB ($\nSUSYEWS$ contributions). 
		We point out that recent literature overlooks pure-$\nSUSYEWSB$ contributions 
		($\propto \mu / M$) that can arise at the same order of perturbation 
		theory as the leading order contribution from $\nSUSYEWS$. 

		We show that there exist realistic radiative seesaw models in which 
		the leading order contribution to $m_\nu$ is proportional to 
		$\nSUSYEWS$. To our knowledge no model with such a feature exists in 
		the literature. We give a complete description of the simplest 
		model-topologies and their leading dependence on $\nSUSY$. We show that in 
		one-loop realisations $L L H H$ operators are suppressed 
		by at least $\mu \, \Msoft / M^3$ or $\Msoft^2 / M^3$. We construct 
		a model example based on a one-loop type-II seesaw. An interesting aspect 
		of these models lies in the fact that the scale of soft-$\nSUSY$ effects 
		generating the leading order $m_\nu$ can be quite small without conflicting 
		with lower limits on the mass of new particles. 
	\end{abstract}

\section{Introduction}

	The large hierarchy between neutrino masses ($m_\nu$) and the electroweak (EW) 
	scale may be regarded a symptom of an hierarchy between the latter and a 
	new mass scale ($M$) that holds lepton number ($L$-number) breaking. 
	The simplest extensions to the Standard Model (SM) that implement 
	this hypothesis (type-I seesaws~\cite{seesaw:I,seesaw:IandII}) generate 
	$L L H H$~\cite{Weinberg:1979sa} with the naively expected dimensionful 
	suppression factor of $1/M$. Both direct~\cite{direct} 
	and indirect~\cite{indirect} bounds on $m_\nu$ suggest $M$ as heavy as 
	$10^{15}~\GeV$ if the underlying parameters are of order one and obey 
	no special relations\footnote{Some special textures in the seesaw parameters 
	allow for relatively large couplings with a smaller $M$, as discussed for 
	e.g. in~\cite{Dev:2013oxa} and references therein.}. 

	One can also conceive that additional mass scales are involved in the making 
	of $L L H H$. If this is the case, a broader class of possibilities emerge 
	that may turn out to yield $M$ within foreseeable experimental reach: 
	\begin{itemize}
		\item[1.] the additional scale is the EW scale ($\sim v$). In this case 
			$L L H H$ is not generated in perturbation theory, 
			but higher dimensional operators are. This replaces 
			the $1/M$ dimensionful suppression by $v^n 
			/ M^{n+1}$, where $5+n$ is the dimension of the leading 
			order (LO) operator. 
			See for example~\cite{Babu:2009aq} for a model in which the LO 
			contribution to $m_\nu$ comes from the dimension-7 operator 
			$L L H H H^\dagger H$. 
 			See also~\cite{Bonnet:2009ej} and references therein. 
		\item[2.] the additional scale ($m$) is an intermediate scale between $m_\nu$ 
			and $M$. In this case $L L H H$ is suppressed by some power 
			of $m / M$. For example, in the inverse seesaw~\cite{inverse} 
			$m$ is connected to some small ($\ll M$) $L$-number breaking 
			scale that is transmitted to the actual leptons by dynamics at the 
			scale $M$. In the type-II seesaw~\cite{seesaw:IandII} $m$ could be 
			the coupling scale of the scalar triplet to the Higgses. Both examples 
			lead to a $m / M^2$ dimensionful suppression. 
	\end{itemize}
	In addition, if $L L H H$ is radiatively generated~\cite{seesaw:oneloop,Babu:1988ki}, loop factors and 
	many coupling dependence may help bringing $M$ close to the TeV scale. 
	This possibility arises naturally in models in which the sector holding $L$-number 
	breaking is charged under a symmetry with respect to (w.r.t.) which $L$ and $H$ are neutral. 
	Such a symmetry may find its motivation connected to the stability of dark matter, 
	as discussed in~\cite{Krauss:2002px,Ma:2006km,Aoki:2008av,Gustafsson:2012vj}. 
	For studies in the space of one-loop seesaw models 
	see~\cite{Ma:1998dn,FileviezPerez:2009ud,Bonnet:2012kz}. 

\medskip 

	Two new scales are introduced by supersymmetric (SUSY) extensions to the SM: 
	the soft SUSY breaking ($\nSUSY$) scale, $\Msoft$; and 
	the scale at which $\nSUSY$ takes place, $M_X$. Naive dimensional analysis 
	gives us grounds to speculate that $M_X$ is much heavier than $\Msoft$, 
	since the strengths of hard- and soft-$\nSUSY$ are related by powers of 
	$\Msoft / M_X$ (see for e.g.~\cite{Martin:1999hc}). The minimal SUSY SM (MSSM) 
	introduces yet another scale: the Higgs bilinear, $\mu$. Though, in general, 
	correct EW symmetry breaking (EWSB) requires $\mu \sim \Msoft$. 
	Do any of these scales play any role in neutrino mass generation? 

	It has been contemplated in~\cite{Frere:1999uv,ArkaniHamed:2000bq,
	Frere:2003ys,Demir:2007dt} that hard-$\nSUSY$ 
	is the source of $L$-number violation, so that $\Msoft / M_X \ll 1$ 
	might be the reason for $m_\nu / v \ll 1$. For example, if $\nSUSY$ 
	generates $\tilde L \tilde L H_u H_u$, then $L L H_u H_u$ arises at 
	one-loop level via a EWino-slepton loop and is suppressed 
	by $\Msoft / M_X$~\cite{Frere:1999uv}. 
	Another possible connection to $\nSUSY$ is in identifying the 
	seesaw mediators with the mediators of $\nSUSY$ 
	to the visible sector~\cite{Joaquim:2006uz,Mohapatra:2008wx,FileviezPerez:2009im}. 

	Holomorphy dictates that tree-level type-I and -III~\cite{Ma:1998dn,seesaw:III} seesaws 
	are superpotential operators that yield $L L H_u H_u$, whereas the 
	tree-level type-II~\cite{Rossi:2002zb} gives, in addition to $L L H_u H_u$ from the superpotential, 
	$L L H_u H^\dagger_d$ from the K\"ahler potential 
	\be
		\frac{1}{M^2_\Delta} L L F_{H_d}^\dagger H^\dagger_d 
		\subset \frac{1}{M^2_\Delta} \int d^4\theta \, \hat L \hat L \hat H^\dagger_d \hat H^\dagger_d 
		\subset \int d^4\theta \, \hat\Delta^\dagger \hat\Delta \,.
	\ee
	Hence, the K\"ahler contribution to neutrino masses is proportional to $\nSUSY$, 
	since it requires $\langle F^\dagger_{H_d} \rangle \neq 0$. 
	If the low energy Higgs sector coincides with that of the MSSM, 
	then $\langle F^\dagger_{H_d} \rangle \simeq \mu \langle H_u \rangle$ 
	which leads to a $L L H_u H^\dagger_d$ operator with a dimensionful suppression 
	factor of $\mu / M^2$. Therefore, the K\"ahler operator is usually disregarded in favour 
	of the superpotential operator which has a $1 / M$ dependence. 
	However, as they involve two different couplings, it is conceivable 
	that the coupling enabling the superpotential operator is sufficiently 
	suppressed so that the K\"ahler operator is the leading one. 
	K\"ahler operators as leading contributions to $m_\nu$ have been 
	studied in~\cite{Casas:2002sn,Brignole:2010nh}. 

\medskip

	Motivated by the SUSY non-renormalisation theorem, which asserts 
	that radiative corrections are $D$-terms, we study how radiative 
	seesaw models are sensitive to different sources of $\nSUSY$ (\footnote{In~\cite{SUSYradiativeHIERARCHY}, 
	the consequences of the SUSY non-renormalisation were explored in the context of radiative 
	corrections as a tentative explanation for the intergenerational mass hierarchy 
	of quarks and charged leptons.}). 
	Although $L$-number breaking can possibly arise from $\nSUSY$, i.e.\ from the 
	VEV of an auxiliary rather than scalar field (see for e.g.~\cite{Borzumati:1999sp}), in 
	here we assume that they are broken separately\footnote{Since SUSY and $L$-number 
	are very different symmetries, that the 
	two are broken separately seems to be a plausible assumption.} so that 
	the non-renormalisation 
	theorem is the only bridge between $m_\nu$ and $\nSUSY$. 
	We thus assume that the radiative seesaw models are realised in the 
	superpotential at a $L$-number breaking scale $M$ 
	that is higher than the scale of soft-$\nSUSY$ effects involving the seesaw mediators. 
	We classify the $\nSUSY$ contributions to neutrino mass operators 
	w.r.t.\ their involvement in EWSB as follows:
	$\nSUSYEWSB$ contributions are those which involve $\nSUSY$ vacuum 
	expectation values (VEVs) of the form 
	\be
		\langle F^\dagger \rangle 
			= \sum_H \mu_H \langle H \rangle + \sum_H \lambda_H \langle H H' \rangle \neq 0 
		~~\text{or}~~ 
		\langle D \rangle = g \sum_H \langle H^\dagger \otimes_H H \rangle \neq 0 \,,
	\ee
	where $H$'s are fields whose VEVs break the EW symmetry (EWS); while $\nSUSYEWS$ 
	contributions correspond to those in which at least one $\nSUSY$ VEV is unrelated 
	to EWSB. We apply the prefix ``pure'' to refer to a contribution 
	in which all $\nSUSY$ VEVs have the same origin in the classification above. For 
	example, the tree-level type-II seesaw K\"ahler operator is a pure-$\nSUSYEWSB$ 
	contribution to neutrino masses. 

	In this context, it is interesting to note that if EWSB is {\it almost} 
	SUSY, in the sense that there is a SUSY vacuum with EWSB~\cite{Batra:2008rc}, 
	and so that only small $\nSUSY$ effects are responsible for lifting its 
	degeneracy with EWS vacua, then $\nSUSYEWSB$ contributions 
	can be quite small due to $\langle F^\dagger \rangle_\text{EWSB} \approx 0$ and 
	$\langle D \rangle_\text{EWSB} \approx 0$ (i.e.\ vanish up to possibly small $\nSUSY$ 
	effects). However, in this work we focus on models with the low energy Higgs 
	sector of the MSSM, and thus, in which $\nSUSYEWSB$ contributions have the form 
	\beald
		& \langle F^\dagger_{H_{u,d}} \rangle = \mu \langle  H_{d,u} \rangle \,,\\
		& \langle D_{U(1)_Y} \rangle = \frac{g'}{2} \left( |\langle H_u \rangle|^2 - |\langle H_d \rangle|^2 \right) \,,~~~~ 
		\langle D^3_{SU(2)_L} \rangle = \frac{g}{2} \left( -|\langle H_u \rangle|^2 + |\langle H_d \rangle|^2 \right) \,.
	\eeald

\medskip

	As we will see in~\secref{sec:radiative_seesaw}, contributions to 
	neutrino mass operators whose dependence on $\nSUSY$ arises entirely 
	by means of $\nSUSY$ sources involved in EWSB are expected to be suppressed by some power of 
	$\mu / M$ or be of dimension higher than $5$ and involve gauge couplings. 
	Exploiting the power of the SUSY non-renormalisation in the space of 
	radiative seesaw models, we then investigate if models exist in which the 
	pure-$\nSUSYEWSB$ contribution to neutrino masses either vanishes or is subleading 
	w.r.t.\ the contribution from $\nSUSYEWS$ (\secref{sec:SOPbreak}).
	We catalogue one-loop model-topologies in which the leading contribution comes from 
	soft-$\nSUSYEWS$ in~\secref{sec:looking}. 
	An explicit model example is presented in~\secref{sec:model_example} and 
	consists of a one-loop type-II seesaw in which the leading pure-$\nSUSYEWSB$ 
	contribution is of dimension-7 -- comprising contributions 
	$\propto \mu / M$ and $\propto g^2$ --, whereas the leading contribution 
	from $\nSUSYEWS$ is of dimension-5 and has the dimensionful dependence 
	$\mu \, \Msoft / M^3$ or $\Msoft^2 / M^3$, the latter corresponding to  
	pure-$\nSUSYEWS$ contributions. 

\medskip

	Our analysis will be carried out using perturbation theory in superspace (supergraph 
	techniques\footnote{
	Extensive details concerning supergraph calculations can be found in chapter 6 of~\cite{Gates:1983nr}.
	}), as it renders the SUSY non-renormalisation theorem a very simple statement 
	and its implications in terms of component fields easier to identify. Points 
	of contact with results in terms of component fields will be established throughout. 
	Another advantage is that perturbation theory in superspace is much simpler than the ordinary QFT treatment. 
	For instance, aside from the algebra of the SUSY covariant derivatives ($\D{\alpha}$ and $\Dbar{\alpha}$), 
	supergraph calculations in a renormalisable SUSY model made of chiral scalar superfields 
	resemble the Feynman diagrammatic approach to an ordinary QFT made of scalars with trilinear 
	interactions. 
	$\nSUSY$ can be parameterised in a manifestly supersymmetric 
	manner by introducing superfields with constant $\theta$-dependent values 
	($\nSUSY$ spurions). Thus, $\nSUSY$ effects will be conveniently taken into 
	account in supergraph calculations by means of considering couplings to 
	external $\nSUSY$ spurions~\cite{Girardello:1981wz}. 
	This allows one to see the $\nSUSYEWS$ contributions to neutrino masses as 
	small $\nSUSY$ effects upon a fundamentally SUSY topology.

\section{Radiative seesaws in SUSY} \label{sec:radiative_seesaw}

\begin{figure}[h!t]
          \vspace{\vspUfig}
          \begin{center}
                \begin{tabular}{cccc}
			\includegraphics[width=40mm]{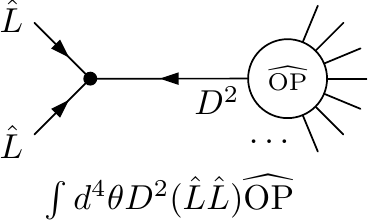} & 
                	\includegraphics[width=40mm]{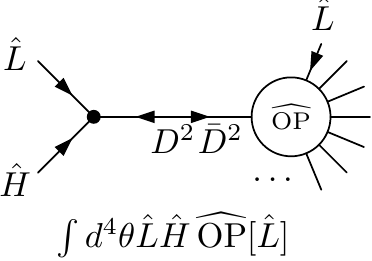} & 
                	\includegraphics[width=40mm]{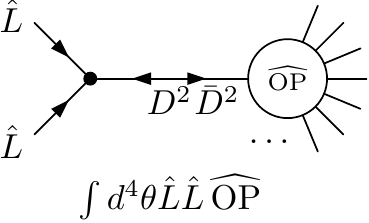} & 
                	\includegraphics[width=20mm]{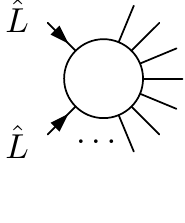} 
		\end{tabular}
                \caption{Characteristic examples of supergraph topologies for radiative seesaws: 
			type-II without a chirality flip (class A), 
			type-I and -III, type-II with a chirality flip and 1PI seesaw, respectively. 
                }\label{fig:SGraph_General_LLHiggses}
          \end{center}
          \vspace{\vspDfig}
        \end{figure}

	Let $\OPnu$ be the set of operators that contribute to neutrino masses once the EW symmetry 
	is broken and $\SOPnu$ be the set of superfield operators (superoperators) that yield at 
	least an $\OP \in \OPnu$. If neutrino masses are radiatively generated the SUSY non-renormalisation 
	theorem asserts that for every $\OP \in \OPnu$ there exists an $\SOP \in \SOPnu$ 
	such that 
	\be
		\OP \subset \int d^4\theta \, \SOP \,.
	\ee
	Hence, as any $\OP \in \OPnu$ is of the form $\OP = L L \otimes \text{Higgses}$, every 
	$\SOP \in \SOPnu$ belongs to one of two classes: 
	\be
		D^2 ( \hat L \hat L \hat H^n ) \otimes \hat A 
		 ~~\text{or}~~ 
		\hat L \hat L \otimes \hat B \,,
	\ee
	with 
	\be
		\int d^4\theta \hat A \supset \text{Higgses} \,,~~~~
		\int d^2\bar\theta \hat B \supset \text{Higgses} \,,
	\ee
	and where $n = 0,1,...$ stands for conceivable insertions of superfields that yield Higgses at 
	$\theta = 0$ (a limit hereafter denoted by $\left.\right|$). 
	Class A superoperators are naturally generated in radiative type-II seesaws in which the one-particle reducible (1PR) 
	propagator does not undergo a chirality flip (i.e.\ is of the form $\hat\Phi\hat\Phi^\dagger$), 
	whereas class B arise in radiative type-I and -III seesaws, radiative type-II seesaws with a chirality 
	flip and one-particle irreducible (1PI) seesaws. See~\figref{fig:SGraph_General_LLHiggses}. We note that 
	type-I and -III without a chirality flip do not yield an $\OP \in \OPnu$ (even in the presence 
	of $\nSUSY$) because 
	\be
		\int d^4\theta \, D^2 ( \hat L \hat H \, \SOP_X ) \, \SOP ~\cancel{\supset}~ L L \, ... \,, \label{eq:TypeIandIII_withoutCHIRALITYflip}
	\ee
	where $\SOP$ is any superoperator containing one $\hat L$ 
	and $\SOP_X$ accounts for conceivable insertions of $\nSUSY$ spurions. In terms of 
	component fields this can be seen to follow from the fact that, without a chirality 
	flip in the 1PR spinor line, the result is always proportional to external momenta ($\pext$). 
	To illustrate this, consider a model in which $\hat L \hat H_u \hat N$ and 
	$\hat N \hat N'$ are superpotential terms and 
	$\hat N^\dagger \hat L \hat H_u \hat\rho$ is radiatively generated. 
	(The coupling $\hat L \hat H_u \hat N'$ can be forbidden by $L$-number 
	conservation, which is spontaneously broken by $\langle \rho \rangle \neq 0$.) 
	In such a model, the type-I (or -III) diagram without a chirality flip arises from the 
	$N N^\dagger$ propagator in conjunction with the terms 
	\be
		L H_u N \subset \int d^2\theta \, \hat L \hat H_u \hat N ~~\text{and}~~ 
		N^\dagger p L H_u \subset \int d^4\theta \, \hat N^\dagger \hat L \hat H_u \langle \hat\rho \rangle \,,
	\ee
	and leads to $L L H_u H_u$ with an overall dependence on 
	$\pext^2$ or, more precisely, $-\Box ( L H_u ) L H_u$. In terms of supergraphs 
	this result follows from 
	\be
		-\Box ( L H_u ) L H_u \subset \int d^4\theta \, D^2 ( \hat L \hat H_u ) \hat L \hat H_u \langle \hat\rho \rangle \,,
	\ee
	which should be compared with~\eqref{eq:TypeIandIII_withoutCHIRALITYflip}. 
	Moreover, $\nSUSY$ insertions into $\hat L \hat N \hat H_u$ and/or  
	$\hat N^\dagger \hat L \hat H_u \hat\rho$ do not change this structure.

	To proceed we assume that only scalar and gauge vector superfields exist. 
	We can then write 
	\beald
		& \hat A \in \hat a \otimes \left\{ \hat H , \hat H^\dagger , D^2 \hat Z , \bar D^2 \hat Z^\dagger , D^2 \bar D^2 \hat V \right\}^n \,,\\
		& \hat B \in {\hat b}^\dagger \otimes \left\{ \hat H , \hat H^\dagger , D^2 \hat Z , \bar D^2 \hat Z^\dagger , D^2 \bar D^2 \hat V \right\}^n \,, \label{eq:ClassAClassBList}
	\eeald
	where $n = 0,1,...$ stands for arbitrary insertions of superfields within the given 
	set (denoted by curly braces), though constrained by internal symmetries. 
	$\hat a$ and $\hat V~\text{(mod $\hat H^\dagger, \hat H$)}$ ($\hat b^\dagger$ and 
	$\hat Z^\dagger~\text{(mod $\hat H^\dagger$)}$) 
	are real (anti-chiral) scalar superfields whose $D$ ($F$) component is a constant 
	or a product of Higgses\footnote{Here and throughout the text, ``mod X'' means 
	modulo insertions of X. For instance, 
	suppose that $\hat V~\text{(mod $\hat H^\dagger, \hat H$)}$ is equal to $\hat U$. Then, 
	this means that the general form of $\hat V$ is $\hat V = \hat U \hat H^{\dagger k} \hat H^{k'}$, 
	where $k,k' = 0,1,...$\,.}.

	\subsection{Pure-$\nSUSYEWSB$ contributions} \label{sec:nSUSYEWSB}

		Superoperators that lead to pure-$\nSUSYEWSB$ contributions are those in which $\hat a$ 
		is a gauge vector superfield $\hat V$ of any symmetry under which Higgses are charged 
		or the real product of $\hat b$ ($\hat b^\dagger \hat b$), and $\hat b^\dagger$ is 
		the anti-chiral projection of $\hat V$ ($D^2 \hat V$), so that\footnote{
		We note that $D := D \bar D^2 D \hat V |$ is equal to 
		$\bar D^2 D^2 \hat V |$ in the Wess-Zumino and Landau gauge, since in this 
		gauge we have $\hat V | = 0$ and $\partial_\mu V^\mu = 0$.} 
		\be
			\left. \bar D^2 D^2 \hat V \right| = D \supset g H^\dagger \otimes H \,, \label{eq:D2D2barV_D_HH}
		\ee
		or any anti-chiral scalar superfield $\hat Z^\dagger$ that has a bilinear with an Higgs or 
		a trilinear with two Higgses, so that 
		\be
			\left. \bar D^2 \hat Z^\dagger \right| = F^\dagger_Z \supset \mu H ~\text{or}~ \lambda H \otimes H' \,. \label{eq:D2barZ_H_HH}
		\ee
		Similarly, $\hat V~\text{(mod $\hat H^\dagger, \hat H$)}$ and $\hat Z^\dagger~\text{(mod $\hat H^\dagger$)}$ 
		in~\eqref{eq:ClassAClassBList} satisfy~\eqref{eq:D2D2barV_D_HH} and \eqref{eq:D2barZ_H_HH}, respectively. 

		\enlargethispage{6pt}

		Under the phenomenologically reasonable assumption of a superpotential mass term for $\hat Z$, the 
		contribution of a trilinear with two Higgses adds up to an overall derivative term of the form $\Box ( H H' )$, 
		as we show in~\appref{app:trilinear}. Moreover\footnote{Here, and throughout the text, a field (or a scalar chiral superfield) with 
		a bar, say $\overline X$ ($\hat{\overline X}$), transforms (under non-$R$-symmetries) in the conjugate 
		representation of $X$ ($\hat X$), so that $X \overline X$ ($\hat X \hat{\overline X}$) is symmetric (i.e.\ 
		invariant under the symmetries of the model). Moreover, the $R$-charges satisfy $Q_R(\hat{\overline X}) + Q_R(\hat X) = 2$ 
		so that $\int d^2\theta \hat X \hat{\overline X}$ is symmetric.}, 
		\be
			\langle F^\dagger_Z \rangle = \mu_Z \langle {\overline Z} \rangle + \lambda \langle H H' \rangle = 0 \,,
		\ee
		up to $\nSUSY$ effects. Hence, and from $\mu_Z \gg \Msoft$, one expects the $\langle F^\dagger_Z \rangle$ 
		contribution to be small due to the cancellation between leading terms. To be precise, one can estimate it as 
		(cf.~\eqref{eq:VEV_FZD_3rdORD} of~\appref{app:trilinear}) 
		\be
			\langle F^\dagger_Z \rangle \simeq \frac{(\Msoft^2)_{\overline Z}}{|\mu_Z|^2} \lambda \langle H H' \rangle \,. \label{eq:FZtri_nSUSY} 
		\ee
		Now, one expects that the EWSB vacuum is not disturbed by $\nSUSY$ effects involving $Z$ or 
		$\overline Z$, since $H$'s operators generated by integrating out $Z$ and $\overline Z$ are suppressed 
		by $\Msoft / \mu_Z \ll 1$ or $\mu / \mu_Z \ll 1$. Therefore, the $\langle F^\dagger_Z \rangle \neq 0$ 
		contribution that arises from a trilinear with two Higgses is more appropriately classified as 
		a $\nSUSYEWS$ contribution. 

\medskip

		Since $D$ is a hypercharge singlet, operators that come from a gauge vector superfield have mass dimension 
		higher than $5$. The least is a dimension-6 operator
		\be
			\int d^4\theta \left\{ \hat V D^2 ( \hat L \hat L ) \,, D^2 \hat V \hat L \hat L \right\} \otimes \hat H' \supset L L H^\dagger H H' \,,
		\ee
		that is conceivable if there exists a hypercharge $+1$ Higgs ($H'$). 
		On the other hand, if the low energy Higgs sector coincides with that of the MSSM, 
		the leading pure-$\nSUSYEWSB$ contributions that are independent of $\langle F^\dagger_Z \rangle$ 
		correspond to the dimension-7 operators 
		\be
			L L \otimes \left\{ H_u H_u \,, H_u H^\dagger_d \,, H^\dagger_d H^\dagger_d \right\} \otimes \left\{ H^\dagger_u H_u \,, H^\dagger_d H_d \right\} \,.
		\ee

		Since realistic SUSY models have Higgs bilinears, be them dynamically generated or 
		otherwise, it is conceivable that in general models there are pure-$\nSUSYEWSB$ 
		contributions to $L L H H$. Indeed, in~\secref{sec:models_literature} we analyse models in the 
		recent literature whose authors missed to identify the presence of such contributions. 

		We then set up to ask a different question. Do Higgs bilinears 
		imply the existence of a pure-$\nSUSYEWSB$ contribution to $m_\nu$? Or are there 
		models in which this implication does not hold? We show that there is always a 
		pure-$\nSUSYEWSB$ contribution (\secref{sec:empty}), however, models exist in which the LO 
		contribution to $m_\nu$ is proportional to $\nSUSYEWS$ (\secref{sec:looking}), 
		as we exemplify in~\secref{sec:model_example}.

\subsection{Models in the literature} \label{sec:models_literature}

	We analyse three recent models~\cite{fm,bmw,kmsy}. The first model is a one-loop type-II seesaw and 
	its superpotential ($\mathcal{W}'$) is defined in Eq.\,(5) of~\cite{fm}. $\mathcal{W}'$ has two continuous 
	Abelian symmetries independent of the hypercharge, and which can be identified with baryon and lepton 
	numbers, and an $R$-symmetry. 
	Once the scalar component of the gauge singlet superfield $\hat\sigma$ acquires a VEV, $L$-number is 
	broken. We will shift the vacuum accordingly by working with the superpotential 
	\be
		\mathcal{W}' + M_{Q'} \hat Q'^c \hat Q' \,.
	\ee
 	As some suitable definition of $L$-number is recovered in the limit in which any coupling of the set 
	$\{ f , f_q , \lambda , y_u \}$ goes to zero, the LO superoperator 
	that breaks $L$-number is a $\hat\Delta$-mediated type-II seesaw (without a chirality 
	flip, cf.~\figref{fig:SGraph_General_LLHiggses}) by means of 
	the one-loop coupling 
	\be
		a \int d^4\theta \, \hat\Delta^\dagger \hat H_u \hat H^\dagger_d \supset a |\mu_H|^2 \Delta^\dagger H_u H^\dagger_d \,, 
	\ee
	as generated by the supergraph of~\figref{fig:SGraph_DeltaHuHd_fm}. ($a$ is some mass dimension $-1$ coefficient whose form 
	will be given below.) On the rightmost diagram 
	we illustrate by means of using 
	auxiliary fields ($F$, depicted by a dotted line with an arrowhead) that the diagram is holomorphy compliant and has an 
	external $F^\dagger - F$ pair. Therefore, a non-vanishing coefficient for that operator is in 
	agreement with the SUSY non-renormalisation theorem.

	\begin{figure}[h!t]
          \vspace{\vspUfig}
          \begin{center}
		\includegraphics[width=120mm]{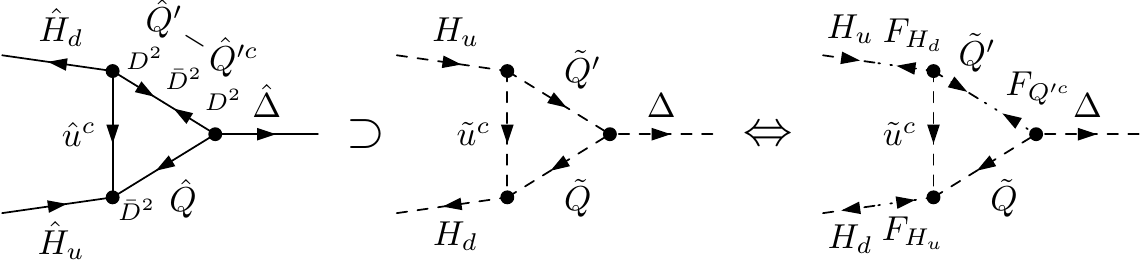} 
		\caption{Leading order supergraph that contributes to the 
			three-scalar coupling $\Delta^\dagger H_u H^\dagger_d$ 
			in the model of~\cite{fm}. 
                }\label{fig:SGraph_DeltaHuHd_fm}
          \end{center}
          \vspace{\vspDfig}
        \end{figure}

	For external neutral Higgses and at $\pext = 0$, $a$ is given by 
	\be
		a = -\frac{y_u^{ii} f^{ji*}_q \lambda^{ji*}}{16 \pi^2 M_{Q'_j}} \left(\frac{-1 + x^2_{ij} - \log x^2_{ij}}{(1-x^2_{ij})^2}\right) \,, ~~~~~~ x_{ij} := \frac{m_{\tilde u_i}}{M_{Q'_j}} \,,
	\ee
	and hence, the pure-$\nSUSYEWSB$ contribution to neutrino masses is 
	\be
		\mathbf{m_\nu^\nSUSYEWSB} \simeq -\frac{\mathbf{f} v c_\beta}{8 \pi^2} \left( \frac{|\mu_H|^2}{M^2_\Delta} \right) \left( \frac{m_t}{M_{Q_i}} f^{i3*}_q \lambda^{i3*} \right) \left(\frac{-1 + x^2_{3i} - \log x^2_{3i}}{(1-x^2_{3i})^2} \right) \,.
	\ee

	At the same order of perturbation theory other holomorphy compliant diagrams for 
	$\Delta^\dagger H_u H^\dagger_d$ can be drawn but none has an external $F^\dagger - F$ pair. 
	Thus, in the $\pext \to 0$ limit the diagrams in such a set add up to zero as mandated 
	by the SUSY non-renormalisation theorem. (This will be better illustrated in the discussion 
	surrounding~\figref{fig:DeltaHH_model_example}.) $\nSUSY$ insertions lift this 
	delicate cancellation, thus leading to $\mu_H$-independent contributions to $m_\nu$. 
	Under the common assumption of $\mu_H \sim \Msoft$, the two contributions are comparable. 

\medskip

	The second model is a one-loop 1PI seesaw. Its superpotential is given in Eq.\,(1) of~\cite{bmw} 
	and we reproduce here the part involved in the generation of $L L H H$: 
	\be
		\frac{M_N}{2} \hat N \hat N + \mu_L \hat H_u \hat H_d 
		+ \mu_{L2} \hat \eta_{L1} \hat \eta_{L2} 
		+ \frac{\mu_{s3}}{2} \hat \zeta_3 \hat \zeta_3 
		+ f_9 \hat H_d \hat \eta_{L2} \hat \zeta_3 
		+ f_{10} \hat H_u \hat \eta_{L1} \hat \zeta_3 
		+ f_{16} \hat L \hat N \hat \eta_{L2} \subset \mathcal{W} \,, \label{eq:SPo_bmw}
	\ee
	where we have made the identifications $\Phi_{L1} \to H_d$, 
	$\Phi_{L2} \to H_u$, $\psi \to L$ and chose a different 
	normalisation for the mass terms. $SU(2)_L$ contractions are defined as in~\eqref{eq:SU2L_contraction_DEF}, 
	except for an overall minus sign in $\mu_L$ and $f_9$ terms. 

\begin{figure}[h!t]
          \vspace{\vspUfig}
          \begin{center}
		\begin{tabular}{ccc}
                	\includegraphics[width=50mm]{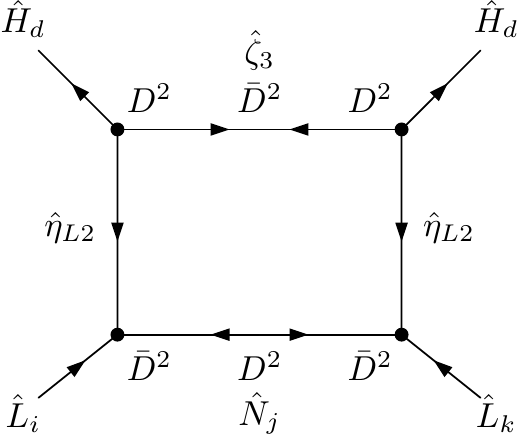} & 
                	\includegraphics[width=50mm]{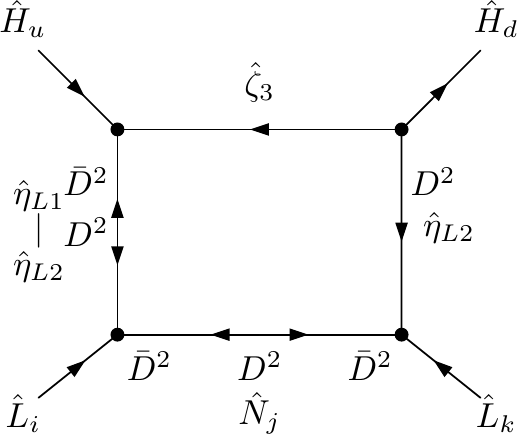} & 
                	\includegraphics[width=50mm]{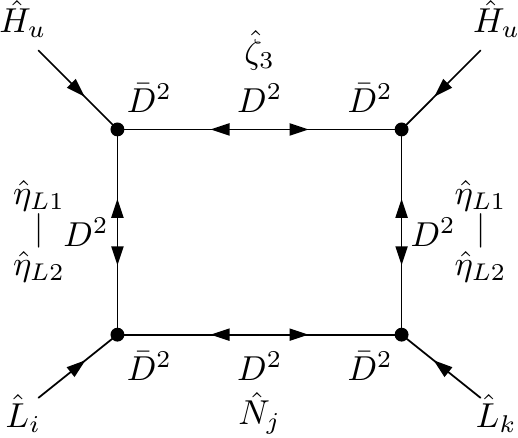}
		\end{tabular}
                \caption{Leading order supergraphs for superoperators 
			$\hat L \hat L \hat H \hat H$ in the model of~\cite{bmw}. 
			The third supergraph vanishes. 
                }\label{fig:LLHH_SOP_bmw}
          \end{center}
          \vspace{\vspDfig}
        \end{figure}

	At (leading) one-loop order three supergraphs with external $\hat L \hat L \hat H \hat H$ 
	are generated, as shown in~\figref{fig:LLHH_SOP_bmw}. By doing the D-algebra we see that 
	the third supergraph vanishes, while the others give the following contribution to 
	the effective Lagrangian: 
	\be
		\frac{f_9^*}{16 \pi^2} \int d^4\theta \, \left( \frac{1}{2} f^*_9 \mu_{s3} \hat H^\dagger_d + f_{10} \mu_{L2} \hat H_u \right) \hat H^\dagger_d \big( \hat L \boldsymbol{\kappa} \hat L \big) 
		\supset \frac{f^*_9 \mu_L}{16 \pi^2} \left( f_9^* \mu_{s3} H^\dagger_d + f_{10} \mu_{L2} H_u \right) H_u \big( L \boldsymbol{\kappa} L \big) \,.
	\ee

	In the $\pext \to 0$ limit $\boldsymbol{\kappa}$ is given by 
	\be
		\boldsymbol{\kappa}_{ik} = (\mathbf{f_{16}})_{ij} M_{N_j} D_0(0,0,0,0,0,0,M^2_{N_j},\mu^2_{L2},\mu^2_{s3},\mu^2_{L2}) (\mathbf{f^T_{16}})_{jk} \,,
	\ee
	where $D_0$ is the scalar one-loop 4-point integral~\cite{Passarino:1978jh}. 
	Hence, upon EWSB the following pure-$\nSUSYEWSB$ contribution 
	to neutrino masses is obtained 
	\be
		\mathbf{m_\nu^\nSUSYEWSB} \simeq -\frac{\mathbf{f_{16}} \mathbf{f_{16}^T} f^*_9}{48 \pi^2} \left( \frac{\mu_L v^2}{M_N^2} \right) \left( f^*_9 c_\beta + f_{10} s_\beta \right) s_\beta \,,
	\ee
	where we have taken the simplifying limit $M_{N_i} = \mu_{s3} = \mu_{L2} = M_N$.

\begin{figure}[h!t]
          \vspace{\vspUfig}
          \begin{center}
                \begin{tabular}{c}
			\includegraphics[width=90mm]{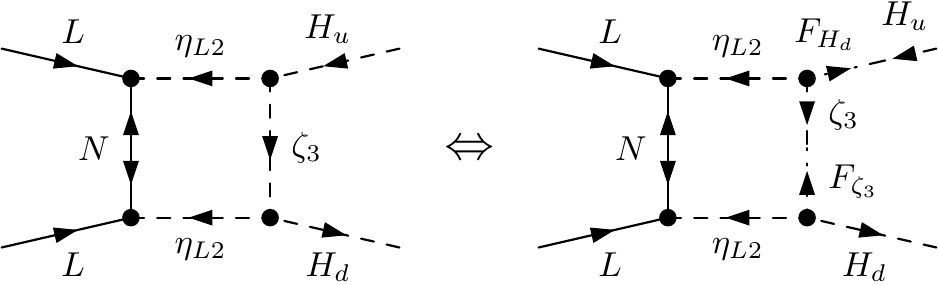} \\
                	\includegraphics[width=90mm]{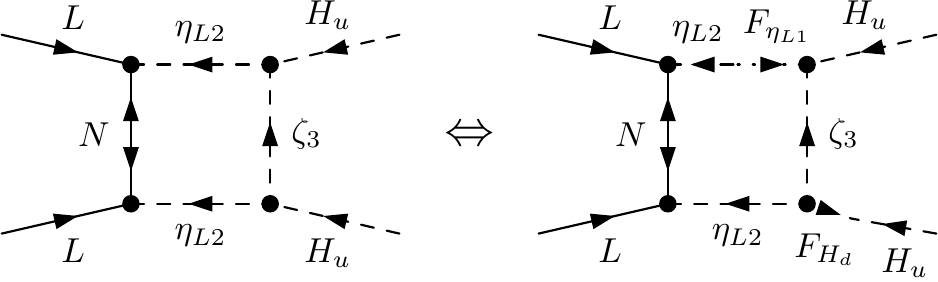}
		\end{tabular}
                \caption{Leading order diagrams generating operators   
		$L L H_u H^\dagger_d$ (upper row) and $L L H_u H_u$ (lower row) in 
		the model of~\cite{bmw}. 
                }\label{fig:LLHuHd_LLHuHu_bmw}
          \end{center}
          \vspace{\vspDfig}
        \end{figure}

	In order to recover this same result working with component fields, we note that the holomorphy of the superpotential 
	dictates that at one-loop order the only possible contributions to $L L H H$ are those displayed 
	in~\figref{fig:LLHuHd_LLHuHu_bmw}. For each diagram we display on the right-hand side its equivalent with auxiliary fields. 
	Contrary to the previous model, in this model all LO holomorphy compliant diagrams have an external $F^\dagger - F$ 
	pair: the $F$ is $L L$ and the $F^\dagger$ is $F^\dagger_{H_d}$. 
	The three-scalar interactions involved can be read from 
	\beald
		& -f^*_9 \mu_L H_u \eta^\dagger_{L2} \zeta^\dagger_3 \subset -f^*_9 F^\dagger_{H_d} \eta^\dagger_{L2} \zeta^\dagger_3 \subset \mathcal{L} \,,\\
		& -f^*_9 \mu_{s3} H^\dagger_d \eta^\dagger_{L2} \zeta_3 \subset -f^*_9 H^\dagger_d \eta^\dagger_{L2} F^\dagger_{\zeta_3} \subset \mathcal{L} \,,\\
		& -f_{10} \mu^*_{L2} H_u \eta^\dagger_{L2} \zeta_3 \subset -f_{10} H_u F_{\eta_{L1}} \zeta_3 \subset \mathcal{L} \,,
	\eeald
	and by means of standard calculations one can confirm the supergraph derivation. 

\begin{figure}[h!t]
          \vspace{\vspUfig}
          \begin{center}
                \begin{tabular}{c}
			\includegraphics[width=90mm]{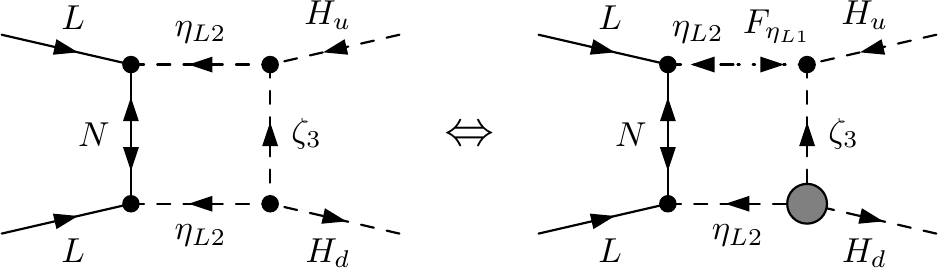} \\
                	\includegraphics[width=90mm]{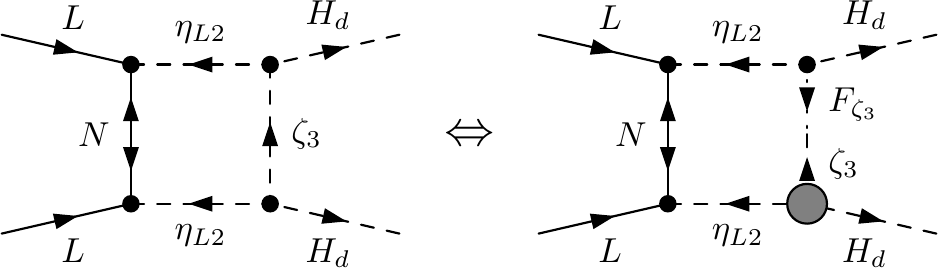}
		\end{tabular}
                \caption{Leading order $A$-term (grey blobs) contribution to $L L H H$ 
			in the model of~\cite{bmw}. 
			We do not display $L L H_u H_u$ since it is subleading 
			as it requires a $B_{\eta_L}$ insertion. 
                }\label{fig:LLHH_Aterms_bmw}
          \end{center}
          \vspace{\vspDfig}
        \end{figure}

	Besides overlooking the pure-$\nSUSYEWSB$ contribution to $m_\nu$, the authors of~\cite{bmw} estimate 
	the $\nSUSYEWS$ contribution as having the dimensionful dependence (cf.~Eq.\,(3) of~\cite{bmw}) 
	\be
		\mathbf{m_\nu^\nSUSYEWS} \propto \frac{v^2 \Msoft^2}{M^3_N} \,,
	\ee
	where we have taken the freedom to identify what they call the $\tilde N \tilde N$ $B$-term by $\Msoft^2$, 
	$\Msoft$ being an overall scale for the soft-$\nSUSY$ parameters. If this were indeed the LO contribution 
	from $\nSUSYEWS$, then $\mathbf{m_\nu^\nSUSYEWSB} \gg \mathbf{m_\nu^\nSUSYEWS}$ under the common assumption of 
	$\mu_L \sim \Msoft$. However, the authors have missed the dominant $\nSUSYEWS$ contribution and which proceeds 
	from the $\eta_{L2} \zeta_3 H_d$ $A$-term, as can be seen in~\figref{fig:LLHH_Aterms_bmw}. To be 
	specific, at LO the $A$-terms lead to 
	\be
		\mathbf{m_\nu^\nSUSYEWS} = \frac{1}{t_\beta} \left(\frac{A^*_9}{\mu_L}\right) \mathbf{m_\nu^\nSUSYEWSB} \,,
	\ee
	where $A_9$ is defined by $f_9 A_9 \eta_{L2} \zeta_3 H_d \subset -\mathcal{L}_\text{soft}$. 
	(Conventions regarding the soft-$\nSUSY$ potential are explained at the beginning of~\appref{app:susy_break}.) 
	On dimensional grounds one would naively expect that, indeed, a dependence of 
	$\Msoft / M^2$ for $L L H H$ would be found, since the underlying, i.e.\ 
	$\int d^4\theta \, \hat L \hat L \hat H \hat H$, has mass dimension $6$. 

	A thorough evaluation of soft-$\nSUSYEWS$ contributions to $L L H H$ up to order $2$ 
	and in the simplifying limit $M_{N_i} = \mu_{s3} = \mu_{L2} = M_N$ is given 
	in~\appref{app:bmw}. 

\medskip

	To end this section let us briefly mention the model of~\cite{kmsy}. It is also a 
	one-loop 1PI seesaw and contains a Higgs bilinear. The model's low-energy superpotential 
	comprises Eq.\,(10) and Eq.\,(12) of~\cite{kmsy}, in addition to MSSM Yukawa couplings. 
	In addition to baryon number, this superpotential has a continuous Abelian symmetry 
	independent of the hypercharge and which is defined by 
	\beald
		& \hat L \to e^{i \phi_L} \hat L \,,~~~~~~ \hat E^c \to e^{-i \phi_L} \hat E^c \,,~~~~~~ \hat \zeta/\hat \eta \to e^{\pm i\phi_L} \hat \zeta/\hat \eta \,,\\
		& \hat\Phi_{u,d} \to e^{\mp i \phi_L} \hat\Phi_{u,d} \,,~~~~~~ \hat\Omega^\pm \to e^{\mp i \phi_L} \hat\Omega^\pm \,,
	\eeald 
	i.e.\ a $L$-number symmetry. 
	The soft-$\nSUSY$ potential of their model (cf.~Eq.\,(11) of~\cite{kmsy}) 
	contains the terms 
	\be
		m^2_{\zeta\eta} \eta^\dagger\zeta + \frac{B^2_\zeta}{2} \zeta^2 + \frac{B^2_\eta}{2} \eta^2 + \Hc 
		\subset -\mathcal{L}_\text{soft} \,,
	\ee
	which explicitly break the $U(1)_L$. (It is noteworthy that these terms are 
	absent from their earlier works~\cite{Kanemura}.)  
	It is thus not surprising that in their model all $L L H H$ operators come from 
	$\nSUSYEWS$. If one adds to the superpotential the analogue of $\zeta^2$ and $\eta^2$ 
	$\nSUSY$-terms, i.e.\ 
	\be
		\frac{M_\zeta}{2}\hat\zeta^2 + \frac{M_\eta}{2}\hat\eta^2 \,,
	\ee
	so that $U(1)_L$ breaking becomes independent of $\nSUSY$, one finds a pure-$\nSUSYEWSB$ 
	contribution to $L L H_u H^\dagger_d$ and $L L H_u H_u$ in striking resemblance to the  
	previous model: $\hat\Phi_{u,d}$ play the role of $\hat \eta_{L2,L1}$, while 
	$\hat\zeta$ (and its the mixture with $\hat\eta$) plays the role of 
	$\hat\zeta_3$ in the generation of $L L H_u H^\dagger_d$ 
	(and $L L H_u H_u$, respectively).

\section{$\nSUSYEWS$ contributions} \label{sec:SOPbreak}

	In the presence of $F$- or $D$-term $\nSUSY$, any operator that comes from $\nSUSYEWS$ 
	is contained in the union of the following cases: 
	\beald
		& \text{a)}~ \int d^4\theta \, \hat X \, \SOP \,; \\
		& \text{b)}~ \int d^4\theta \, \hat X^\dagger \, \SOP \,; \\
		& \text{c)}~ \int d^4\theta \, \hat Y \, \SOP \,; 
	\eeald
	modulo $D^2 \hat X$, $\bar D^2 \hat X^\dagger$ and $D^2 \bar D^2 \hat Y$ insertions, and where 
	$\hat X$ and $\hat Y$ are $F$- and $D$-term $\nSUSY$ spurions, respectively. 
	Under the common assumption that $\nSUSY$ is blind to the internal 
	symmetries of the visible sector, it is conceivable the existence of models in which 
	both $\{\hat X, \hat X^\dagger, \hat Y\} \SOP$ (cases a, b and c, respectively) 
	and $\SOP$ are generated up to some order in perturbation theory. 
	We can now ask ourselves which instances of $\SOP \in \SOPnu$ do 
	not yield an $\OP \in \OPnu$ in the absence of $\nSUSY$ spurions\footnote{To simplify 
	the discussion, from now on any $\SOP \in \SOPnu$ is defined 
	modulo $\nSUSY$ insertions.}. The general answer is: 
	\beald
		& \text{1.}~ \SOP = D^2 ( \hat L \hat L \hat H^n ) \otimes \Big(\text{a superoperator whose $D$-term is zero at $\pext = 0$}\Big) \,; \\
		& \text{2.}~ \SOP = \hat L \hat L \otimes \Big(\text{a superoperator whose $F^\dagger$-term is zero at $\pext = 0$}\Big) \,.
	\eeald
	In the following, let $\hat Z^\dagger$ and $\hat V$ denote any superfields whose $\hat Z^\dagger~\text{(mod $\hat H^\dagger$)}$ 
	and $\hat V~\text{(mod $\hat H, \hat H^\dagger$)}$ parts satisfy~\eqref{eq:D2barZ_H_HH} 
	and~\eqref{eq:D2D2barV_D_HH}, respectively. 
	Type-1 superoperators that only give $\OP \in \OPnu$ from $\nSUSYEWS$ according to a, b and c, are: 
	\beald
		& \text{1.a)}~ D^2 ( \hat L \hat L \hat H^n ) \otimes \left\{ \hat Z^\dagger , D^2 \hat V \right\}  \otimes \left\{ \hat H^\dagger , \bar D^2 \hat Z^\dagger , D^2 \hat Z , D^2 \bar D^2 \hat V \right\}^{n'} \,; \\
		& \text{1.b)}~ D^2 ( \hat L \hat L \hat H^n ) \otimes \left\{ \hat Z , \bar D^2 \hat V \right\} \otimes \left\{ \hat H , \bar D^2 \hat Z^\dagger, D^2 \hat Z , D^2 \bar D^2 \hat V \right\}^{n'} \,; \\
		& \text{1.c)}~ D^2 ( \hat L \hat L \hat H^n ) \otimes \left\{ (\hat H^\dagger)^k , (\hat H)^k \right\} \otimes \left\{ \bar D^2 \hat Z^\dagger , D^2 \hat Z , D^2 \bar D^2 \hat V \right\}^{n'} \,; 
		\label{eq:Type1SUSYbreak}
	\eeald
	where $n, n', k = 0,1,...$ stand for any number of insertions, though constrained by internal 
	symmetries. 
	Type-2 $\SOP$'s that only give $\OP \in \OPnu$ from $\nSUSYEWS$ can only proceed from b: 
	\beald
		& \text{2.b)}~ \hat L \hat L \otimes \left\{ \hat H , D^2 \hat Z , \bar D^2 \hat Z^\dagger , D^2 \bar D^2 \hat V \right\}^n \,. 
		\label{eq:Type2SUSYbreak}
	\eeald

\medskip

	If at low energy the only Higgses are MSSM's, then the superoperators of lowest dimension that 
	only give $\OP \in \OPnu$ from $\nSUSYEWS$ are 
	\beald
		& \text{1.a)}~ D^2 ( \hat L \hat L ) \hat H^\dagger_d \otimes \left\{ \hat H^\dagger_d , \bar D^2 \hat H^\dagger_d , D^2 \hat H_u \right\} \cup D^2 ( \hat L \hat L \hat H_u ) \hat H^\dagger_d \,; \\
		& \text{1.b)}~ D^2 ( \hat L \hat L ) \hat H_u \otimes \left\{ \hat H_u , \bar D^2 \hat H^\dagger_d , D^2 \hat H_u \right\} \cup D^2 ( \hat L \hat L \hat H_u ) \hat H_u \,; \\
		& \text{1.c)}~ D^2 ( \hat L \hat L ) \otimes \bead[t] \Big\{ 
			& D^2 \hat H_u \otimes \left\{ D^2 \hat H_u , \bar D^2 \hat H^\dagger_d \right\} , \bar D^2 \hat H^\dagger_d \bar D^2 \hat H^\dagger_d , \\
			& D^2 ( \hat H_u \hat H_u ) , \bar D^2 ( \hat H^\dagger_d \hat H^\dagger_d ) \Big\} \cup D^2 ( \hat L \hat L \hat H_u \hat H_u ) \cup \text{1.a} \cup \text{1.b} \,; 
		\eead \\
		& \text{2.b)}~ \hat L \hat L \otimes \bead[t]
			\Big\{ & \hat H_u \otimes \left\{ \hat H_u , D^2 \hat H_u , \bar D^2 \hat H^\dagger_d \right\} , D^2 \hat H_u \otimes \left\{ D^2 \hat H_u , \bar D^2 \hat H^\dagger_d \right\} , \\
			& \bar D^2 \hat H^\dagger_d \bar D^2 \hat H^\dagger_d , D^2 ( \hat H_u \hat H_u ) , \bar D^2 ( \hat H^\dagger_d \hat H^\dagger_d ) \Big\} \,. 
		\eead
		\label{eq:HuHd_dim5_candidates}
	\eeald

\subsection{Are there models in which the pure-$\nSUSYEWSB$ subset of $\OPnu$ is empty?} \label{sec:empty}

	Since every $\SOP \in \SOPnu$ has $U(1)_Y$ and $SU(2)_L$ charges 
	flowing in internal lines, one might be tempted to think that this alone suffices to show that 
	the subset is always non-empty. Indeed, as insertions of external $\hat V_{U(1)_Y}$ and $\hat V^\alpha_{SU(2)_L}$ 
	into internal lines are allowed, and in particular into loop lines, it is conceivable 
	that any $\SOP \in \SOPnu$ can be promoted to a superoperator 
	that yields a pure-$\nSUSYEWSB$ $\OP \in \OPnu$ by means of judicious appendages 
	of gauge vector superfields $\hat V$ and their chiral projections $D^2 \hat V$ and $\bar D^2 \hat V$. 
	An example of this that we will encounter in~\secref{sec:model_example} is  
	\be
		D^2 ( \hat L \hat L ) \hat H_u \hat H_u \to D^2 ( \hat L \hat L ) \hat H_u \hat H_u \hat V \,, 
	\ee
	which yields dimension-7 operators of the form 
	\bea
		L L H_u H_u H^\dagger H \in \OPnu \,.
	\eea
	However, even though supergraphs with any given number of 
	external $\hat V$'s can be constructed from any underlying $\SOP \in \SOPnu$, 
	the so obtained $\SOP \in \SOPnu$ may vanish as the supergraphs add up to zero. 
	In fact, this happens whenever all charge carrying internal lines undergo a chirality 
	flip that is symmetric w.r.t.\ the local symmetry of which $\hat V$ is 
	the gauge superfield. More generally, $\hat V$'s insertions can be seen to 
	correspond to terms in the $\hat V$-expansion of gauge completed superoperators\footnote{
	For example, $D^2 ( \hat L \hat L ) \hat H_u \hat H_u \hat V_{U(1)_Y}$ is a term 
	in the $\hat V$-expansion of 
	$D^2 ( \hat L \hat L e^{-2 g' Y_L \hat V_{U(1)_Y}} ) \hat H_u \hat H_u e^{-2 g' Y_{H_u} \hat V_{U(1)_Y}}$.}. 

\medskip

\begin{figure}[h!t]
          \vspace{\vspUfig}
          \begin{center}
		\includegraphics[width=160mm]{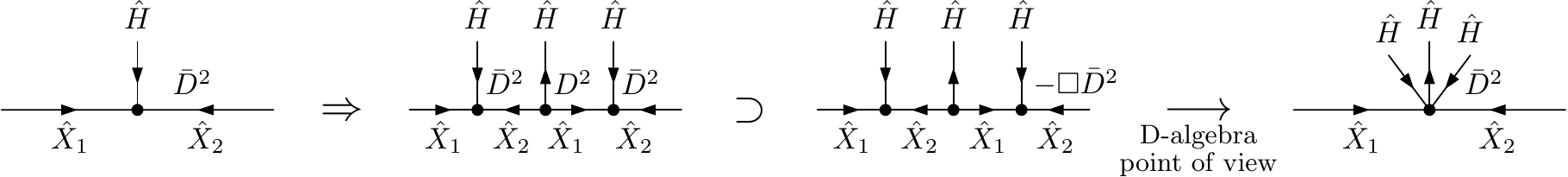}
                \caption{A $\hat H \hat X_1 \hat X_2$ vertex (leftmost diagram) 
			implies a non-vanishing $\hat H \hat H^\dagger \hat H \hat X_1 \hat X_2$ 
			interaction that is local in $\theta$, i.e.\ ``a vertex'' from the D-algebra 
			point of view (rightmost diagram). 
                }\label{fig:SGraph_H_Hdagger_insertion}
          \end{center}
          \vspace{\vspDfig}
        \end{figure}

	Regarding models in which there exists a Higgs bilinear. Pick a $\SOP \in \SOPnu$. 
	Each supergraph contributing to $\SOP$ belongs to one of the following two classes: 
	\begin{itemize}
		\item[a)] at least one external Higgs $\hat H$ (or $\hat H^\dagger$) is locally 
			connected to loop superfields, i.e.\ at least one external Higgs is 1PI; 
		\item[b)] all external Higgses are connected to the loop(s) by means of 
			1PR propagators, i.e.\ all external Higgses are 1PR. 
	\end{itemize}
	Without loss of generality, say that for a particular supergraph belonging to class-a 
	the vertex is $\hat H \hat X_1 \hat X_2$, where $\hat X$'s are loop superfields. One can then 
	see (cf.~\figref{fig:SGraph_H_Hdagger_insertion}) that an insertion 
	of $\hat H^\dagger$ ($\hat H$) followed by an insertion of $\hat H$ ($\hat H^\dagger$) leads to 
	a supergraph for the superoperator 
	\be
		\hat H^\dagger \hat H \, \SOP \,.
	\ee 

\begin{figure}[h!t]
          \vspace{\vspUfig}
          \begin{center}
		\includegraphics[width=144.871mm]{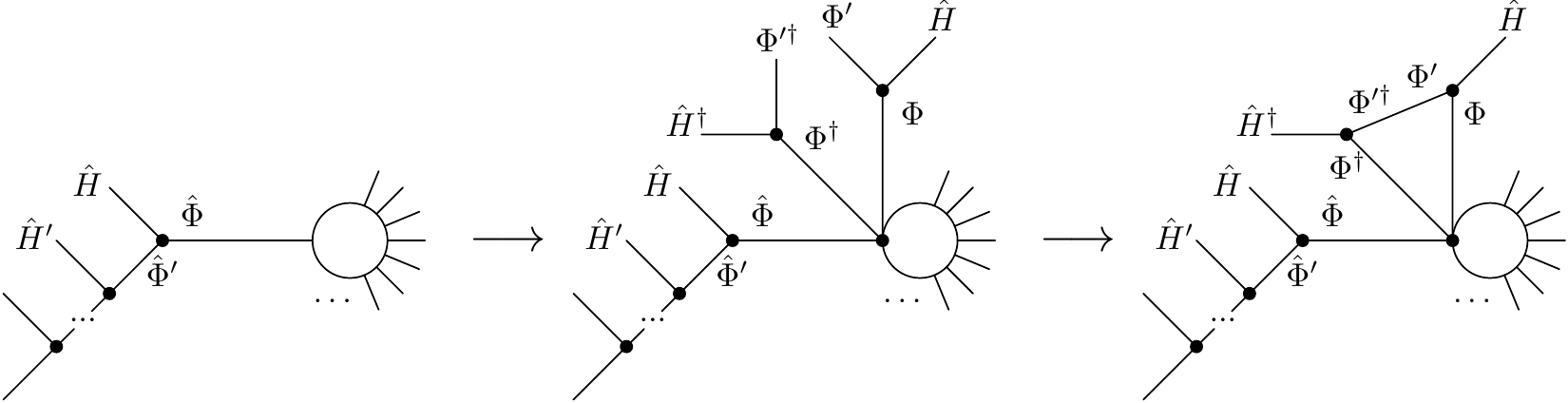}
                \caption{Schematic of a procedure to go from a class-b supergraph for $\SOP$ (leftmost diagram) to a supergraph for 
		$\hat H^\dagger \hat H \, \SOP$ (rightmost diagram) by means of a double insertion in the loop line to which 
		the 1PR leg is attached (middle diagram). 
		The dot at which the lines of $\hat\Phi^\dagger$ and the two $\hat\Phi$'s meet is a vertex in the sense 
		of~\figref{fig:SGraph_H_Hdagger_insertion}. In order to describe all conceivable assignments of chiralities to 
		external and internal superfields, the chiralities of $\hat H$, $\hat H'$, $\hat\Phi$ and $\hat\Phi'$ are left unspecified. 
		However, $\hat H$, $\hat\Phi$ and $\hat\Phi'$ have the same chirality, as is implied by the vertex. 
		Moreover, and so that all conceivable propagators are described, we also do not specify 
		how $\hat\Phi$ is connected to the loop(s) (depicted by the circle), nor how $\hat\Phi'$ is connected 
		to $\hat H'$. 
                }\label{fig:SGraph_H_Hdagger_1PRinsertion}
          \end{center}
          \vspace{\vspDfig}
        \end{figure}

	Each class-b supergraph can also be transformed into a supergraph for $\hat H^\dagger \hat H \, \SOP$, as we proceed to 
	show. Choose some 1PR leg. To be completely general, we take the Higgses along that leg to be $\hat H$, 
	$\hat H'$, ..., where $\hat H$ is attached to the loop(s) by one 1PR propagator, $\hat H'$ by two, and 
	so on along the leg, and the chiralities are left unspecified (for e.g.\ $\hat H$ and $\hat H'$ need not 
	have the same chirality, and $\hat H$ can be either chiral or anti-chiral). This is depicted in the left-hand side 
	supergraph of~\figref{fig:SGraph_H_Hdagger_1PRinsertion}. Let $\hat H \hat \Phi \hat \Phi'$ be the vertex that connects $\hat H$ 
	to the leg, and where $\hat\Phi$ is the superfield that connects $\hat H$ to the loop(s) (depicted by a circle) by 
	either a $\hat\Phi\hat\Phi^\dagger$ or a $\hat\Phi\hat{\overline\Phi}$ propagator. Now, in the same way 
	as a $\hat H^\dagger \hat H$ insertion is performed in~\figref{fig:SGraph_H_Hdagger_insertion}, one can 
	make an insertion of $\hat\Phi^\dagger\hat\Phi$ (or $\hat{\overline\Phi}{}^\dagger\hat{\overline\Phi}$, depending on 
	how $\hat\Phi$ is connected to the loop(s)) in the 
	the loop line to which $\hat\Phi^\dagger$ (or $\hat{\overline\Phi}$) is locally connected. 
	Then, take $\hat\Phi^\dagger$ (or $\hat{\overline\Phi}$) to propagate via $\hat\Phi\hat\Phi^\dagger$ (or $\hat\Phi\hat{\overline \Phi}$) 
	to $\hat\Phi' \hat H$, so that 
	the insertion leads to two additional legs: one with $\hat\Phi' \hat H$ and the other with $\hat\Phi'^\dagger \hat H^\dagger$, 
	as shown in the middle supergraph of~\figref{fig:SGraph_H_Hdagger_1PRinsertion}. 
	Now, by contracting $\hat\Phi'$ with $\hat\Phi'^\dagger$ we arrive at a supergraph 
	(see right-hand side of~\figref{fig:SGraph_H_Hdagger_1PRinsertion}) for the superoperator 
	$\hat H^\dagger \hat H \, \SOP $. 

	The procedures described above can be applied to each class-a or -b supergraph of the set contributing to $\SOP$ 
	up to any given order of perturbation theory. Hence, if class-a or -b supergraphs for superoperator $\SOP$ do 
	not add up to zero, the transformed ones do not add up to zero for $\hat H^\dagger \hat H \, \SOP$ either. Now, if there exists 
	a Higgs bilinear, $\hat H^\dagger \hat H \, \SOP$ yields a pure-$\nSUSYEWSB$ $\OP \in \OPnu$ 
	regardless of $\SOP \in \SOPnu$. We will illustrate this for a particular model
	in~\secref{sec:model_example}. 

\medskip

	On dimensional grounds one expects that the strength of a pure-$\nSUSYEWSB$ $\OP \in \OPnu$ 
	obtained from $\SOP$ by an insertion of $\hat V$ compares to the 
	strength of a pure-$\nSUSYEWSB$ $\OP' \in \OPnu$ obtained from the same superoperator 
	by an insertion of $\hat H^\dagger \hat H$ as 
	\be
		g^2 : \lambda^2 \left(\frac{\mu}{M_X}\right)^{2~\text{or}~1} \,,
	\ee
	for class A or B superoperators, respectively, and where $\lambda$ is the coupling strength of 
	$\hat H$'s to the loop(s). Moreover, if the leading supergraphs for $\SOP$ 
	are of class-b, and the model is such that the only feasible $\hat H^\dagger \hat H$ insertion is 
	by means of the procedure described in~\figref{fig:SGraph_H_Hdagger_1PRinsertion}, then the 
	$\propto \mu / M$ contribution comes with an additional loop suppression factor.

\section{Models in which the leading order subset of $\OPnu$ is proportional to $\nSUSYEWS$} \label{sec:looking}

	A possible strategy to construct models of this kind is the following. Pick a set of superoperators 
	that cannot yield a pure-$\nSUSYEWSB$ $\OP \in \OPnu$ (cf.~\eqref{eq:Type1SUSYbreak} and~\eqref{eq:Type2SUSYbreak}). 
	Choose the LO topologies at which these operators appear. 
	Write the necessary superfields and couplings. As a final step, pick an internal symmetry group 
	that precludes, at least up to the same order of perturbation theory, all superoperators that yield 
	a pure-$\nSUSYEWSB$ $\OP \in \OPnu$. In particular, it is essential that the ``wrong'' Higgs 
	does not communicate (at least up to the same order as the ``right'' Higgs) to the sector that holds 
	$L$-number breaking. To illustrate this, consider for example the one-loop realisation of 1PI 
	$\hat L \hat L \hat H_u \hat H_u$. $\hat H_u$ couples to, say, $\hat X_1 \hat X_2$, where $\hat X_{1,2}$ 
	have mass terms. Without loss of generality let the mass terms be $\hat{\overline X}_i \hat X_i$. 
	Hence, $\hat{\overline X}_1 \hat{\overline X}_2 \hat H_d$ is invariant under non-$R$-symmetries 
	in this phase. If such a term exists in the superpotential, this same model generates the supergraph 
	topology shown in the middle panel of~\figref{fig:LLHH_SOP_bmw}, leading to $\hat L \hat L \hat H_u \hat H^\dagger_d$ 
	which yields a pure-$\nSUSYEWSB$ $\OP \in \OPnu$. 

	We cannot think of any serious obstruction that would compromise this procedure for constructing general models 
	of this kind. In fact, in the next section we give a proof of existence based on a one-loop type-II 
	seesaw, also showing that this kind of models need not be complicated. 

\medskip

\begin{figure}[h!t]
          \vspace{\vspUfig}
          \begin{center}
                \includegraphics[height=60mm]{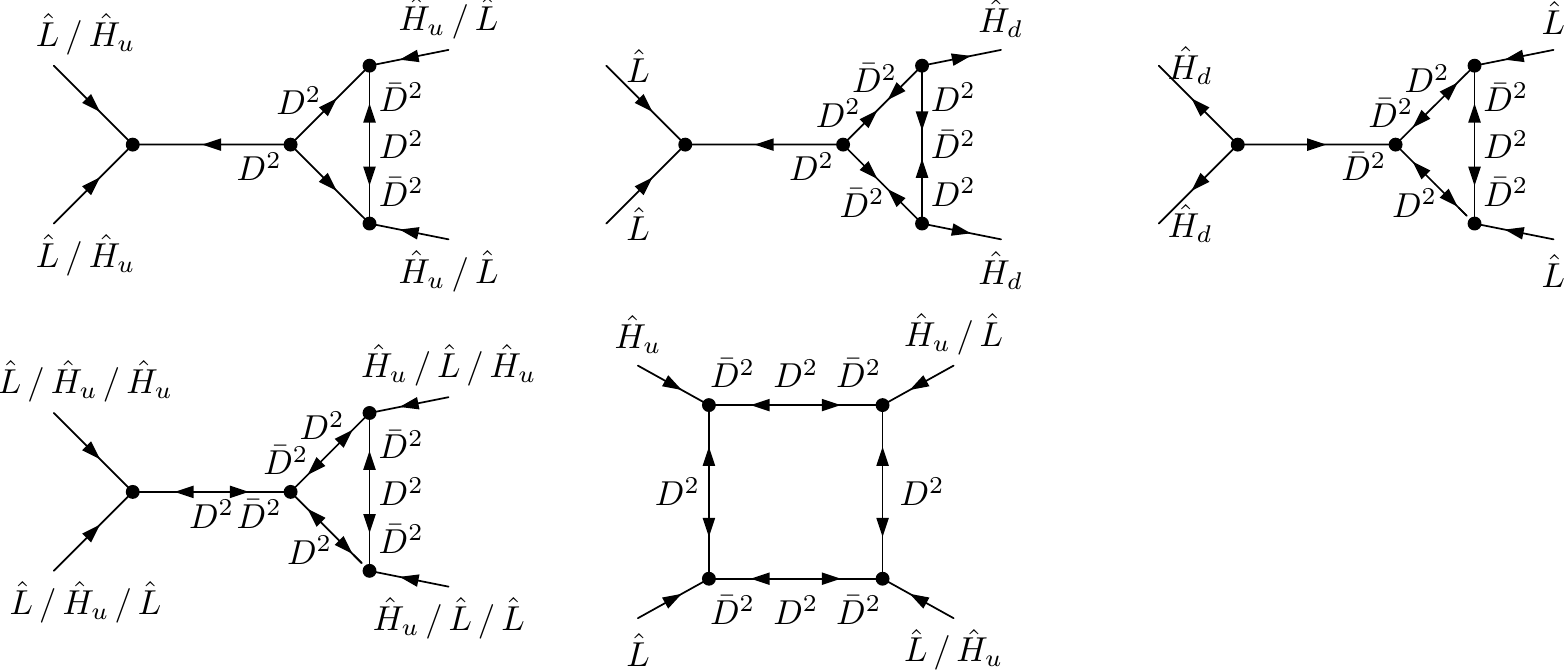} 
        	\caption{One-loop supergraph topologies that are identified in the text. From left to right: 
			$D^2 ( \hat L \hat L ) \hat H_u \hat H_u$ or $\hat L \hat L D^2 ( \hat H_u \hat H_u )$, 
			$D^2 ( \hat L \hat L ) \hat H^\dagger_d \hat H^\dagger_d$, 
			$\hat L \hat L \bar D^2 ( \hat H^\dagger_d \hat H^\dagger_d )$, 
			$\hat L \hat L \hat H_u \hat H_u$ (1PR) and 
			$\hat L \hat L \hat H_u \hat H_u$ (1PI). 
                }\label{fig:SGraph_1LoopTop_LLHH}
          \end{center}
          \vspace{\vspDfig}
        \end{figure}

	Under the assumption of a standard set of Higgses ($\hat H_{u,d}$), the simplest models of this kind 
	are those that generate, at the one-loop order, superoperators that were identified in~\eqref{eq:HuHd_dim5_candidates}. 
	From D-algebra considerations, and relegating topologies with self-energies to~\appref{app:susy_break_self}, 
	one obtains the following list of possibilities\footnote{
	A systematic method to derive this list is the following. 
	The class of one-loop 4-point supergraph topologies with a one-loop vertex can be partitioned w.r.t.\ the $4$ possible types of 
	1PR propagators: $\hat\Phi\hat\Phi^\dagger$, its $\Hc$, $\hat\Phi\hat{\overline \Phi}$ and 
	its $\Hc$. Of these topologies, only $3+1+3+1$ (partitioned as mentioned) can underlie an $\SOP \in \SOPnu$ 
	as a consequence of requiring at least two external chiral lines that will be identified as a pair of 
	$\hat L$'s. Of these, only $2+1+1+0$ can underlie a superoperator listed in~\eqref{eq:HuHd_dim5_candidates}. 
	These $2+1+1+0$ topologies can be identified by the superoperators 
	$D^2 ( \hat A \hat B ) \hat C \hat D$, $D^2 ( \hat A \hat B ) \hat C^\dagger \hat D^\dagger$, 
	$\bar D^2 ( \hat A^\dagger \hat B^\dagger ) \hat C \hat D$ and $\hat A \hat B \hat C \hat D$, respectively. 
	Regarding irreducible topologies: only $3$ have at least two external chiral lines and, of these, 
	only $1$ can underlie a superoperator listed in~\eqref{eq:HuHd_dim5_candidates}.}:
	\begin{itemize}
		\item $D^2 ( \hat L \hat L ) \hat H_u \hat H_u$, $\hat L \hat L D^2 ( \hat H_u \hat H_u )$, 
		$D^2 ( \hat L \hat L ) \hat H^\dagger_d \hat H^\dagger_d$ and $\hat L \hat L \bar D^2 ( \hat H^\dagger_d \hat H^\dagger_d )$ \\ -- type-II without a chirality flip;
		\item $\hat L \hat L \hat H_u \hat H_u$ (1PR) \\ -- type-II with a chirality flip, type-I and -III;
		\item $\hat L \hat L \hat H_u \hat H_u$ (1PI). 
	\end{itemize}
	The corresponding supergraph topologies are depicted in~\figref{fig:SGraph_1LoopTop_LLHH}. Notice that we 
	populate the supergraphs with $D$'s in a manner that makes the non-trivial 1PI part separable. Moreover, when doing the 
	D-algebra, we integrate by parts the $D$'s in a way that avoids crossing over the non-trivial 1PI part. The usefulness of 
	this procedure is in allowing to associate superoperators to whole 1PR supergraphs, even when the result of some 
	of their 1PI parts is zero in the SUSY limit. This works by extending the $d^4\theta$ integration of the non-trivial 1PI part 
	to a $d^4\theta$ integration that encompasses all external superfields. To illustrate what we mean, consider the second 
	supergraph topology, and let $\hat\Phi\hat\Phi^\dagger$ be the 1PR propagator. If, after doing the 
	loop's D-algebra, we integrated by parts the $D^2$ that lies over the 1PR line to the right, we would obtain 
	$\hat L \hat L D^2 ( \hat H^\dagger_d \hat H^\dagger_d ) = 0$. However, as we integrate it to 
	the left, we end up with $D^2 ( \hat L \hat L ) \hat H^\dagger_d \hat H^\dagger_d$. With this procedure 
	the zero of the non-trivial 1PI part, i.e.\ $\int d^4\theta \, \hat\Phi^\dagger \hat H^\dagger_d \hat H^\dagger_d = 0$, 
	is transferred to $\int d^4\theta \, D^2 ( \hat L \hat L ) \hat H^\dagger_d \hat H^\dagger_d = 0$. 

	The subcase $\hat L \hat L D^2 ( \hat H_u \hat H_u )$ of the first topology, i.e.\ in which $\hat H_u \hat H_u$ 
	is coupled to the 1PR propagator (say $\hat\Phi\hat\Phi^\dagger$), contains an example of the trilinear case 
	discussed in~\secref{sec:nSUSYEWSB}. To be precise, its non-trivial 1PI part gives 
	\be
		\frac{1}{M} \int d^4\theta \hat L \hat L \hat\Phi^\dagger \supset \frac{1}{M} L L F^\dagger_\Phi \,,
	\ee
	and since (cf.~\eqref{eq:VEV_FZD_3rdORD} and let $\lambda$ be the $\hat\Phi\hat H_u \hat H_u$ superpotential coupling) 
	\be
		\langle F^\dagger_\Phi \rangle = \frac{\Msoft^2}{M_\Phi^2} \lambda \langle H_u H_u \rangle + \frac{\lambda B^*}{M_\Phi^3} \left( A \langle H_u H_u \rangle + 2 \mu^* \langle H_u H^\dagger_d \rangle \right) + \mathcal{O}\left(\frac{\Msoft^4}{M^4_\Phi}\right) \,, \label{eq:VEV_FPhiD}
	\ee
	it effectively generates $L L H_u H_u$ and $L L H_u H^\dagger_d$. 

	To study how $\nSUSY$ effects upon these topologies can generate an $\SOP$ which yields an $\OP \in \OPnu$, 
	we include soft-$\nSUSY$ in supergraph calculations 
	by means of the following\footnote{
	We disregard non-holomorphic soft-$\nSUSY$ trilinears as naive dimensional 
	analysis indicates that they are suppressed by $\Msoft / M_X$ w.r.t.\ $A$, $\sqrt B$ and 
	$\Msoft$. 
	} non-chiral vertices with $\nSUSY$ spurions ($\hat X \sim M_X^2 \theta^2$): 
	\beald
		& \lambda A \phi_1 \phi_2 \phi_3 = \lambda \int d^4\theta \left(\frac{X^\dagger X}{M_X^3}\right)_A \hat\Phi_1 \hat\Phi_2 \hat\Phi_3 \,, \\
		& B \phi_1 \phi_2 = \int d^4\theta \left(\frac{X^\dagger X}{M_X^2}\right)_B \hat\Phi_1 \hat\Phi_2 \,, \\
		& \Msoft^2 \phi_1^\dagger \phi_1 = \int d^4\theta \left(\frac{X^\dagger X}{M_X^2}\right)_{\Msoft^2} \hat\Phi_1^\dagger \hat\Phi_1 \,. \label{eq:SoftSUSYbreak_vertices}
	\eeald
	We note that this form for $A$- and $B$-terms is equivalent to (d) and (b) of~\cite{Girardello:1981wz}, respectively, 
	since $\int d^4\theta \hat X^\dagger \hat X \, \SOP \sim \int d^2\theta \hat X \, \SOP$ (\footnote{ 
	In spite of this, one could still be suspicious on whether our parameterisation for 
	holomorphic soft-$\nSUSY$ is actually soft, since the $A$-term vertex gives three factors 
	of $\bar D^2$, whereas only a maximum of four $\D{\alpha}$ or $\Dbar{\alpha}$ is compatible 
	with the renormalisability criterion for softness. To see that it is, notice that any sub-graph 
	in which one of these $\bar D^2$ is not absorbed by $\hat X^\dagger$ vanishes identically 
	as there is a $\bar D^2$ factor on every internal line attached to the vertex. Similarly, 
	non-vanishing sub-graphs with a $B$-term are those in which the $B$ is seen to introduce 
	only a factor of $\bar D^2$.}). 
	The complete list of $\nSUSY$ insertions that yield an $\OP \in \OPnu$ reads 
	\beald
		& \int d^4\theta \left[ D^2 ( \hat X^\dagger \hat X )~\text{or}~ \hat X^\dagger \hat X \right] \left\{ D^2 \bar D^2 ( \hat X^\dagger \hat X ) \right\}^n D^2 ( \hat L \hat L ) \hat H_u \hat H_u \,, \\
		& \int d^4\theta \, \left\{ D^2 \bar D^2 ( \hat X^\dagger \hat X ) \right\}^n \hat L \hat L \, \Big\{ \bead[t] 
				& D^2 ( \hat X^\dagger \hat X \hat H_u \hat H_u ) , \\
				& D^2 ( \hat X^\dagger \hat X ) \left[ D^2 ( \hat H_u \hat H_u ) ~\text{or}~ D^2 ( \hat H_u \hat H_u \bar D^2 ( \hat X^\dagger \hat X ) ) \right] \Big\} \,, \eead \\
		& \int d^4\theta \left[ \bar D^2 ( \hat X^\dagger \hat X )~\text{or}~ \hat X^\dagger \hat X \right] \left\{ D^2 \bar D^2 ( \hat X^\dagger \hat X ) \right\}^n D^2 ( \hat L \hat L ) \hat H_d^\dagger \hat H_d^\dagger \,, \\
		& \int d^4\theta \, D^2 ( \hat X^\dagger \hat X ) \left\{ D^2 \bar D^2 ( \hat X^\dagger \hat X ) \right\}^n \hat L \hat L \left[ \bar D^2 ( \hat H_d^\dagger \hat H_d^\dagger )~\text{or}~ \bar D^2 ( \hat H_d^\dagger \hat H_d^\dagger D^2 ( \hat X^\dagger \hat X ) ) \right] \,, \\
		& \int d^4\theta \, D^2 ( \hat X^\dagger \hat X ) \left\{ D^2  \bar D^2 ( \hat X^\dagger \hat X ) \right\}^n \hat L \hat L \hat H_u \hat H_u \,, \label{eq:full_XdaggerX_insertions}
	\eeald
	where $n = 0,1,...$ stands for the number of insertions of $D^2 \bar D^2 ( \hat X^\dagger \hat X )$. 

	A soft-$\nSUSY$ insertion into a (anti-)chiral vertex, i.e.\ an $A$-term, introduces an extra $\hat X$ ($\hat X^\dagger$, 
	respectively) factor in the corresponding supergraph. Hence, D-algebra considerations reveal that a single soft-$\nSUSY$ 
	insertion of an $A$-term can generate an $\OP \in \OPnu$ only in the case of a 
	type-II seesaw without a chirality flip, i.e.\ the first topology of~\figref{fig:SGraph_1LoopTop_LLHH}, 
	and which leads to 
	\be
		\frac{A^* \mu^*}{M^3} L L H_u H^\dagger_d \subset \frac{1}{M^3} \int d^4\theta \, D^2 \left( \frac{\hat X^\dagger \hat X}{M_X^3} \right)_{A^*} \left[ D^2 ( \hat L \hat L ) \hat H_u \hat H_u ~\text{or}~ \hat L \hat L D^2 ( \hat H_u \hat H_u ) \right] \,.
	\ee
	For a detailed catalogue up to order $3$ in the scale of soft-$\nSUSY$ ($\Msoft$) 
	see~\appref{app:susy_break}. It is important to notice that $\nSUSY$-insertions 
	into the supergraph underlying the superoperator $\hat L \hat L D^2 ( \hat H_u \hat H_u )$ 
	do yield the $\langle F^\dagger_\Phi \rangle$ contribution mentioned in~\eqref{eq:VEV_FPhiD}. 
	Indeed, the terms in~\eqref{eq:VEV_FPhiD} correspond respectively to the following entries 
	of~\tbref{tb:SUSY_breaking_insertions_BDEL_MDEL2_D2LLHuHu_LLD2HuHu}: the 5th row of the second 
	table and the 4th and 1st rows of the first table. 

	From the tables in~\appref{app:susy_break} three different kinds of leading dimensionful suppression factors are found: 
	\begin{itemize}
		\item $\mu \, \Msoft / M^3$ or $\Msoft^2 / M^3$ -- $D^2 ( \hat L \hat L ) \hat H_u \hat H_u$ and $\hat L \hat L D^2 ( \hat H_u \hat H_u )$; 
		\item $\mu \, \Msoft^2 / M^4$ or $\Msoft^3 / M^4$ -- $\hat L \hat L \bar D^2 ( \hat H^\dagger_d \hat H^\dagger_d )$; 
		\item $\Msoft^2 / M^3$ -- $D^2 ( \hat L \hat L ) \hat H^\dagger_d \hat H^\dagger_d$ and $\hat L \hat L \hat H_u \hat H_u$ (both 1PR and 1PI). 
	\end{itemize}
	The absence of a contribution linear in $\Msoft$ for some topologies is most easily seen to 
	stem from the fact that one-loop topologies for $\hat L \hat L \hat H_u \hat H_u$, as well as the 
	one-loop 1PI parts of $D^2 ( \hat L \hat L ) \hat H^\dagger_d \hat H^\dagger_d$ and 
	$\hat L \hat L \bar D^2 ( \hat H^\dagger_d \hat H^\dagger_d )$, use vertices of a single chirality. 
	Moreover, and in regard to $\hat L \hat L \bar D^2 ( \hat H^\dagger_d \hat H^\dagger_d )$, the 
	leading contributions from the $\bar D^2 ( \hat H^\dagger_d \hat H^\dagger_d )$ piece are 
	$\mu H_u H^\dagger_d$ and $A^* H^\dagger_d H^\dagger_d$. 

	In~\appref{app:susy_break_self}, where we conduct a similar analysis for one-loop realisations with self-energies, we 
	find that these too have leading dimensionful suppression factors that range from $\mu \, \Msoft / M^3$ or $\Msoft^2 / M^3$ 
	to $\mu \, \Msoft^2 / M^4$ or $\Msoft^3 / M^4$. 

	If we take $\mu \sim \Msoft$, we can conclude that in one-loop models of 
	this kind $L L H H$ operators have a dimensionful suppression of at least $\Msoft^2 / M^3$. 
	This result is naively expected for type-II seesaws without a chirality flip, since 
	$\int d^4\theta D^2 ( \hat L \hat L ) \hat H \hat H$ has mass dimension $7$. 
	For other realisations this dependence is not trivial, since for an underlying 
	superoperator $\hat L \hat L \hat H \hat H$ one in general expects a 
	$\Msoft / M^2$ dependence, as was indeed found in~\secref{sec:models_literature}. 

\begin{figure}[h!t]
          \vspace{\vspUfig}
          \begin{center}
		\begin{tabular}{lcr}
                	\includegraphics[width=46.5mm]{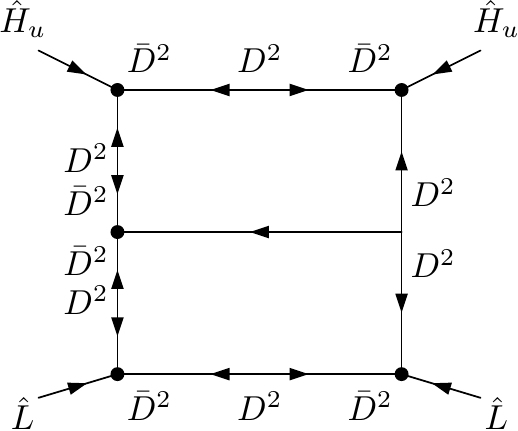} & \hspace{0.2in} & 
                	\includegraphics[width=100mm]{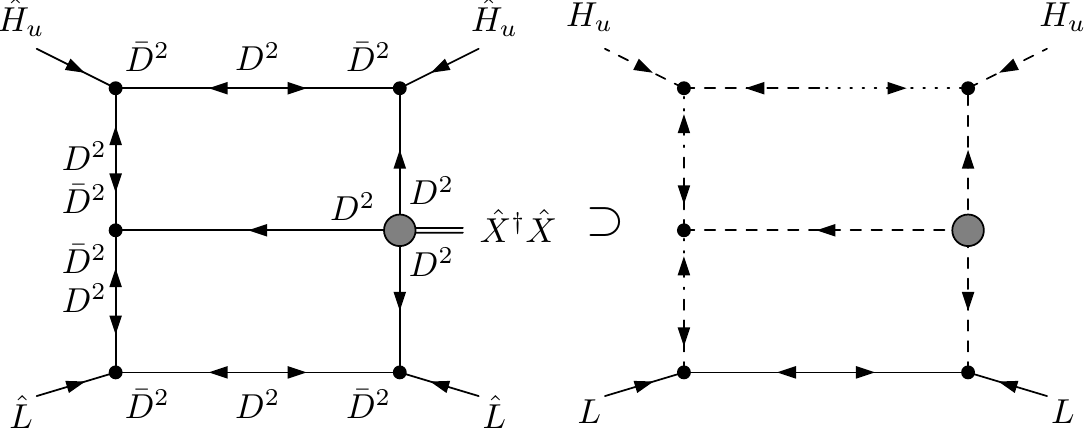}
		\end{tabular}
                \caption{Example of a two-loop supergraph for superoperator $\hat L \hat L \hat H_u \hat H_u$ (left) 
		which yields an $\OP \in \OPnu$ by means of a single $A$-term insertion (right). 
                }\label{fig:SGraphLLHuHu_twoloopEg}
          \end{center}
          \vspace{\vspDfig}
        \end{figure}

	The dimensionful suppression $\mu \, \Msoft / M^3$ or $\Msoft^2 / M^3$ does not hold at 
	higher loops. For instance, consider $\hat L \hat L \hat H_u \hat H_u$ generated 
	by the 1PI two-loop topology shown in the left-hand side of~\figref{fig:SGraphLLHuHu_twoloopEg}. A single 
	$A$-term insertion (depicted as a grey blob, on the right) leads to 
	\be
		\frac{1}{M^2} \int d^4\theta \, D^2 \left( \frac{\hat X^\dagger \hat X}{M_X^3} \right)_{A^*} \hat L \hat L \hat H_u \hat H_u 
		\supset \frac{A^*}{M^2} L L H_u H_u \,.
	\ee

\section{A model example} \label{sec:model_example}

	Looking at the one-loop topology for $D^2 ( \hat L \hat L ) \hat H_u \hat H_u$ 
	(cf.~\figref{fig:SGraph_1LoopTop_LLHH}) we see that the 
	most general set of scalar superfields and superpotential terms involved is $7$ and 
	$5$ ($4$ trilinears and $1$ bilinear), respectively. The subset of $U(1)^7$ (acting independently on each scalar superfield)  
	under which the $5$ terms are invariant consists of the hypercharge and a new 
	$U(1)_X$ charge carried by the superfields in the loop (say $\hat X$'s). These 
	are responsible for communicating $L$-number breaking to the SM leptons via 
	the exchange of a type-II seesaw mediator, $\hat \Delta$. 

	Since $\hat \Delta$ must be massive, the only way 
	by which the coupling $\hat \Delta^\dagger \hat H_u \hat H_u$ can be made to be genuinely 
	radiative is by linking it to the VEV of a superoperator of at least dimension $4$ in 
	superfields. One simple example is 
	\be
		\hat \rho^\dagger \hat \Delta^\dagger \hat H_u \hat H_u \to 
		\langle \rho^\dagger \rangle \hat \Delta^\dagger \hat H_u \hat H_u + \hat \rho^\dagger \hat \Delta^\dagger \hat H_u \hat H_u \,.
	\ee
	This is similar to the procedure described in~\cite{Bonnet:2012kz} 
	to prevent a 1PR seesaw from having a tree-level contribution 
	and which in an ordinary QFT only works for type-I and -III topologies. It 
	can be successfully applied to the type-II topology in a SUSY setting because 
	renormalisable four-scalar interactions can be genuinely radiative 
	in SUSY (see~\appref{app:radiative_couplings}).	
	To understand this result, we note the following. In order for the $\chi\chi\phi$ 
	interaction to be genuinely radiative, and thus realise a radiative type-I 
	or -III seesaw, it must arise from some symmetric operator that is not present at 
	tree-level in the UV complete model. Only non-renormalisable operators satisfy this 
	criterion. Thus, if one builds a model in which $\chi\chi\phi\phi'$ is not generated 
	at tree-level (this can always be done) and $\phi'$ gets a symmetry breaking VEV, in the 
	broken phase we obtain the so desired radiative coupling. 
	(The way by which this is done in~\cite{Bonnet:2012kz} is to consider 
	that $\phi'$ is attached to an internal spinor line of an underlying 1PI one-loop 
	topology for $\chi\chi\phi\phi'$.) 
	In an ordinary QFT this cannot work for a target $\phi^3$ from a symmetric $\phi^3\phi'$ 
	because $\phi^3\phi'$, being renormalisable, must be present at tree-level in the UV 
	complete model. 

\begin{figure}[h!t]
          \vspace{\vspUfig}
          \begin{center}
                \begin{tabular}{ccc}
			\includegraphics[width=45mm]{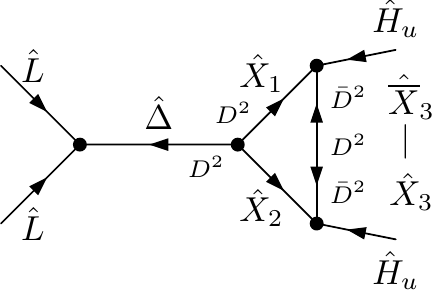} & \hspace{0.3in} & 
        	        \includegraphics[width=45mm]{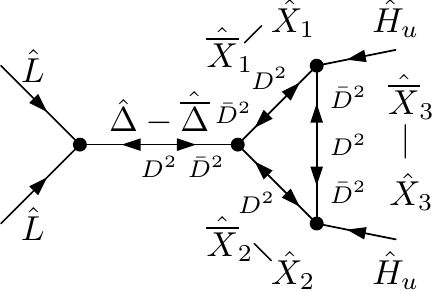} 
		\end{tabular}
                \caption{Leading order subset of $\SOPnu$ 
			in the model example. 
                }\label{fig:SGraph_LLHH_model_example}
          \end{center}
          \vspace{\vspDfig}
        \end{figure}

	We will assume that this is achieved by a $U(1)$ $L$-number symmetry that is 
	broken by the VEV of the scalar component of $\hat\rho$. Since $L$-number breaking 
	is communicated by $X$'s, the simplest choice is to consider that they 
	couple directly to $\hat\rho$. We remain agnostic as to what drives $\langle \rho \rangle \neq 0$. 
	Furthermore, the simplest holomorphy compliant 
	choice is to make a $\hat\rho^\dagger$ insertion in the loop line where chirality 
	flips, so that the mass term originates from $L$-number breaking. We thus arrive 
	at the left-hand side diagram of~\figref{fig:SGraph_LLHH_model_example}. Even though the topology 
	does not require $\hat X_1$ and $\hat X_2$ to have mass terms, we will assume that 
	they do have $\hat X \hat{\overline X}$ mass terms already at the $L$-number symmetric phase. 

	The model is thus summarised in~\tbref{tb:model_example} and its most general renormalisable 
	superpotential reads\footnote{Although not relevant to our analysis, for definiteness we assume that the 
	$\hat u^c \hat d^c \hat d^c$ term is forbidden by, for instance, R-parity or baryon number conservation.}
	\bea
		\mathcal{W} & := & \mathcal{W}_\text{MSSM} 
		+ M_\Delta \hat \Delta \hat{\overline \Delta} 
		+ \sum_{i = 1}^2 M_{X_i} \hat X_i \hat{\overline X}_i 
		+ \lambda \hat \rho \hat X_3 \hat{\overline X}_3 \nonumber\\
		&& + \hat H_u \left( \lambda_1 \hat X_1 \hat{\overline X}_3 + \lambda_2 \hat X_2 \hat X_3 \right) 
		+ \hat \Delta \left( \lambda_L \hat L \hat L + \lambda_X \hat X_1 \hat X_2 \right) 
		+ \bar \lambda_X \hat{\overline \Delta} \, \hat{\overline X}_1 \hat{\overline X}_2 \,. 
		\label{eq:SPot_model_example}
	\eea
	(Conventions regarding $SU(2)_L$ contractions are given in~\appref{app:model_example}.) 
	In the absence of the last term the model acquires the $R$-symmetry shown in the last column 
	of~\tbref{tb:model_example}. This term allows for a chirality flipped type-II seesaw 
	of superoperator $\hat L \hat L \hat H_u \hat H_u$, as shown in the right-hand side supergraph 
	of~\figref{fig:SGraph_LLHH_model_example}. The broken $L$-number phase corresponds to 
	\be
		\lambda \hat\rho \hat X_3 \hat{\overline X}_3 \to M_{X_3} \hat X_3 \hat{\overline X}_3 + \lambda \hat\rho \hat X_3 \hat{\overline X}_3 \,,~~~ M_{X_3} := \lambda \langle \rho \rangle \,.
	\ee

	\begin{table}[h!t]
		\centering
		\renewcommand{\arraystretch}{1.4}
		\begin{tabular}{|c|c|c|c|c|c|} \hline
					& $SU(2)_L \otimes U(1)_Y$ 	& $U(1)_X$ 	& $U(1)_L$ 	& $U(1)_R$ 	\\\hline
			$\hat \Delta$	& $(\mathbf{3},1)$		& $0$		& $-2$ 		& $4$		\\\hline
			$\hat \rho$	& $(\mathbf{1},0)$		& $0$		& $2$		& $0$		\\\hline
			$\hat X_1$ 	& $(\mathbf{2},-1/2)$		& $1$		& $1$		& $-2$		\\\hline
			$\hat X_2$	& $(\mathbf{2},-1/2)$		& $-1$		& $1$		& $0$		\\\hline
			$\hat X_3$	& $(\mathbf{1},0)$		& $1$		& $-1$		& $0$		\\\hline
			$\hat{\overline X}_3$ & $(\mathbf{1},0)$	& $-1$		& $-1$		& $2$		\\\hline
		\end{tabular}
		\caption{Extension of the MSSM in the model example. 
			We omitted the conjugates of $\hat \Delta$ and $\hat X_{1,2}$. $U(1)_R$ stands for an $R$-symmetry 
			that is acquired as $\bar\lambda_X \to 0$.}\label{tb:model_example} 
	\end{table}

	It is now convenient to notice that, as any coupling in $\{\lambda_1,\lambda_2,\lambda_L\}$, or both 
	$\lambda_X$ and any in $\{\bar\lambda_X, M_\Delta, M_{X_1}, M_{X_2}\}$, goes to zero the model recovers 
	a $L$-number symmetry, any superoperator that breaks $L$-number must be proportional to 
	\be
		\mathbf{a} := \lambda_1\lambda_2\boldsymbol{\lambda_L}\lambda_X^* ~~\text{or}~~ M_\Delta M_{X_1} M_{X_2} \, \mathbf{b} := \lambda_1\lambda_2\boldsymbol{\lambda_L}\bar\lambda_X M_\Delta M_{X_1} M_{X_2} \,. 
	\ee
	Hence, the set of LO (w.r.t.\ perturbation theory only, i.e.\ disregarding hypothetical 
	hierarchies among couplings or masses) superoperators that break 
	$L$-number proceed from the two supergraphs of~\figref{fig:SGraph_LLHH_model_example} (and 
	no others) and are 
	\bea
		D^2 ( \hat L \hat L ) \hat H_u \hat H_u \,, ~~~ \hat L \hat L \hat H_u \hat H_u \,.
	\eea
	In the $\pext \to 0$ limit the LO coefficients are given by 
	\be
		-\left(\frac{\mathbf{a} M_{X_3}}{32 \pi^2 M_\Delta^2}\right) C_0 \,,~~~~ 
		\left(\frac{\mathbf{b} M_{X_3} M_{X_1} M_{X_2}}{32 \pi^2 M_\Delta}\right) D_{0,0} \,,
	\ee
	respectively, and where $C_0$ and $D_0$ are abbreviations of scalar one-loop 3- and 4-point integrals, 
	respectively, as defined in~\appref{app:model_example}.
	In the SUSY limit LO $L$-number breaking is thus 
	\bea
		\int d^4\theta \, D^2 ( \hat L \hat L ) \hat H_u \hat H_u 
			& = & -\Box ( \tilde L \tilde L ) \left[ \tilde H_u \tilde H_u + 2 F_{H_u} H_u \right] 
				- \Box ( H_u H_u ) \left[ L L + 2 F_L \tilde L \right] \nonumber\\
			&& + 4 \, (p_L + p_{\tilde L})^2 \, L \tilde H_u \tilde L H_u \,, 
	\eea
	while $\int d^4\theta \hat L \hat L \hat H_u \hat H_u = 0$. Hence, we see that there is 
	no pure-$\nSUSYEWSB$ contribution to neutrino masses. 
	An equivalent way to arrive at this conclusion is the following. 
	Of the two supergraphs, only the first has a non-vanishing (non-trivial) 1PI part. 
	It reads 
	\be
		\int d^4\theta \, \hat\Delta^\dagger \hat H_u \hat H_u = 
			2 \tilde\Delta^{\dagger\dot\alpha} ( p_{\tilde H_u}+p_{H_u} )_{\beta\dot\alpha} {\tilde H}_u^\beta H_u 
			+ F^\dagger_\Delta \left( \tilde H_u \tilde H_u + 2 F_{H_u} H_u \right) 
			- \Delta^\dagger \Box ( H_u H_u ) \,.
	\ee
	Then, by adding to the classical Lagrangian these operators, one sees 
	that $\langle F_{H_u} \rangle = \mu^* \langle H^\dagger_d \rangle \neq 0$ generates 
	a tadpole contribution to $F^\dagger_\Delta \supset M_\Delta {\overline \Delta}$. 
	Thus, ${\overline \Delta}$ acquires a VEV. However, as there is no mixing between 
	${\overline \Delta}$ and $\Delta$, this VEV is inconsequential for neutrino masses. 
	On the other hand, when $\nSUSYEWS$ contributions are considered, 
	$\langle {\overline \Delta} \rangle \neq 0$ will give a contribution to neutrino masses 
	by means of the soft-$\nSUSY$ term $B_\Delta \Delta {\overline \Delta}$. We will comment 
	on this below. 

\begin{figure}[h!t]
          \vspace{\vspUfig}
          \begin{center}
		\includegraphics[width=150mm]{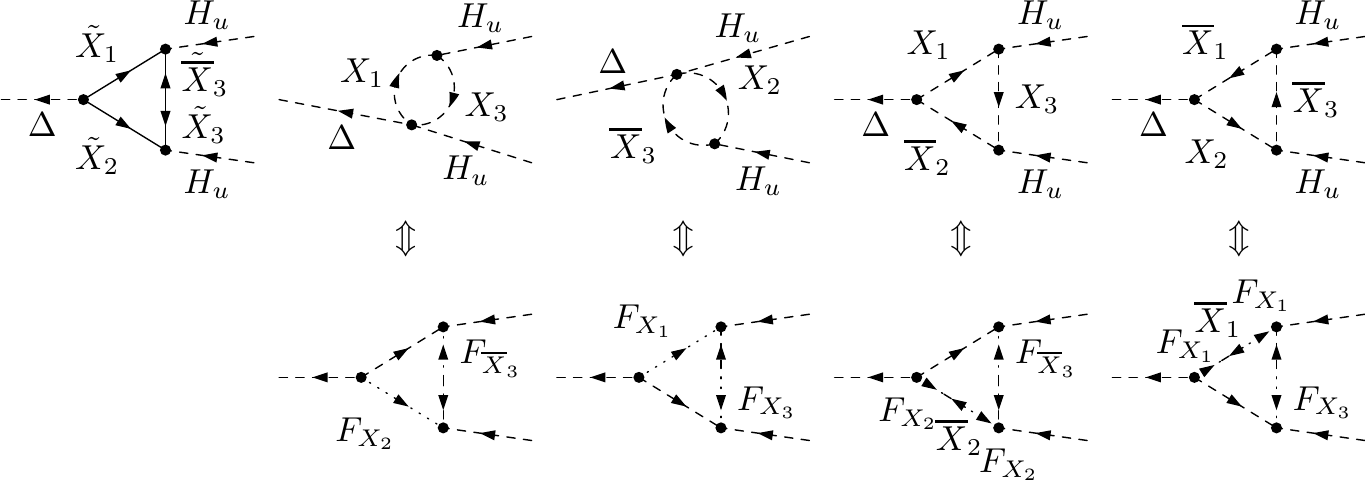}  
                \caption{Leading order diagrams for $\Delta^\dagger H_u H_u$ 
			in the model example. 
                }\label{fig:DeltaHH_model_example}
          \end{center}
          \vspace{\vspDfig}
        \end{figure}

	It is instructive to illustrate in terms of component fields why there is no 
	pure-$\nSUSYEWSB$ contribution to $L L H H$. 
	In order to yield $L L H H$, the first supergraph of~\figref{fig:SGraph_LLHH_model_example} necessitates the 
	three-scalar coupling $\Delta^\dagger H_u H_u$. There are three topologies 
	contributing to this coupling at LO: two with scalars in the loop and the other 
	with spinors (see \figref{fig:DeltaHH_model_example}). In the $\pext \to 0$ limit 
	the latter cancels the former exactly. 
 	Another way to look at this result is the following. If one draws diagrams 
	for $\Delta^\dagger H_u H_u$ using auxiliary fields -- so that holomorphy becomes 
	more transparent -- one concludes that there does not exist a single diagram that 
	is simultaneously holomorphy compliant and has at least an external $F^\dagger - F$ pair. 
	Moreover, all such diagrams that are holomorphy compliant can be paired in sets in such a 
	way that a set with scalar loops is matched to a set with spinor loops and an exact cancellation 
	in the $\pext \to 0$ limit is operative. 
	Regarding the second supergraph, it necessitates $F_{\overline\Delta} H_u H_u$ but 
	no holomorphy compliant diagram for $F_{\overline\Delta} H_u H_u$ can be drawn.

\medskip

\begin{figure}[h!t]
		\vspace{\vspUfig}
		\begin{center}
                	\begin{tabular}{cccc}
				\includegraphics[width=40mm]{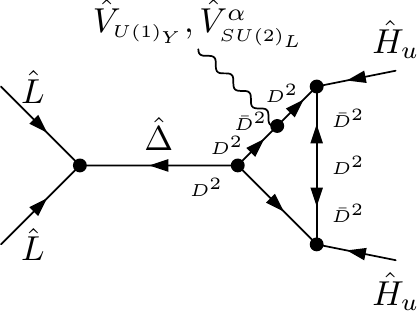} & 
        		        \includegraphics[width=40mm]{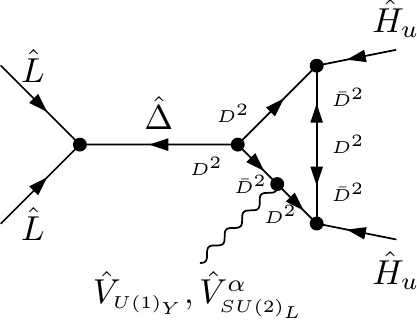} &  
        		        \includegraphics[width=40mm]{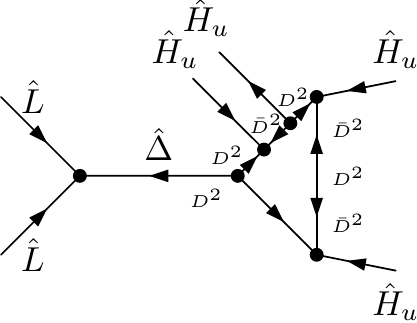} &
        		        \includegraphics[width=40mm]{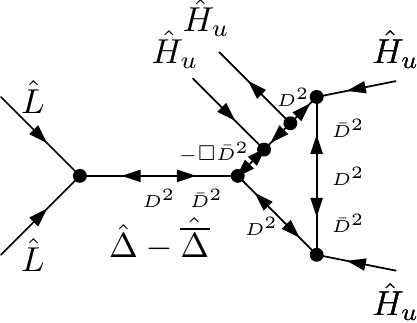} \\
        		        \includegraphics[width=40mm]{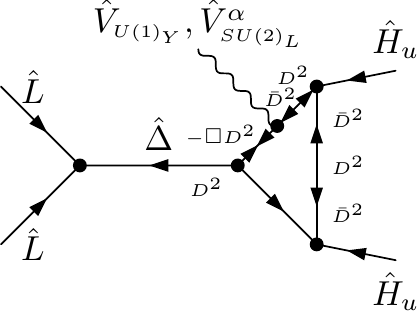} & 
				\includegraphics[width=40mm]{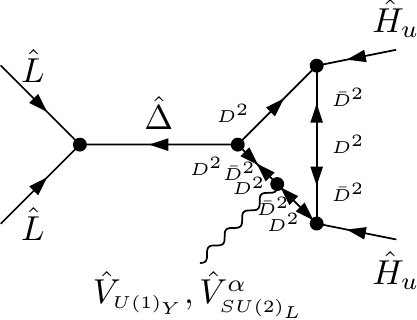} & 
				\includegraphics[width=40mm]{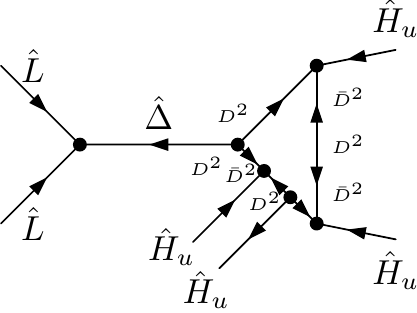} &
				\includegraphics[width=40mm]{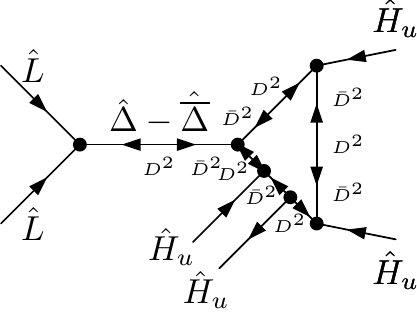}
			\end{tabular}
			\caption{Leading order supergraphs for the pure-$\nSUSYEWSB$ 
				subset of $\OPnu$ in the model example. 
			}\label{fig:SuperGraph_Gauge_dim7_model_example}
		\end{center}
		\vspace{\vspDfig}
	\end{figure}

	By recalling the discussion in~\secref{sec:empty}, one can see that the pure-$\nSUSYEWSB$ subset of $\OPnu$ comprises 
	at LO the dimension-7 operators generated by the supergraphs depicted in~\figref{fig:SuperGraph_Gauge_dim7_model_example}. 
	(Insertions of gauge vector superfields into the second supergraph of~\figref{fig:SGraph_LLHH_model_example} have been omitted as they add 
	up to zero, cf.~\secref{sec:empty}.) They generate the superoperators 
	\be
		D^2 ( \hat L \hat L ) \hat H_u \hat H_u \hat V_{U(1)_Y} \,,~~ D^2 ( \hat L \hat L ) \hat H_u \hat H_u \hat V_{SU(2)_L} \,,~~ 
		D^2 ( \hat L \hat L ) \hat H_u \hat H_u \hat H^\dagger_u \hat H_u \,,~~ \hat L \hat L \hat H_u \hat H_u \hat H_u^\dagger \hat H_u \,,
	\ee
	with LO coefficients 
	\beald
		& \frac{g' \mathbf{a} M_{X_3} C_0}{32 \pi^2 M_\Delta^2} \,, ~~~~ 
		\frac{g \mathbf{a} M_{X_3} C_0}{16 \pi^2 M_\Delta^2} \,, \\
		& -\frac{\mathbf{a} M_{X_3}}{32 \pi^2 M_\Delta^2} \sum_{i = 1}^2 |\lambda_i|^2 
			\left( D_{0,3} + M_{X_i}^2 E_{0,i} \right) \,, ~~~~ 
		\frac{\mathbf{b} M_{X_3} M_{X_1} M_{X_2}}{32 \pi^2 M_\Delta} \sum_{i = 1}^2 |\lambda_i|^2 E_{0,i} \,,
	\eeald
	respectively. More explicit expressions are given in~\appref{app:model_example_dim7}, in 
	particular~\eqref{eq:SOP_LLHHV_app} and~\eqref{eq:SOP_LLHHHH_app}. 
	Hence, the LO pure-$\nSUSYEWSB$ subset of $\OPnu$ is 
	\beal
		-\frac{1}{64 \pi^2 M_\Delta^2 M_X} \Bigg( 
			& \mathbf{a} \Bigg[ 
				\bead[t] 
					& \frac{g^2}{2 c_w^2} \left( L H_u \right) \left( L H_u \right) H^\dagger_u H_u 
						+ \left( \frac{g^2 c_{2 w}}{2 c_w^2} + \frac{|\mu|^2 (|\lambda_1|^2+|\lambda_2|^2)}{6 M_X^2} \right) \left( L H_u \right) \left( L H_u \right) H^\dagger_d H_d \\
					& + \left( g^2 - \frac{|\mu|^2 (|\lambda_1|^2+|\lambda_2|^2)}{3 M_X^2} \right) \left( L H_u \right) \left( H_u H_d \right) H^\dagger_d L \Bigg] 
				\eead \nonumber\\
			& + \frac{\mathbf{b} M_\Delta \mu \, (|\lambda_1|^2+|\lambda_2|^2)}{6 M_X^2} \left( L H_u \right) \left( L H_u \right) \left( H_u H_d \right) 
		\Bigg) \,,
	\eeal
	where we have taken the simplifying limit 
	$M_{X_{1,2,3}} = M_X$ (cf.~\eqref{eq:gLLHHHH_app} and~\eqref{eq:muLLHHHH_app}). 
	From this expression we can see that the gauge couplings' contribution to 
	neutrino masses, which reads 
	\be
		\mathbf{m^\nSUSYEWSB_\nu} \supset -\frac{g^2 \mathbf{a}}{64 \pi^2 c_w^2} \frac{v^4}{M_\Delta^2 M_X} c_{2\beta} s_\beta^2\,,
	\ee
	vanishes at $v_u = v_d$. This agrees with the fact that the contribution is $\propto \langle D \rangle$ 
	since $v_u = v_d$ corresponds to the $D$-flat direction of the scalar potential. 

	To understand, in terms of component fields, how these insertions are enablers of 
	contributions to $\OPnu$ consider the following. 
	As the insertion of an external auxiliary component of a gauge vector superfield ($D$) into a scalar line preserves chirality 
	(or, diagrammatically, the arrowhead's direction), 
	any holomorphy compliant diagram with a $D$ attached has a corresponding (underlying) 
	holomorphy compliant diagram without that $D$. Since in our example we are 
	considering a single $D$ insertion, the LO underlying diagrams 
	are the ones depicted in~\figref{fig:DeltaHH_model_example}, and no others. Once an external $D$ is attached 
	to an internal scalar line, the spinor loop diagram does not contribute and the sum of the others 
	need not vanish anymore to respect the SUSY non-renormalisation theorem. 
	Regarding the $\hat H^\dagger_u \hat H_u$ insertion, one can see that it allows for holomorphy compliant diagrams with 
	an external $F^\dagger - F$ pair by means of attaching $F^\dagger_{H_u}$ and $F_{H_u}$ to the scalar loop. 

\medskip

	The LO subset of $\OPnu$ is composed of dimension-5 operators that come from $\nSUSYEWS$. 
	Complete expressions for these operators up to order $3$ in $\Msoft$ are given 
	in~\appref{app:model_example_dim5}. Here we take the simplifying limits 
	$M_{X_{1,2,3}} = M_X$, $(\Msoft^2)_{X_{1,2,3}} = (\Msoft^2)_{\overline X_{1,2,3}} = \Msoft^2$, 
	$A_{1,2} = A$ and $B_{X_{1,2,3}} = B_X$. \eqref{eq:model_example_full3SUSYbreak} then reads 
	\bea
		&& \frac{1}{64 \pi^2 M_\Delta^2} \left( 
			\mathbf{a} \left[ \frac{2 \Msoft^2}{M_X} + \frac{2 A}{M_X} \left( A^*_X - \frac{B_\Delta}{M_\Delta} \right) - \frac{A^*_X B_X}{M_X^2} \right] 
			+ \mathbf{b} M_\Delta \frac{B_X}{M_X^2} \right) L L H_u H_u \nonumber\\
		&& -\frac{\mathbf{a}}{32 \pi^2 M_\Delta^2} \left(\frac{\mu^*}{M_X}\right) \left[ A^*_X \left( 1 - \frac{\Msoft^2}{M_X^2} - \frac{(\Msoft^2)_\Delta}{M_\Delta^2} \right) - \frac{B_\Delta}{M_\Delta} \right] L L H_u H^\dagger_d \nonumber\\
		&& -\frac{\mathbf{a}}{192 \pi^2 M_\Delta^2} \left(\frac{\mu^*}{M_X} \right)^2 \frac{A^*_X B_X}{M_X^2} L L H_d^\dagger H^\dagger_d \,.
	\eea
	The discussion surrounding~\figref{fig:DeltaHH_model_example} already suggested that one type of $\nSUSY$ 
	contribution would come from the mass splittings within components of chiral scalar superfields, 
	as induced by $\Msoft^2$ and $B_X$, since they introduce a mismatch in the cancellation between spinor and 
	scalar loops. However, unlike $\Msoft^2$, $B$ insertions reverse chirality. Thus, while a single 
	chirality flip in a scalar line makes holomorphy compliant diagrams for $F_{\overline \Delta} H_u H_u$ 
	possible -- and that is why there is a $B_X$-term contribution from the second supergraph (identified by the 
	$\mathbf{b}$ dependence in the expression above) --, a single $\nSUSY$ insertion of a $B_X$ disables holomorphy 
	compliant diagrams for $\Delta^\dagger H_u H_u$ and hence the absence of a single $B_X$-term 
	contribution proportional to $\mathbf{a}$ for $L L H_u H_u$ (cf.~\eqref{eq:model_example_full3SUSYbreak}). 
	For $L L H_u H^\dagger_d$ such a contribution can be holomorphy compliant\footnote{
	It does not appear in the expression above due to a fortuitous cancellation in the simplifying limit we have taken, 
	cf.~\eqref{eq:model_example_full3SUSYbreak}.} due to an external $F$ ($F_{H_u} \to \mu^* H^\dagger_d$). 
	Concerning contributions proportional to $B_\Delta$, they rely on the fact that EWSB induces, at the one-loop 
	level, a VEV for ${\overline \Delta}$ which, through $B_\Delta$, induces a VEV for 
	$\Delta$ and hence $L L \langle \Delta \rangle \subset \int d^2\theta \, \mathcal{W}$. In fact, one can 
	confirm that the dependence of $L L H_u H^\dagger_d$ on $B_\Delta$ is what one obtains from 
	$L L \langle \Delta \rangle$, where $\langle \Delta \rangle$ is computed by following the route 
	\be
		\langle H \rangle \xrightarrow[\int d^4\theta \, \hat\Delta^\dagger \hat H_u \hat H_u]{}
		\langle {\overline\Delta} \rangle \xrightarrow[B_\Delta \Delta {\overline \Delta}]{}
		\langle \Delta \rangle \,. 
	\ee
	In order to obtain the $B_\Delta$ dependence of $L L H_u H_u$, one must take into account the shift 
	in $\langle {\overline\Delta} \rangle$ induced by $\nSUSY$. To leading order, this shift is 
	proportional to $A_1 + A_2$.

\section{Conclusions}

	While the smallness of $m_\nu$ points towards an high seesaw scale $M$, 
	the resolution of the hierarchy problem suggests that the scale of 
	soft-$\nSUSY$ should lie close to the TeV scale. 
	It is then tempting to conceive that $\Msoft / M$ is partially responsible 
	for $m_\nu \ll v$. Since in the SUSY limit there are no radiative corrections 
	to the superpotential, models in which neutrino masses arise at the loop level 
	provide a scenario in which such a connection is natural. 
	How $m_\nu$ is proportional to $\nSUSY$ depends on the particular radiative 
	seesaw model or, more specifically, on the form of the leading $L$-number 
	breaking superoperators. 

\medskip

	By classifying the dependence on $\nSUSY$ according to their involvement 
	in EWSB, we identified a subset of model-topologies in which the leading 
	contributions to $m_\nu$ depend on $\nSUSY$ sources that are not involved in EWSB. 
	In a first stage, we argued in favour of this by showing that, 
	of all superoperators that can possibly contribute to neutrino masses, there is a subset 
	which does it only by means of insertions of $\nSUSY$ spurions. 
	Then, in a second stage, we gave a complete description of the simplest 
	model-topologies in which all leading superoperators were of this type, 
	and calculated their dependence on soft-$\nSUSY$ up to order $3$. 
	We found that all one-loop realisations generated $L L H H$ operators with a leading 
	dimensionful dependence that ranged from 
	$\mu \, \Msoft / M^3$ or $\Msoft^2 / M^3$ to $\mu \, \Msoft^2 / M^4$ or $\Msoft^3 / M^4$.

\medskip

	Even though the majority of all conceivable model-topologies do in fact generate contributions 
	to $m_\nu$ proportional to $\nSUSYEWS$, we pointed out that all models in the 
	literature\footnote{Barring those in which $L$-number is a symmetry of the superpotential that is 
	broken by the $\nSUSY$ sector.} that we are aware of generate at least one leading topology that 
	gives a contribution in which all $\nSUSY$ sources are involved in EWSB. 
	To serve as a proof of existence of models in which $m_\nu$ is proportional to $\nSUSYEWS$ at 
	leading order, we built a model in which the leading neutrino mass operators were of 
	dimension-5 and came from $\nSUSYEWS$, whereas the pure-$\nSUSYEWSB$ ones had dimension-7. 

\medskip

	One phenomenologically interesting aspect of these models is that soft-$\nSUSY$ effects generating 
	the leading order $m_\nu$ can be quite small without conflicting with lower limits 
	on the mass of new particles. This is due to the fact that these effects involve states that can 
	possess superpotential mass terms in the EWS phase, as we have seen in the model example. This is 
	in contrast with models that contain pure-$\nSUSYEWSB$ contributions to $m_\nu$ at leading order, 
	because $\mu$ and the soft-$\nSUSY$ effects driving EWSB provide the dominant contribution to the 
	mass of the corresponding states, and are therefore severely constrained by present lower limits 
	on sparticle masses. 

	If one conceives the leading order $m_\nu$ to be small as a result of some small 
	scale (say $m$) in the underlying soft-$\nSUSY$ effects, its explanatory value for the 
	smallness of $m_\nu$ must be confronted with the size of next-to-leading order 
	contributions that are insensitive to $m$. These next-to-leading contributions 
	do appear at the same loop level in the form of operators of higher dimension, 
	but can also appear as higher-loop contributions to operators of leading dimension.  
	For instance, in the model example the former were dimension-7 operators proportional to 
	$\mu / M$ or $g^2$, whereas the latter arise as two-loop contributions to dimension-5 operators. 
	These are proportional to: 
	\begin{itemize}
		\item $\mu^2 / M^2$ (and $\mu A^*_\ell / M^2$), due to superpotential terms 
			involving the ``wrong'' Higgs. To be specific, $\hat L \hat L \hat H_u \hat H^\dagger_d$ is generated 
			by a 1PI two-loop topology that is constructed from the 1-loop topology in the 
			left-hand side of~\figref{fig:SGraph_LLHH_model_example} by means of the coupling 
			$Y_\ell \hat L \hat e^c \hat H_d \subset \mathcal{W}$\,; 
		\item $m_{\widetilde{\text{\tiny EW}}} / M$, due to topologies with internal EW 
			gauge vector superfields in which a EWino mass term ($m_{\widetilde{\text{\tiny EW}}}$) is inserted. 
	\end{itemize}
	In this particular model, and taking $\mu \sim 2~\TeV$, one 
	can obtain $0.1~\eV \lesssim m_\nu \lesssim 1~\eV$ with seesaw mediators ($\hat\Delta$'s 
	and $\hat X$'s) lying at $\sim 10~\TeV$ and order $0.1$ couplings, provided $m \lesssim 100~\GeV$. 

	The parameter space of these models is quite rich as there are many couplings and masses involved in 
	the generation $m_\nu$. From a qualitative point of view, one can identify two overlapping regions of 
	parameter space of potential phenomenological interest. An interesting region is the one in which both 
	$\mu$ and $m$ are particularly small w.r.t.\ $M$, while higher-order contributions to $m_\nu$ that are 
	independent of both $\mu$ and $m$ remain subleading. In this region a small $m_\nu / v$ can be generated 
	with even larger couplings and/or lighter seesaw mediators. 
	Since $m_\nu$ is sensitive to at least the fourth power of couplings involved in $L$-number breaking, 
	another possibly interesting region comprises a lighter $M$ at the expense of slightly weaker couplings. 
	For instance, in the model of~\secref{sec:model_example}, decreasing all the couplings by a factor of 
	$1/2$ allows to decrease $M_X$ by a factor of $1/10$ while keeping $m_\nu$ fixed. 
	A detailed phenomenological analysis of this model will be presented in a future publication. 

\medskip

	To summarise, we have shown that there exist radiative seesaw models in which $m_\nu / v \ll 1$ can be 
	explained by $\Msoft / M \ll 1$ with $M$ not very far above the EW scale. Under the assumption of $L$-number 
	breaking at the superpotential level and low $M$, this explanation can be regarded to be more natural than that 
	of tree-level seesaws in the sense that it does not require very small superpotential couplings (as canonical 
	seesaws do) nor does it require two very different superpotential mass scales (as inverse seesaws do).

\section*{Acknowledgements}

This work has been partially funded by {\it Funda\c c\~ao para a Ci\^encia e a 
Tecnologia} (FCT) through the fellowship SFRH/BD/64666/2009. 
We also acknowledge the partial support from the projects EXPL/FIS-NUC/0460/2013 
and PEST-OE/FIS/UI0777/2013 financed by FCT. 

\appendix

\section{Trilinear with two Higgses} \label{app:trilinear}

\begin{figure}[h!t]
          \vspace{\vspUfig}
          \begin{center}
		\includegraphics[width=150mm]{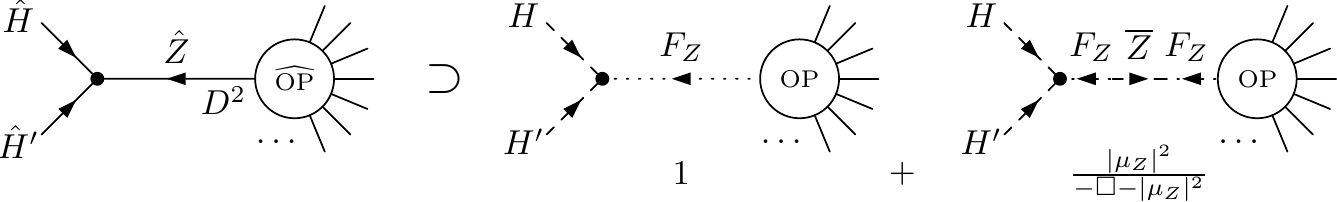} 
                \caption{Supergraph containing the trilinear contribution ($F^\dagger_Z \supset \lambda H H'$) to $\OP \otimes \text{Higgses}$. 
                }\label{fig:SGraph_Trilinear}
          \end{center}
          \vspace{\vspDfig}
        \end{figure}

	Let $\hat Z$ be involved in a trilinear with two Higgses ($\hat H, \hat H'$) 
	and $\hat{\overline Z}$ be the conjugate of $\hat Z$, as specified by the 
	following superpotential terms 
	\be
		\lambda \hat Z \hat H \hat H' + \mu_Z \hat Z \hat{\overline Z} \,.
	\ee
	Now suppose that the $F^\dagger_Z$ component of $\hat Z^\dagger$ is involved 
	in the generation of some operator $\OP$, i.e.\ 
	\be
		\OP \, F^\dagger_Z \subset \int d^4\theta \, \SOP \, \hat Z^\dagger \,, 
	\ee
	for some suitable $\SOP$. 
	The terms of the effective Lagrangian involving $F_Z$ or ${\overline Z}$ are then 
	\be
		-{\overline Z}^\dagger \Box Z + F_Z^\dagger F_Z - 
		\left( -\OP^\dagger F_Z + \lambda F_Z H H' + \mu_Z F_Z {\overline Z} + \Hc \right) \,,
	\ee
	apart from other possible interactions involving $F_Z$ or ${\overline Z}$ that are 
	not relevant for the following. 
	Using the equations of motion for $F_Z$ gives 
	\be
		-{\overline Z}^\dagger \Box {\overline Z} - \left| -\OP^\dagger + \lambda H H' + \mu_Z {\overline Z} \right|^2 \subset \mathcal{L}_\text{eff} \,. 
	\ee
	Now, by using the equations of motion for ${\overline Z}$ one sees that the terms involving $\lambda \, \OP \, H H'$ add 
	up as follows 
	\be
		\OP \, \lambda H H' + \OP \, \frac{|\mu_Z|^2}{-\Box-|\mu_Z|^2} \lambda H H' = \lambda \, \OP \frac{-\Box}{-\Box-|\mu_Z|^2} H H' \subset \mathcal{L}_\text{eff} \,,
	\ee
	as we wanted to show. An easier way to obtain this result is by evaluating the supergraph 
	depicted in~\figref{fig:SGraph_Trilinear}. One finds, 
	\be
		\lambda \int d^4\theta \, \SOP \, \frac{1}{-\Box-|\mu_Z|^2} D^2 ( \hat H \hat H' ) \supset \lambda \, \OP \, \frac{-\Box}{-\Box-|\mu_Z|^2} ( H H' ) \,.
	\ee

	We now note that ${\overline Z}$ is an Higgs in its own right, since $\langle H H' \rangle \neq 0$ 
	gives a tadpole for ${\overline Z}$. Thus, it seems that there is a contribution to 
	$\OP \otimes \text{Higgses}$ which is non-derivative in Higgses 
	\be
		\OP \, \left( \mu_Z {\overline Z} + \lambda H H' \right) = \OP \, F^\dagger_Z \,.
	\ee 
	However, $\langle \mu_Z {\overline Z} + \lambda H H' \rangle = 0$ up to $\nSUSY$ effects. 
	In the following we evaluate the effects of soft-$\nSUSY$ on $\langle F^\dagger _Z \rangle \neq 0$, 
	and, as a result, on the generation of a non-derivative $\OP \otimes \text{Higgses}$ 
	which upon EWSB yields $\OP$. 

	We take the VEVs of $H$'s to be, for all practical purposes, fixed. Then, 
	$\langle F^\dagger _Z \rangle$ is proportional to the shift in $\langle {\overline Z} \rangle$ 
	induced by soft-$\nSUSY$ terms involving $Z$ or ${\overline Z}$. 
	The relevant part of the scalar potential reads 
	\bea
		&& m^2_Z |Z|^2 + m^2_{\overline Z} |{\overline Z}|^2 + \left( B_Z Z {\overline Z} + \lambda A Z H H' + \Hc \right) \nonumber\\
		&& + \left( \lambda \mu_Z^* H H' {\overline Z}^\dagger + \Hc \right) + |\lambda|^2 \left( |H|^2 + |H'|^2 \right) |Z|^2 \nonumber\\
		&& + \left( \lambda \mu^* Z H' {\overline H}^\dagger + \lambda \mu'^* Z H {\overline H}'^\dagger + \Hc \right) \subset \mathcal{V} \,,
	\eea
	where $\mu$ and $\mu'$ are conceivable $\hat H \hat {\overline H}$ and $\hat H' \hat{\overline H}{}'$ 
	superpotential bilinears, and 
	\be
		m^2_{Z,{\overline Z}} := |\mu_Z|^2 + (\Msoft^2)_{Z,{\overline Z}} \,.
	\ee
	One then finds 
	\be
		\langle F^\dagger_Z \rangle = \lambda \langle H H' \rangle \left( 1 + \frac{\mu_Z \left( A B_Z^* - m'^2_Z \mu_Z^* \right)}{m'^2_Z m^2_{\overline Z} - |B_Z|^2} \right) + \lambda \mu_Z B_Z^* \left( \frac{\mu^* \langle {\overline H}^\dagger H' \rangle + \mu'^* \langle H {\overline H}'^\dagger \rangle}{m'^2_Z m^2_{\overline Z} - |B_Z|^2} \right) \,,
	\ee
	where $m'^2_Z := m^2_Z + |\lambda|^2 \left( |\langle H \rangle|^2 + |\langle H' \rangle|^2 \right)$. 
	Expanding this expression up to order $3$ in $\Msoft$ gives 
	\be
		\langle F^\dagger_Z \rangle \simeq \frac{(\Msoft^2)_{\overline Z}}{|\mu_Z|^2} \lambda \langle H H' \rangle 
			+ \frac{\lambda \mu_Z B_Z^*}{|\mu_Z|^4} \left( A \langle H H' \rangle + \mu^* \langle {\overline H}^\dagger H' \rangle + \mu'^* \langle H {\overline H}'^\dagger \rangle \right) \,. \label{eq:VEV_FZD_3rdORD}
	\ee

\section{Soft SUSY breaking insertions} \label{app:susy_break}

	Our conventions regarding soft-$\nSUSY$ are the following. For superpotential bilinears 
	normalised as 
	\be
		M \hat\Phi_1 \hat\Phi_2 \,,~ \frac{M}{2} \hat\Phi^2 \subset \mathcal{W} \,,
	\ee
	so that $M$ are canonical tree-level masses, the corresponding 
	soft-$\nSUSY$ bilinears are 
	\be
		(\Msoft^2)_i \Phi_i^\dagger \Phi_i + \Big( B \Phi_1 \Phi_2 + \Hc \Big) \,,~
		\Msoft^2 \Phi^\dagger \Phi + \left( \frac{B}{2} \Phi^2 + \Hc \right) \subset -\mathcal{L} \,.
	\ee
	Regarding holomorphic soft-$\nSUSY$ trilinears, for each superpotential trilinear 
	\be
		\lambda \hat\Phi_1 \hat\Phi_2 \hat\Phi_3 \subset \mathcal{W} \,,
	\ee
	we define the so-called $A$-terms by factoring out $\lambda$, i.e.\ 
	\be
		\lambda A \Phi_1 \Phi_2 \Phi_3 \subset -\mathcal{L} \,.
	\ee
	Gaugino mass terms are not relevant to our analysis. 
	Regarding non-holomorphic soft-$\nSUSY$ trilinears, we disregard them as 
	they are expected to be very suppressed w.r.t.\ the others. 
	As to mass terms for the spinor component of chiral scalar superfields, they can be reabsorbed into a redefinition of 
	superpotential mass terms, $\Msoft^2$ and non-holomorphic trilinears. 

\medskip 

	Soft-$\nSUSY$ effects are taken into account in supergraph calculations by 
	means of considering the vertices given in~\eqref{eq:SoftSUSYbreak_vertices}. 
	As perturbation theory in superspace is simpler than the ordinary QFT treatment, 
	this approach is preferable as long as $\Msoft / M$ is small. 

\begin{figure}[h!t]
          \vspace{0.2in}
          \begin{center}
		\begin{tabular}{c}
			\includegraphics[width=80mm]{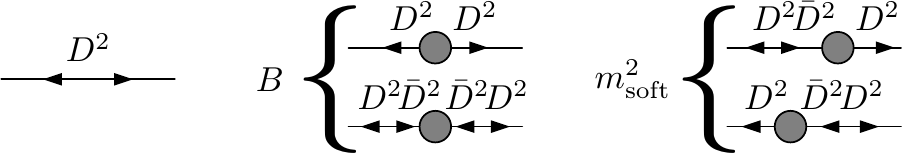} \\
			\includegraphics[width=80mm]{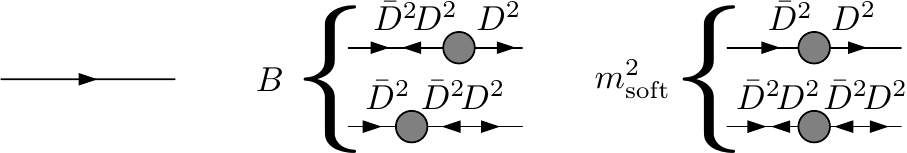} 
		\end{tabular}
                \caption{$B$ and $\Msoft^2$ insertions into $\hat\Phi\hat{\overline \Phi}$ (up row) and 
			$\hat\Phi\hat\Phi^\dagger$ (down row) propagators. 
                }\label{fig:SGraph_M2_B_insertions}
          \end{center}
          \vspace{-0.2in}
        \end{figure}

	Soft-$\nSUSY$ insertions have the following diagrammatic representation. 
	An $A$-term insertion is vertex of definite chirality promoted to a 
	grey blob. $\Msoft^2$- and $B$-terms are grey blobs inserted into propagators. 	
	For each type of propagator ($\hat\Phi\hat{\overline \Phi}$ and $\hat\Phi\hat\Phi^\dagger$) 
	there are two possibilities as we proceed to explain. A (anti-)chiral $B$-term 
	introduces either a $\bar D^2$ ($D^2$) or a $D^2$ ($\bar D^2$) and two $\bar D^2$ ($D^2$), 
	corresponding to the replacement of a $\hat\Phi^\dagger\hat{\overline \Phi}{}^\dagger$ ($\hat\Phi\hat{\overline \Phi}$) 
	propagator by a $B$-term blob or to an insertion into a $\hat\Phi\hat\Phi^\dagger$ propagator 
	by adjoining a $\hat\Phi\hat{\overline \Phi}$ ($\hat\Phi^\dagger\hat{\overline \Phi}{}^\dagger$) propagator, respectively. 
	The insertion of $\Msoft^2$ introduces a $D^2$ and a $\bar D^2$ or two $\bar D^2 D^2$, corresponding 
	to a simple insertion or an insertion adjoined by propagators $\hat\Phi\hat{\overline \Phi}$ and $\hat\Phi^\dagger\hat{\overline \Phi}{}^\dagger$. 
	All these possibilities are summarised in~\figref{fig:SGraph_M2_B_insertions}.

\medskip 

	In the following tables we list the soft-$\nSUSY$ insertions 
	up to order $3$ in $\Msoft$ for the topologies identified 
	in~\figref{fig:SGraph_1LoopTop_LLHH}. For each insertion set we give the D-algebra 
	result -- abbreviating $\nSUSY$ spurions by 
	\be
		\hat K := \hat X^\dagger \hat X 
	\ee
	 -- and whether it yields an $\text{OP} \in \text{OP}_\nu$ -- 
	if yes, we identify the operator and its dependence on soft-$\nSUSY$. 
	We have simplified the D-algebra results by taking advantage of 
	the fact that $\hat K$'s are pure-spurions, i.e.~$\hat K \sim \theta^2 \bar\theta^2$. 
	In particular, and since the result is local in $\theta$, expressions with too many 
	$\theta$'s from $\hat K$'s vanish. An unassigned D-algebra result (denoted by an horizontal 
	line) differs from a zero in the sense that it vanishes even if $\hat K$'s are not pure-spurions. 

	We do not display insertions that are redundant due to some symmetry of the supergraph. 
	For example, 
	consider the topology analysed in~\tbref{tb:SUSY_breaking_insertions_A_D2LLHuHu_LLD2HuHu}. 
	Since this supergraph topology is symmetric under the interchange of the 
	two chiral vertices of the triangle, the insertion of an 
	$A$-term into the upper chiral vertex leads to the same result as an 
	insertion into the lower chiral vertex. 

	We also do not display insertions into the 1PR propagator when the non-trivial 1PI 
	part has a definite chirality, as in this case the result is trivially zero up 
	to order $3$ in $\Msoft$. Thus, the only topology whose insertions 
	into the 1PR propagator we display is the one underlying 
	both $D^2 ( \hat L \hat L ) \hat H_u \hat H_u$ and $\hat L \hat L D^2 ( \hat H_u \hat H_u )$ 
	(see~\tbref{tb:SUSY_breaking_insertions_BDEL_MDEL2_D2LLHuHu_LLD2HuHu}).

	To see that the results in the following tables agree 
	with~\eqref{eq:full_XdaggerX_insertions}, we note that 
	\beald
		& \int d^4\theta \, D^2 \hat K \bar D^2 \hat K \, \SOP = \int d^4\theta \, \hat K D^2 \bar D^2 \hat K \, \SOP \,,\\
		& \int d^4\theta \, \bar D^2 \hat K D^2 \hat K \left\{ D^2 \SOP ~\text{or}~ \bar D^2 \SOP \right\} = \int d^4\theta \, \left\{ D^2 \hat K ~\text{or}~\bar D^2 \hat K \right\} D^2 \bar D^2 \hat K \, \SOP \,.
	\eeald

	\begin{table}[h!t]
		\centering
		\renewcommand{\arraystretch}{1.2}
	\begin{tabular}{cc}
	\begin{tabular}{| m{20mm} |c|c|}
		\hline
		Supergraph & D-algebra result & $\text{OP} \in \text{OP}_\nu$ \\\hline
		\vspace{1.8mm}\multirow{2}{*}{\includegraphics[width=20mm]{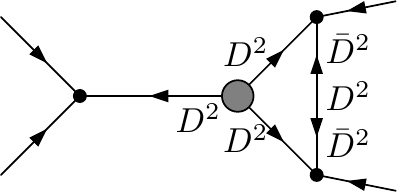}} & \multirow{2}{*}{$D^2 \hat K$} & $( A^* \mu^* )$ \\
			& & $L L H_u H_d^\dagger$ \\\hline
		\vspace{2mm}\includegraphics[width=20mm]{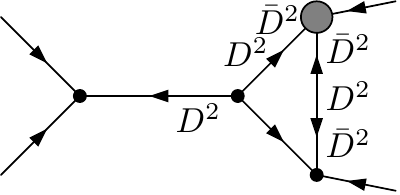} & $\bar D^2 \hat K$ & ----- \\\hline
		\vspace{2mm}\includegraphics[width=20mm]{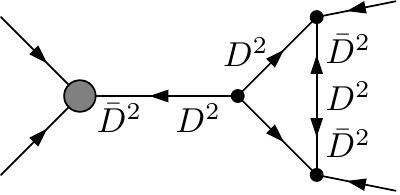} & $D^2 ( \bar D^2 \hat K \hat A \hat A )$ & ----- \\\hline
		\vspace{1.8mm}\multirow{2}{*}{\includegraphics[width=20mm]{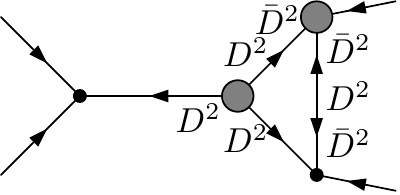}} & \multirow{2}{*}{$D^2 \hat K \bar D^2 \hat K$} & i: $( A^* A )$ \\
			& & $L L H_u H_u$ \\\hline
		\vspace{2mm}\includegraphics[width=20mm]{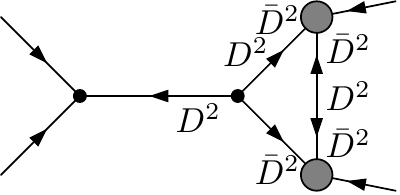} & $0$ & ----- \\\hline
	\end{tabular} & 
	\begin{tabular}{| m{20mm} |c|c|}
		\hline
		Supergraph & D-algebra result & $\text{OP} \in \text{OP}_\nu$ \\\hline
		\vspace{2mm}\includegraphics[width=20mm]{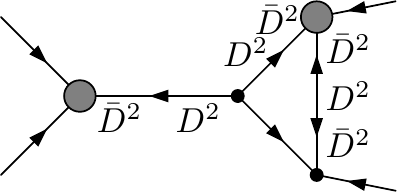} & $\bar D^2 \hat K D^2 ( \bar D^2 \hat K \hat A \hat A )$ & ----- \\\hline
		\vspace{1.8mm}\multirow{2}{*}{\includegraphics[width=20mm]{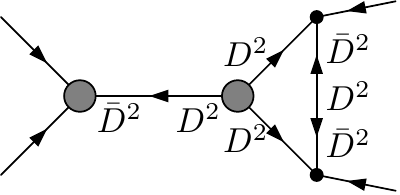}} & \multirow{2}{*}{$D^2 \hat K D^2 ( \bar D^2 \hat K \hat A \hat A )$} & ii: $( A^* A )$ \\
			& & $L L H_u H_u$ \\\hline
		\vspace{2mm}\includegraphics[width=20mm]{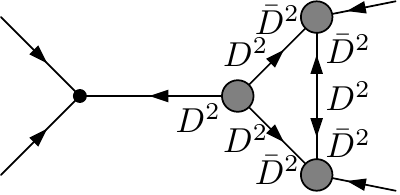} & $0$ & ----- \\\hline
		\vspace{2mm}\includegraphics[width=20mm]{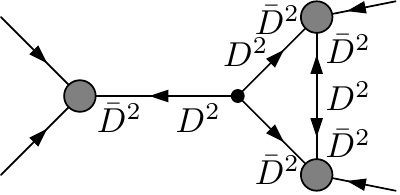} & $0$ & ----- \\\hline
		\vspace{2mm}\includegraphics[width=20mm]{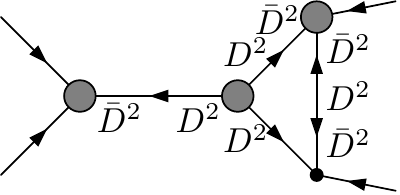} & $D^2 \hat K \bar D^2 \hat K D^2 ( \bar D^2 \hat K \hat A \hat A )$ & ----- \\\hline
	\end{tabular}
	\end{tabular}
		\caption{$A$-term insertions up to order $3$ in the 
		soft-$\nSUSY$ scale for the one-loop topology underlying both 
		$D^2 ( \hat L \hat L ) \hat H_u \hat H_u$ (``i'') and $\hat L \hat L D^2 ( \hat H_u \hat H_u )$ (``ii'') 
		superoperators. $\hat A$ is given by $\hat L$ or $\hat H_u$, depending 
		on whether the superoperator under evaluation is ``i'' or ``ii'', respectively. 
		When a given $\OP \in \OPnu$ entry stands for only one of the superoperators, we 
		identify it by starting with ``i'' or ``ii''.
		}\label{tb:SUSY_breaking_insertions_A_D2LLHuHu_LLD2HuHu} 
	\end{table}

	\begin{table}[h!t]
		\centering
		\renewcommand{\arraystretch}{1.2}
	\begin{tabular}{cc}
	\begin{tabular}{| m{20mm} |c|c|}
		\hline
		Supergraph & D-algebra result & $\text{OP} \in \text{OP}_\nu$ \\\hline
		\vspace{1.8mm}\multirow{2}{*}{\includegraphics[width=20mm]{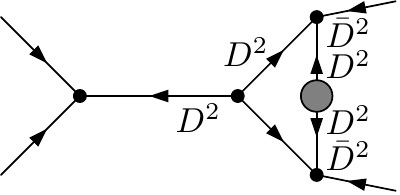}} & $\underline{D^2 \hat K}$ {\small (*)} & $( B^* \mu^* )$ \\ 
			& $D^2 \hat B \bar D^2 D^2 \hat K$ & $L L H_u H_d^\dagger$ \\\hline
		\vspace{1.8mm}\multirow{2}{*}{\includegraphics[width=20mm]{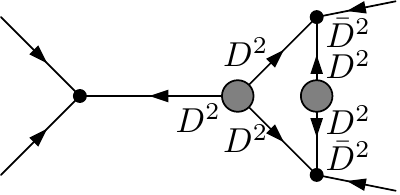}} & \multirow{2}{*}{$D^2 \hat B D^2 \bar D^2 \hat K D^2 \hat K$} & i: $( A^* B^* (\mu^*)^2 )$ \\ 
			& & $L L H^\dagger_d H^\dagger_d$ \\\hline
		\vspace{1.8mm}\multirow{2}{*}{\includegraphics[width=20mm]{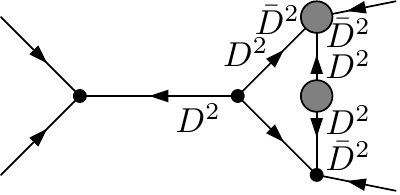}} & $\underline{\bar D^2 \hat K D^2 \hat K}$ {\small (*)} & i: $( A B^* )$ \\ 
			& $D^2 ( \hat B \bar D^2 \hat K ) \bar D^2 D^2 \hat K$ & $L L H_u H_u$ \\\hline
		\vspace{1.8mm}\multirow{2}{*}{\includegraphics[width=20mm]{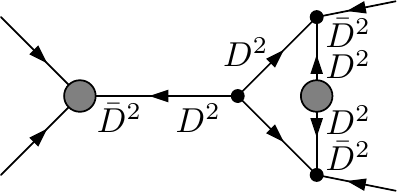}} & $\{\text{\underline{$D^2 \hat K$}}, D^2 \hat B \bar D^2 D^2 \hat K \}$ & ii: $( A B^* )$ \\ 
			&  $\times D^2 ( \bar D^2 \hat K \hat A \hat A )$ {\small (*)} & $L L H_u H_u$ \\\hline

		\vspace{2mm}\includegraphics[width=20mm]{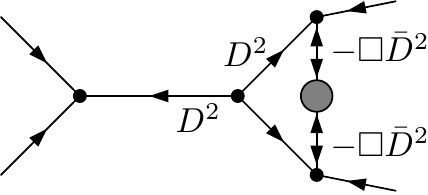} & $\bar D^2 \hat K$ & ----- \\\hline
		\vspace{1.8mm}\multirow{2}{*}{\includegraphics[width=20mm]{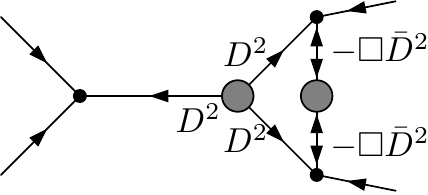}} & \multirow{2}{*}{$D^2 \hat K \bar D^2 \hat K$} & i: $( A^* B )$ \\
			& & $L L H_u H_u$ \\\hline
		\vspace{2mm}\includegraphics[width=20mm]{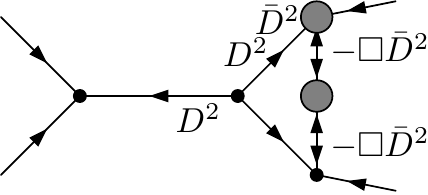} & $0$ & ----- \\\hline
		\vspace{2mm}\includegraphics[width=20mm]{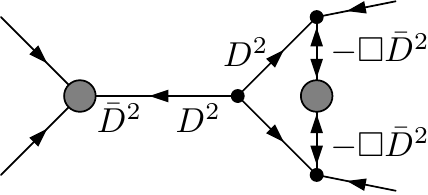} & $\bar D^2 \hat K D^2 ( \bar D^2 \hat K \hat A \hat A )$ & ----- \\\hline
	\end{tabular} & 
	\begin{tabular}{| m{20mm} |c|c|}
		\hline
		Supergraph & D-algebra result & $\text{OP} \in \text{OP}_\nu$ \\\hline
		\vspace{1.8mm}\multirow{2}{*}{\includegraphics[width=20mm]{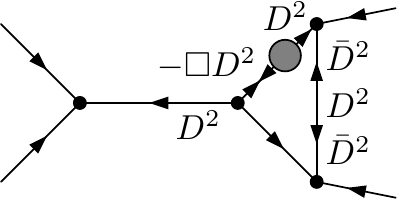}} & \multirow{2}{*}{$D^2 \hat K$} & $( B^* \mu^* )$ \\ 
			& & $L L H_u H_d^\dagger$ \\\hline
		\vspace{2mm}\includegraphics[width=20mm]{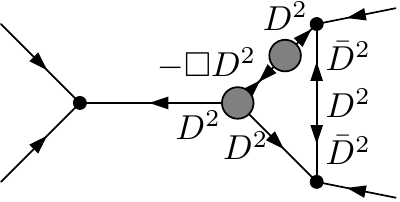} & $0$ & ----- \\\hline
		\vspace{1.8mm}\multirow{2}{*}{\includegraphics[width=20mm]{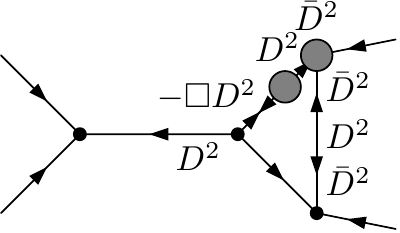}} & \multirow{2}{*}{$D^2 \hat K \bar D^2 \hat K$} & i: $( A B^* )$ \\ 
			& & $L L H_u H_u$ \\\hline
		\vspace{1.8mm}\multirow{2}{*}{\includegraphics[width=20mm]{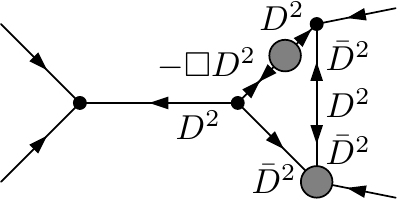}} & \multirow{2}{*}{$D^2 \hat K \bar D^2 \hat K$} & i: $( A B^* )$ \\ 
			& & $L L H_u H_u$ \\\hline
		\vspace{1.8mm}\multirow{2}{*}{\includegraphics[width=20mm]{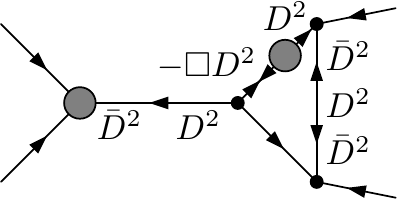}} & \multirow{2}{*}{$D^2 \hat K D^2 ( \bar D^2 \hat K \hat A \hat A )$} & ii: $( A B^* )$ \\ 
			& & $L L H_u H_u$ \\\hline

		\vspace{2mm}\includegraphics[width=20mm]{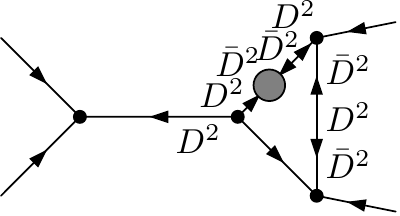} & $\bar D^2 \hat K$ & ----- \\\hline
		\vspace{1.8mm}\multirow{2}{*}{\includegraphics[width=20mm]{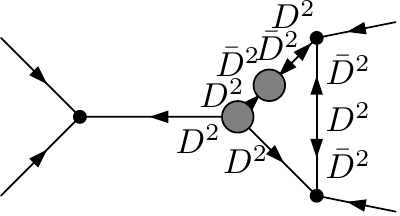}} & \multirow{2}{*}{$D^2 \hat K \bar D^2 \hat K$} & i: $( A^* B )$ \\
			& & $L L H_u H_u$ \\\hline
		\vspace{2mm}\includegraphics[width=20mm]{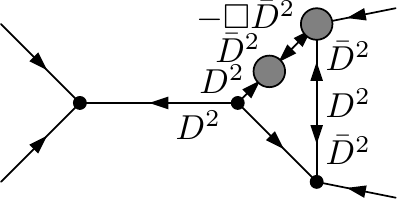} & $0$ & ----- \\\hline
		\vspace{2mm}\includegraphics[width=20mm]{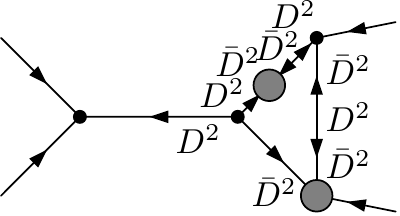} & $0$ & ----- \\\hline
		\vspace{2mm}\includegraphics[width=20mm]{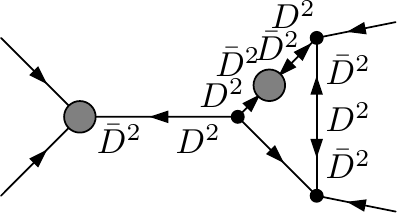} & $\bar D^2 \hat K D^2 ( \bar D^2 \hat K \hat A \hat A )$ & ----- \\\hline
	\end{tabular} 
	\end{tabular}
	\caption{Same as in~\tbref{tb:SUSY_breaking_insertions_A_D2LLHuHu_LLD2HuHu} but now for insertions of 
		$B$ and $A \times B$ into the non-trivial 1PI part. $\hat B$ is given by $\hat H_u$ or $\hat L$, depending 
		on whether the superoperator under evaluation is $D^2 ( \hat L \hat L ) \hat H_u \hat H_u$ (``i'') 
		or $\hat L \hat L D^2 ( \hat H_u \hat H_u )$ (``ii''), respectively. 
		When the D-algebra returns several results, we underline the one which yields an 
		$\text{OP} \in \text{OP}_\nu$. (*) stands for omitted terms that vanish as $\pext \to 0$. 
		}\label{tb:SUSY_breaking_insertions_B_D2LLHuHu_LLD2HuHu} 
	\end{table}

	\begin{table}[h!t]
		\centering
		\renewcommand{\arraystretch}{1.2}
	\begin{tabular}{cc}
	\begin{tabular}{| m{20mm} |c|c|}
		\hline
		Supergraph & D-algebra result & $\text{OP} \in \text{OP}_\nu$ \\\hline
		\vspace{1.8mm}\multirow{2}{*}{\includegraphics[width=20mm]{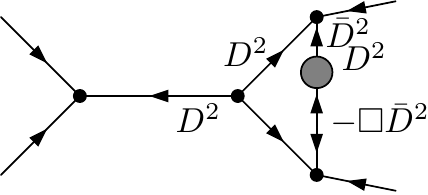}} & $\underline{\hat K}$ {\small (*)} & i: $( \Msoft^2 )$ \\ 
			& $D^2 \hat B \bar D^2 \hat K$ & $L L H_u H_u$ \\\hline
		\vspace{1.8mm}\multirow{2}{*}{\includegraphics[width=20mm]{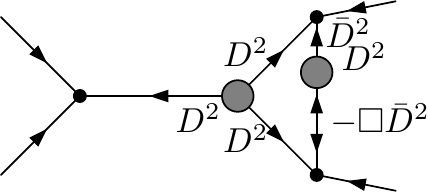}} & \multirow{2}{*}{$D^2 \hat B \bar D^2 \hat K D^2 \hat K$} & i: $( A^* \Msoft^2 \mu^* )$ \\ 
			& & $L L H_u H_d^\dagger$ \\\hline
		\vspace{2mm}\includegraphics[width=20mm]{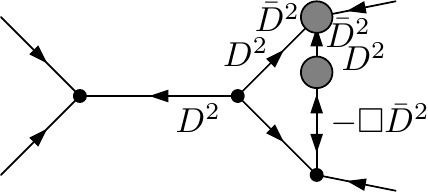} & $D^2 \bar D^2 \hat K \bar D^2 \hat K$ & ----- \\\hline
		\vspace{1.8mm}\multirow{2}{*}{\includegraphics[width=20mm]{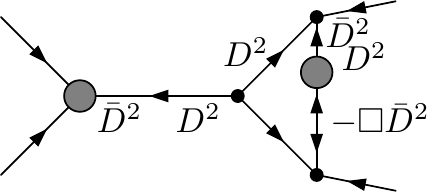}} & $\{\hat K, D^2 \hat B \bar D^2 \hat K\}$ & \multirow{2}{*}{-----} \\ 
			& $\times D^2 ( \bar D^2 \hat K \hat A \hat A )$ {\small (*)} & \\\hline

		\vspace{1.8mm}\multirow{2}{*}{\includegraphics[width=20mm]{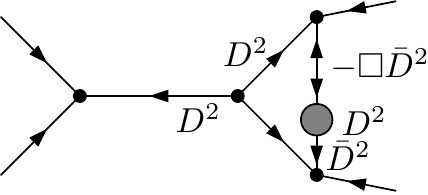}} & $\underline{\hat K}$ {\small (*)} & i: $( \Msoft^2 )$ \\ 
			& $D^2 \hat B \bar D^2 \hat K$ & $L L H_u H_u$ \\\hline
		\vspace{1.8mm}\multirow{2}{*}{\includegraphics[width=20mm]{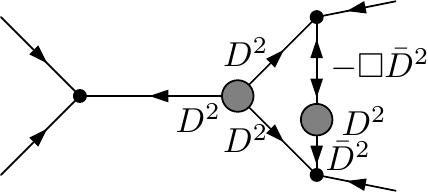}} & \multirow{2}{*}{$D^2 \hat B \bar D^2 \hat K D^2 \hat K$} & i: $( A^* \Msoft^2 \mu^* )$ \\ 
			& & $L L H_u H_d^\dagger$ \\\hline
		\vspace{2mm}\includegraphics[width=20mm]{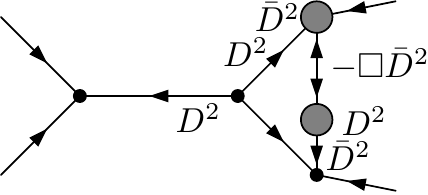} & $0$ & ----- \\\hline
		\vspace{1.8mm}\multirow{2}{*}{\includegraphics[width=20mm]{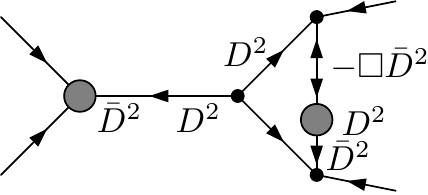}} & $\{ \hat K, D^2 \hat B \bar D^2 \hat K \}$ & \multirow{2}{*}{-----} \\ 
			& $\times D^2 ( \bar D^2 \hat K \hat A \hat A )$ {\small (*)} & \\\hline
	\end{tabular} & 
	\begin{tabular}{| m{20mm} |c|c|}
		\hline
		Supergraph & D-algebra result & $\text{OP} \in \text{OP}_\nu$ \\\hline
		\vspace{1.8mm}\multirow{2}{*}{\includegraphics[width=20mm]{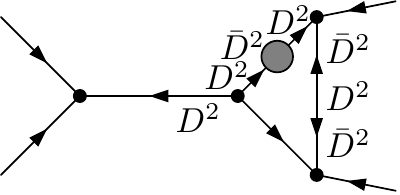}} & $\underline{\hat K}$ {\small (*)} & i: $( \Msoft^2 )$ \\ 
			& $D^2 \bar D^2 \hat K$ & $L L H_u H_u$ \\\hline
		\vspace{1.8mm}\multirow{2}{*}{\includegraphics[width=20mm]{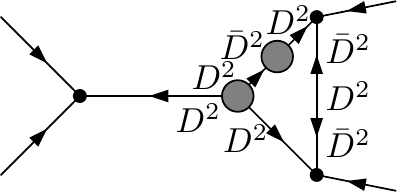}} & \multirow{2}{*}{$D^2 \hat K D^2 \bar D^2 \hat K$} & $( A^* \Msoft^2 \mu^* )$ \\ 
			& & $L L H_u H_d^\dagger$ \\\hline
		\vspace{2mm}\includegraphics[width=20mm]{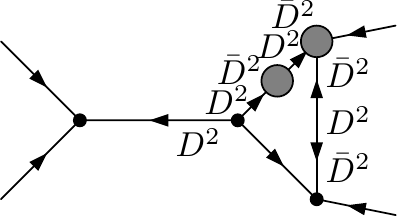} & $\bar D^2 \hat K D^2 \bar D^2 \hat K$ & ----- \\\hline 
		\vspace{2mm}\includegraphics[width=20mm]{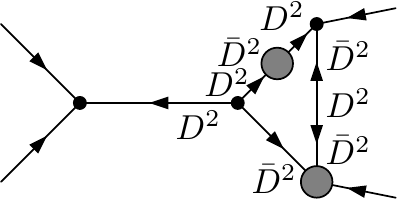} & $\bar D^2 \hat K D^2 \bar D^2 \hat K$ & ----- \\\hline
		\vspace{1.8mm}\multirow{2}{*}{\includegraphics[width=20mm]{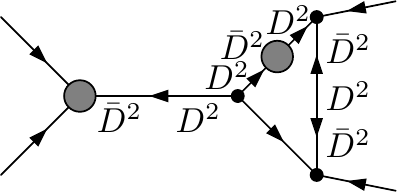}} & $\{ \hat K, D^2 \bar D^2 \hat K \}$ & \multirow{2}{*}{-----} \\ 
			& $\times D^2 ( \bar D^2 \hat K \hat A \hat A )$ {\small (*)} & \\\hline

		\vspace{1.8mm}\multirow{2}{*}{\includegraphics[width=20mm]{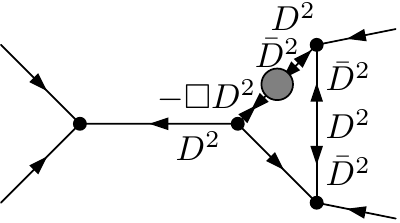}} & \multirow{2}{*}{$\hat K$} & i: $( \Msoft^2 )$ \\ 
			& & $L L H_u H_u$ \\\hline
		\vspace{2mm}\includegraphics[width=20mm]{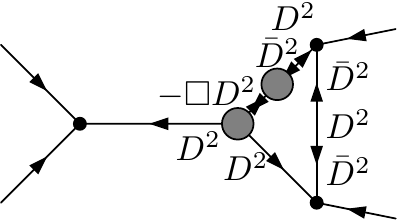} & $0$ & ----- \\\hline
		\vspace{2mm}\includegraphics[width=20mm]{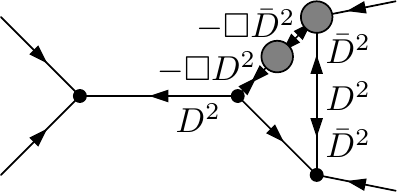} & $0$ & ----- \\\hline 
		\vspace{2mm}\includegraphics[width=20mm]{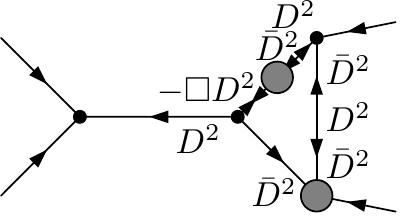} & $0$ & ----- \\\hline 
		\vspace{2mm}\includegraphics[width=20mm]{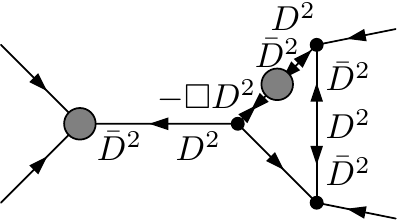} & $\hat K D^2 ( \bar D^2 \hat K \hat A \hat A )$ & ----- \\\hline
	\end{tabular} 
	\end{tabular}
	\caption{Same as in~\tbref{tb:SUSY_breaking_insertions_B_D2LLHuHu_LLD2HuHu} but now for insertions of 
		$\Msoft^2$ and $A \times \Msoft^2$. 
		}\label{tb:SUSY_breaking_insertions_M2_D2LLHuHu_LLD2HuHu} 
	\end{table}

	\begin{table}[h!t]
		\centering
		\renewcommand{\arraystretch}{1.2}
	\begin{tabular}{cc}
	\begin{tabular}{| m{20mm} |c|c|}
		\hline
		Supergraph & D-algebra result & $\text{OP} \in \text{OP}_\nu$ \\\hline
		\vspace{1.8mm}\multirow{2}{*}{\includegraphics[width=20mm]{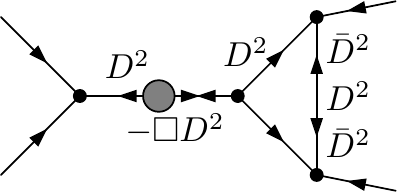}} & \multirow{2}{*}{$D^2 \hat K$} & $( B^* \mu^* )$ \\
			& & $L L H_u H^\dagger_d$ \\\hline 
		\vspace{2mm}\includegraphics[width=20mm]{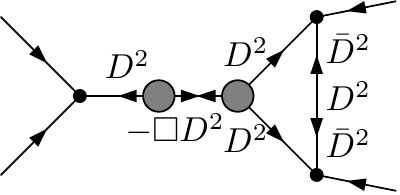} & $0$ & ----- \\\hline
		\vspace{1.8mm}\multirow{2}{*}{\includegraphics[width=20mm]{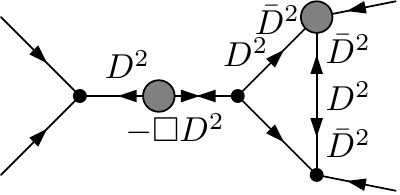}} & \multirow{2}{*}{$D^2 \hat K \bar D^2 \hat K$} & i: $( A B^* )$ \\
			& & $L L H_u H_u$ \\\hline
		\vspace{1.8mm}\multirow{2}{*}{\includegraphics[width=20mm]{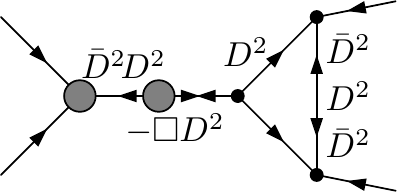}} & \multirow{2}{*}{$D^2 \hat K D^2 ( \bar D^2 \hat K \hat A \hat A )$} & ii: $( A B^* )$ \\
			& & $L L H_u H_u$ \\\hline

		\vspace{2mm}\includegraphics[width=20mm]{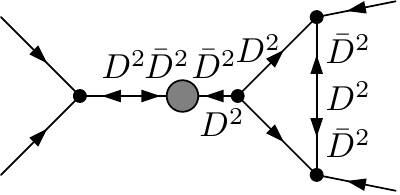} & $D^2 ( \bar D^2 \hat K \hat A \hat A )$ & ----- \\\hline
		\vspace{2mm}\includegraphics[width=20mm]{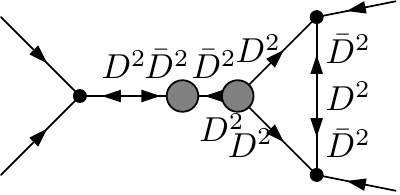} & $D^2 \hat K D^2 ( \bar D^2 \hat K \hat A \hat A )$ & ----- \\\hline
		\vspace{2mm}\includegraphics[width=20mm]{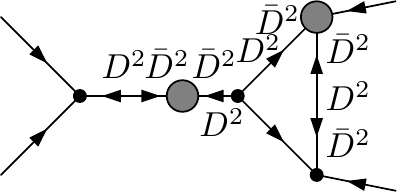} & $\bar D^2 \hat K D^2 ( \bar D^2 \hat K \hat A \hat A )$ & ----- \\\hline
		\vspace{2mm}\includegraphics[width=20mm]{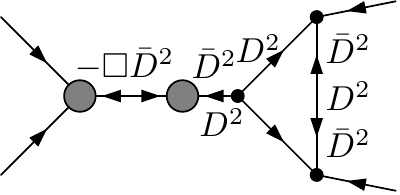} & $0$ & ----- \\\hline
	\end{tabular} & 
	\begin{tabular}{| m{20mm} |c|c|}
		\hline
		Supergraph & D-algebra result & $\text{OP} \in \text{OP}_\nu$ \\\hline
		\vspace{2mm}\includegraphics[width=20mm]{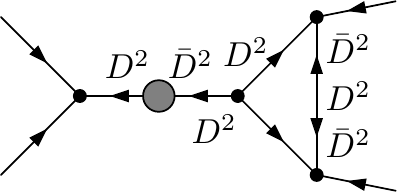} & $D^2 \bar D^2 ( D^2 ( \hat A \hat A ) \hat K )$ & ----- \\\hline
		\vspace{1.8mm}\multirow{2}{*}{\includegraphics[width=20mm]{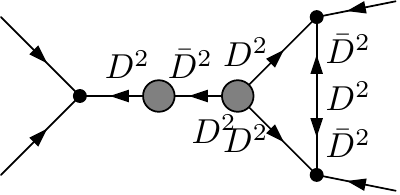}} & \multirow{2}{*}{$D^2 \hat K D^2 \bar D^2 ( D^2 ( \hat A \hat A ) \hat K )$} & $( A^* \Msoft^2 \mu^* )$ \\
			& & $L L H_u H^\dagger_d$ \\\hline
		\vspace{2mm}\includegraphics[width=20mm]{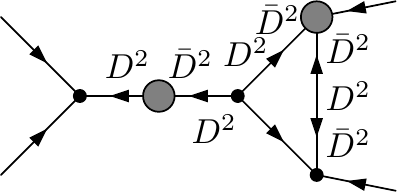} & $\bar D^2 \hat K D^2 \bar D^2 ( D^2 ( \hat A \hat A ) \hat K )$ & ----- \\\hline
		\vspace{2mm}\includegraphics[width=20mm]{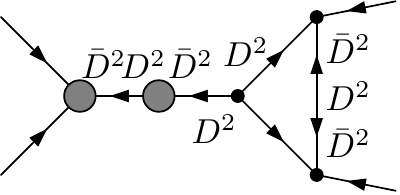} & $D^2 \bar D^2 ( D^2 ( \bar D^2 \hat K \hat A \hat A ) \hat K )$ & ----- \\\hline

		\vspace{1.8mm}\multirow{2}{*}{\includegraphics[width=20mm]{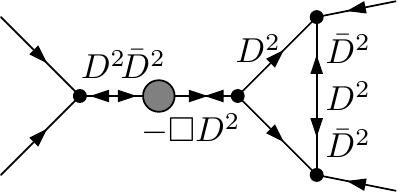}} & \multirow{2}{*}{$D^2 ( \hat K \hat A \hat A )$} & ii: $( \Msoft^2 )$ \\
			& & $L L H_u H_u$ \\\hline
		\vspace{2mm}\includegraphics[width=20mm]{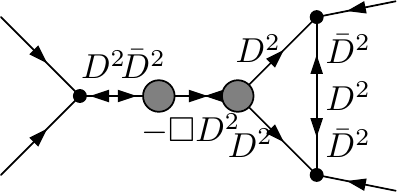} & $0$ & ----- \\\hline
		\vspace{2mm}\includegraphics[width=20mm]{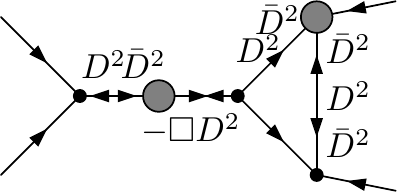} & $\bar D^2 \hat K D^2 ( \hat K \hat A \hat A )$ & ----- \\\hline
		\vspace{2mm}\includegraphics[width=20mm]{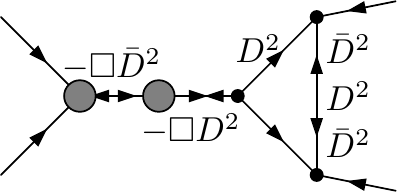} & $0$ & ----- \\\hline
	\end{tabular} 
	\end{tabular}
	\caption{Same as in~\tbref{tb:SUSY_breaking_insertions_B_D2LLHuHu_LLD2HuHu} and~\tbref{tb:SUSY_breaking_insertions_M2_D2LLHuHu_LLD2HuHu} 
		but now for $B$ and $\Msoft^2$ insertions into the 1PR propagator. 
		}\label{tb:SUSY_breaking_insertions_BDEL_MDEL2_D2LLHuHu_LLD2HuHu} 
	\end{table}

	\begin{table}[h!t]
		\centering
		\renewcommand{\arraystretch}{1.2}
	\begin{tabular}{cc}
	\multicolumn{2}{c}{\begin{tabular}{| m{20mm} |c|c|}
		\hline
		Supergraph & D-algebra result & $\text{OP} \in \text{OP}_\nu$ \\\hline
		\vspace{2mm}\includegraphics[width=20mm]{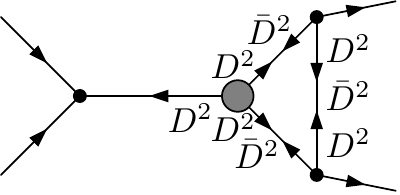} & ----- & ----- \\\hline
		\vspace{2mm}\includegraphics[width=20mm]{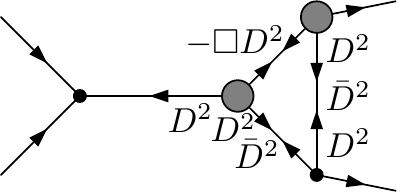} & ----- & ----- \\\hline
		\vspace{2mm}\includegraphics[width=20mm]{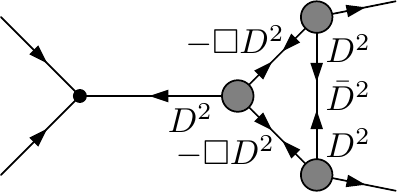} & ----- & ----- \\\hline
	\end{tabular}} 
	\\
	\begin{tabular}{| m{20mm} |c|c|}
		\hline
		Supergraph & D-algebra result & $\text{OP} \in \text{OP}_\nu$ \\\hline
		\vspace{1.8mm}\multirow{2}{*}{\includegraphics[width=20mm]{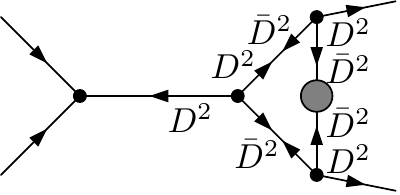}} & \multirow{2}{*}{$\bar D^2 \hat K$} & $( B \mu )$ \\
			& & $L L H_u H^\dagger_d$ \\\hline
		\vspace{1.8mm}\multirow{2}{*}{\includegraphics[width=20mm]{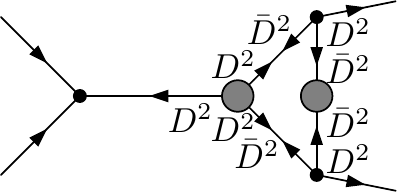}} & \multirow{2}{*}{$D^2 \hat K \bar D^2 \hat K$} & $( A^* B )$ \\
			& & $L L H^\dagger_d H^\dagger_d$ \\\hline
		\vspace{1.8mm}\multirow{2}{*}{\includegraphics[width=20mm]{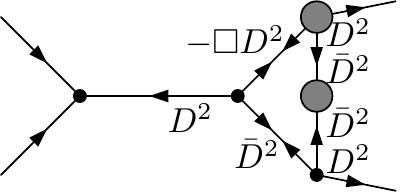}} & \multirow{2}{*}{$D^2 \hat K \bar D^2 \hat K$} & $( A^* B )$ \\
			& & $L L H^\dagger_d H^\dagger_d$ \\\hline

		\vspace{2mm}\includegraphics[width=20mm]{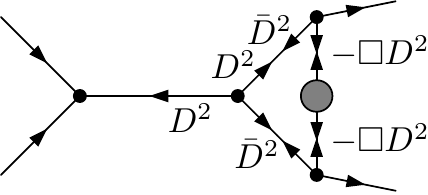} & ----- & ----- \\\hline
		\vspace{2mm}\includegraphics[width=20mm]{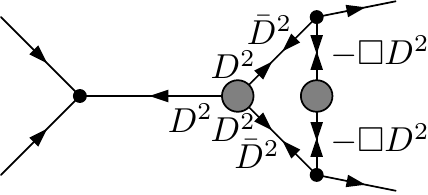} & ----- & ----- \\\hline
		\vspace{2mm}\includegraphics[width=20mm]{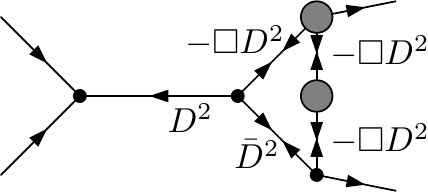} & ----- & ----- \\\hline
	\end{tabular} & 
	\begin{tabular}{| m{20mm} |c|c|}
		\hline
		Supergraph & D-algebra result & $\text{OP} \in \text{OP}_\nu$ \\\hline
		\vspace{1.8mm}\multirow{2}{*}{\includegraphics[width=20mm]{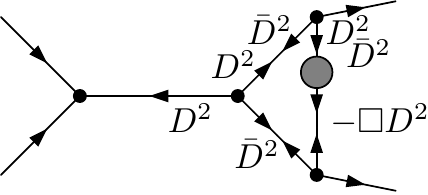}} & \multirow{2}{*}{$\hat K$} & $( \Msoft^2 )$ \\
			& & $L L H_d^\dagger H^\dagger_d$ \\\hline
		\vspace{2mm}\includegraphics[width=20mm]{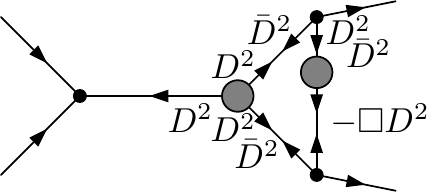} & $0$ & ----- \\\hline
		\vspace{2mm}\includegraphics[width=20mm]{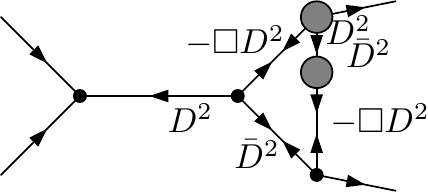} & $0$ & ----- \\\hline

		\vspace{1.8mm}\multirow{2}{*}{\includegraphics[width=20mm]{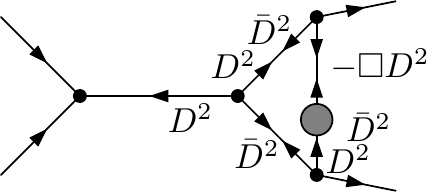}} & \multirow{2}{*}{$\hat K$} & $( \Msoft^2 )$ \\
			& & $L L H_d^\dagger H^\dagger_d$ \\\hline
		\vspace{2mm}\includegraphics[width=20mm]{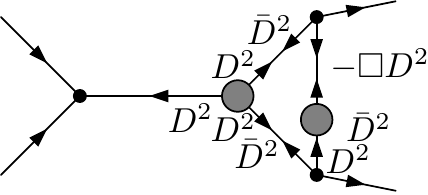} & $0$ & ----- \\\hline
		\vspace{2mm}\includegraphics[width=20mm]{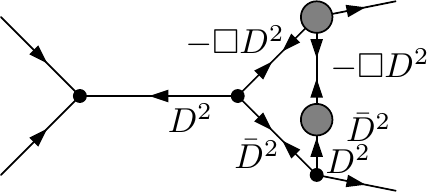} & $0$ & ----- \\\hline
	\end{tabular}
	\end{tabular}
		\caption{Soft-$\nSUSY$ insertions up to order $3$ in the soft-$\nSUSY$ 
		scale for one-loop $D^2 ( \hat L \hat L ) \hat H_d^\dagger \hat H_d^\dagger$. 
		}\label{tb:SUSY_breaking_insertions_D2LLHdHd} 
	\end{table}

	\begin{table}[h!t]
		\centering
		\renewcommand{\arraystretch}{1.2}
	\begin{tabular}{cc}
	\multicolumn{2}{c}{\begin{tabular}{| m{20mm} |c|c|}
		\hline
		Supergraph & D-algebra result & $\text{OP} \in \text{OP}_\nu$ \\\hline
		\vspace{2mm}\includegraphics[width=20mm]{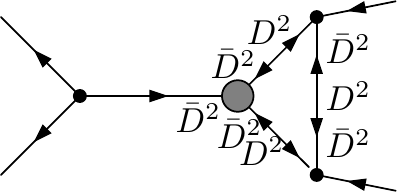} & ----- & ----- \\\hline
		\vspace{2mm}\includegraphics[width=20mm]{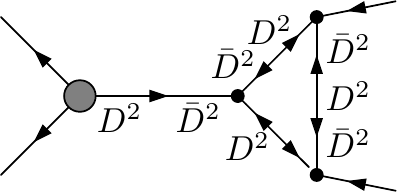} & $\bar D^2 ( D^2 \hat K \hat H^\dagger_d \hat H^\dagger_d )$ & ----- \\\hline
		\vspace{2mm}\includegraphics[width=20mm]{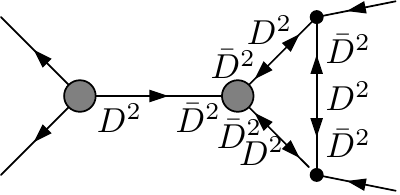} & ----- & ----- \\\hline
	\end{tabular}} 
	\\
	\begin{tabular}{| m{20mm} |c|c|}
		\hline
		Supergraph & D-algebra result & $\text{OP} \in \text{OP}_\nu$ \\\hline
		\vspace{1.8mm}\multirow{2}{*}{\includegraphics[width=20mm]{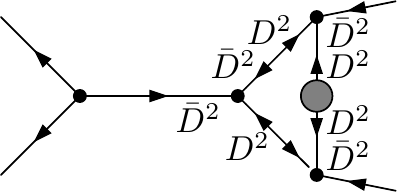}} & \multirow{2}{*}{$D^2 \hat K$} & $( B^* \mu )$ \\
			& & $L L H_u H^\dagger_d$ \\\hline
		\vspace{2mm}\includegraphics[width=20mm]{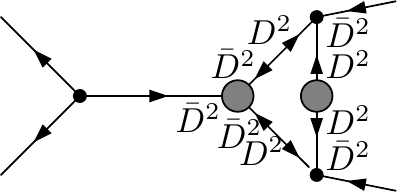} & $\bar D^2 \hat K D^2 \hat K$ & ----- \\\hline
		\vspace{1.8mm}\multirow{2}{*}{\includegraphics[width=20mm]{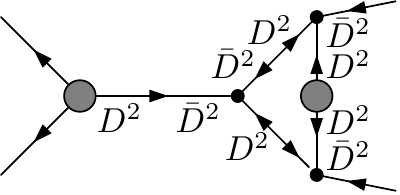}} & \multirow{2}{*}{$D^2 \hat K \bar D^2 ( D^2 \hat K \hat H^\dagger_d \hat H^\dagger_d )$} & $( A^* B^* )$ \\
			& & $L L H^\dagger_d H^\dagger_d$ \\\hline

		\vspace{2mm}\includegraphics[width=20mm]{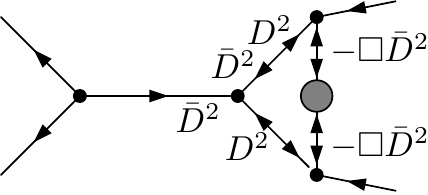} & ----- & ----- \\\hline
		\vspace{2mm}\includegraphics[width=20mm]{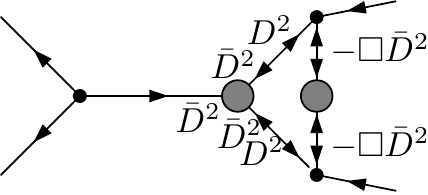} & ----- & ----- \\\hline
		\vspace{2mm}\includegraphics[width=20mm]{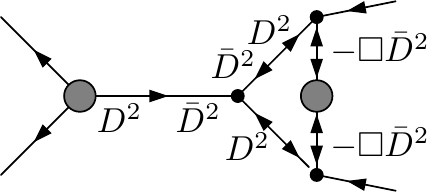} & ----- & ----- \\\hline
	\end{tabular} & 
	\begin{tabular}{| m{20mm} |c|c|}
		\hline
		Supergraph & D-algebra result & $\text{OP} \in \text{OP}_\nu$ \\\hline
		\vspace{2mm}\includegraphics[width=20mm]{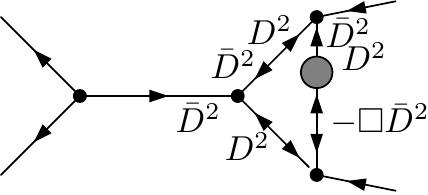} & $\hat K$ & ----- \\\hline
		\vspace{2mm}\includegraphics[width=20mm]{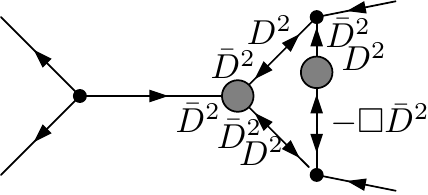} & $0$ & ----- \\\hline
		\vspace{2mm}\includegraphics[width=20mm]{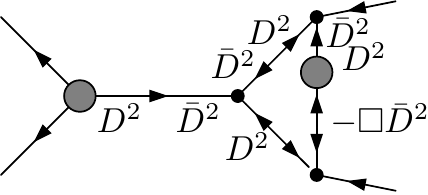} & $\hat K \bar D^2 ( D^2 \hat K \hat H^\dagger_d \hat H^\dagger_d )$ & ----- \\\hline

		\vspace{2mm}\includegraphics[width=20mm]{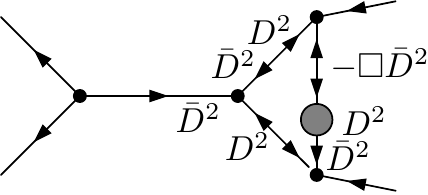} & $\hat K$ & ----- \\\hline
		\vspace{2mm}\includegraphics[width=20mm]{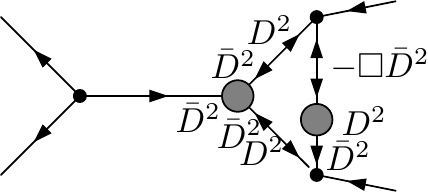} & $0$ & ----- \\\hline
		\vspace{2mm}\includegraphics[width=20mm]{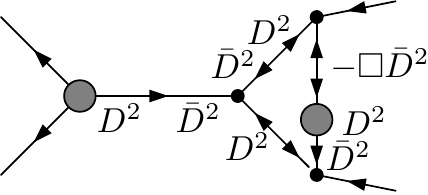} & $\hat K \bar D^2 ( D^2 \hat K \hat H^\dagger_d \hat H^\dagger_d )$ & ----- \\\hline
	\end{tabular}
	\end{tabular}
		\caption{Soft-$\nSUSY$ insertions up to order $3$ in the soft-$\nSUSY$ 
		scale for one-loop $\hat L \hat L \bar D^2 ( \hat H_d^\dagger \hat H_d^\dagger )$. 
		}\label{tb:SUSY_breaking_insertions_LLD2barHdHd} 
	\end{table}

	\begin{table}[h!t]
		\centering
		\renewcommand{\arraystretch}{1.2}
	\begin{tabular}{cc}
	\begin{tabular}{| m{20mm} |c|c|}
		\hline
		Supergraph & D-algebra result & $\text{OP} \in \text{OP}_\nu$ \\\hline
		\vspace{2mm}\includegraphics[width=20mm]{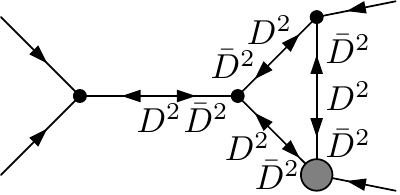} & ----- & ----- \\\hline
		\vspace{2mm}\includegraphics[width=20mm]{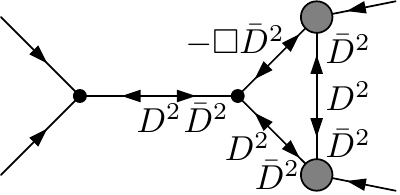} & ----- & ----- \\\hline
		\vspace{2mm}\includegraphics[width=20mm]{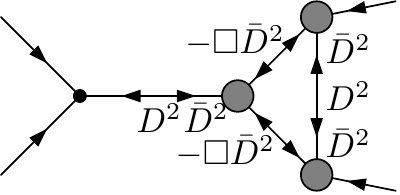} & ----- & ----- \\\hline
	\end{tabular} & 
	\begin{tabular}{| m{20mm} |c|c|}
		\hline
		Supergraph & D-algebra result & $\text{OP} \in \text{OP}_\nu$ \\\hline
		\vspace{2mm}\includegraphics[width=20mm]{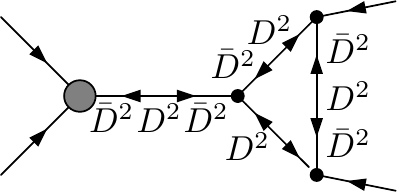} & $\bar D^2 \hat K$ & ----- \\\hline
		\vspace{2mm}\includegraphics[width=20mm]{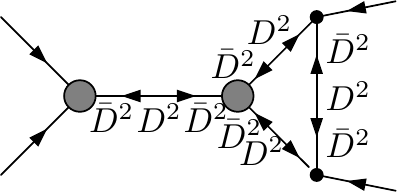} & $0$ & ----- \\\hline
	\end{tabular} \\
	\begin{tabular}{| m{20mm} |c|c|}
		\hline
		Supergraph & D-algebra result & $\text{OP} \in \text{OP}_\nu$ \\\hline
		\vspace{1.8mm}\multirow{2}{*}{\includegraphics[width=20mm]{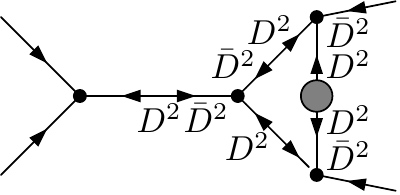}} & \multirow{2}{*}{$D^2 \hat K$} & $( B^* )$ \\
			& & $L L H_u H_u$ \\\hline
		\vspace{2mm}\includegraphics[width=20mm]{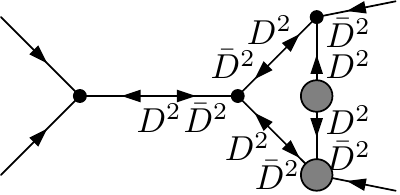} & $D^2 \hat K \bar D^2 \hat K$ & ----- \\\hline
		\vspace{2mm}\includegraphics[width=20mm]{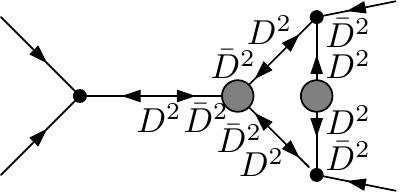} & $D^2 \hat K \bar D^2 \hat K$ & ----- \\\hline
		\vspace{2mm}\includegraphics[width=20mm]{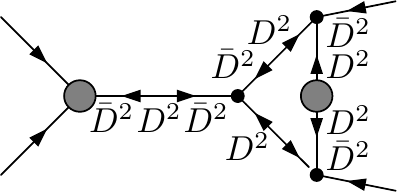} & $D^2 \hat K \bar D^2 \hat K$ & ----- \\\hline

		\vspace{2mm}\includegraphics[width=20mm]{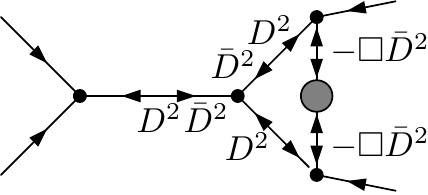} & ----- & ----- \\\hline
		\vspace{2mm}\includegraphics[width=20mm]{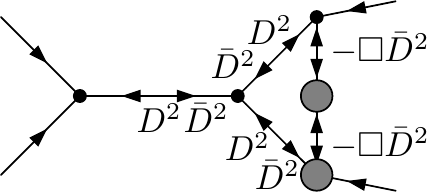} & ----- & ----- \\\hline
		\vspace{2mm}\includegraphics[width=20mm]{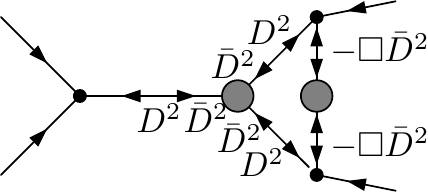} & ----- & ----- \\\hline
		\vspace{2mm}\includegraphics[width=20mm]{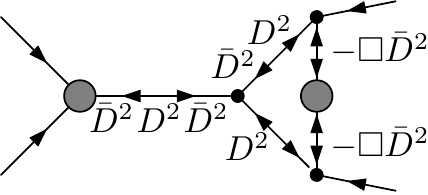} & ----- & ----- \\\hline
	\end{tabular} & 
	\begin{tabular}{| m{20mm} |c|c|}
		\hline
		Supergraph & D-algebra result & $\text{OP} \in \text{OP}_\nu$ \\\hline
		\vspace{2mm}\includegraphics[width=20mm]{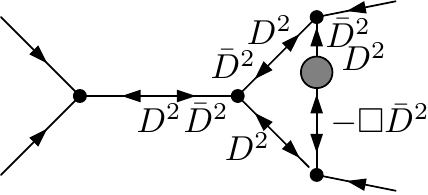} & $\hat K$ & ----- \\\hline
		\vspace{2mm}\includegraphics[width=20mm]{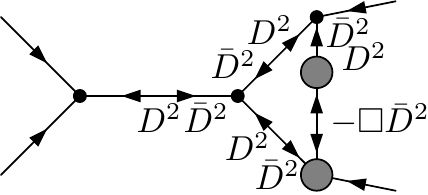} & $0$ & ----- \\\hline
		\vspace{2mm}\includegraphics[width=20mm]{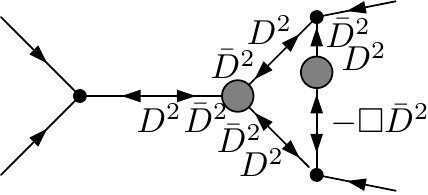} & $0$ & ----- \\\hline
		\vspace{2mm}\includegraphics[width=20mm]{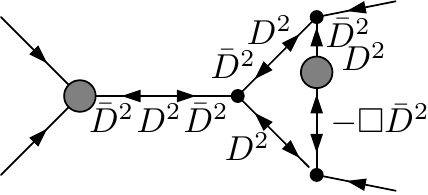} & $0$ & ----- \\\hline

		\vspace{2mm}\includegraphics[width=20mm]{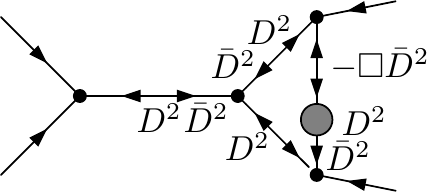} & $\hat K$ & ----- \\\hline
		\vspace{2mm}\includegraphics[width=20mm]{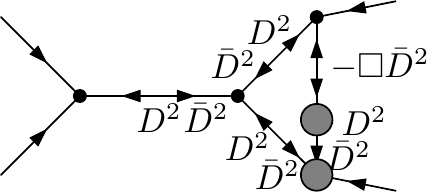} & $0$ & ----- \\\hline
		\vspace{2mm}\includegraphics[width=20mm]{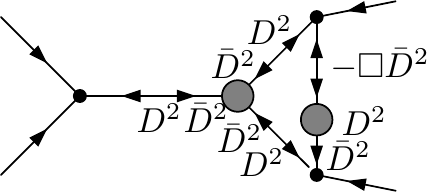} & $0$ & ----- \\\hline
		\vspace{2mm}\includegraphics[width=20mm]{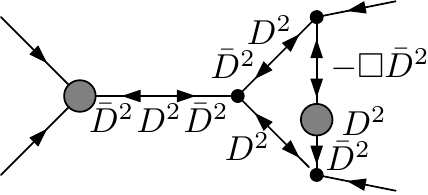} & $0$ & ----- \\\hline
	\end{tabular} 
	\end{tabular}
		\caption{Soft-$\nSUSY$ insertions up to order $3$ in the soft-$\nSUSY$ 
		scale for one-loop 1PR $\hat L \hat L \hat H_u \hat H_u$. 
		}\label{tb:SUSY_breaking_insertions_1PR_LLHuHu} 
	\end{table}

	\begin{table}[h!t]
		\centering
		\renewcommand{\arraystretch}{1.2}
	\begin{tabular}{cc}
	\begin{tabular}{| m{20mm} |c|c|}
		\hline
		Supergraph & D-algebra result & $\text{OP} \in \text{OP}_\nu$ \\\hline
		\vspace{2mm}\includegraphics[width=20mm]{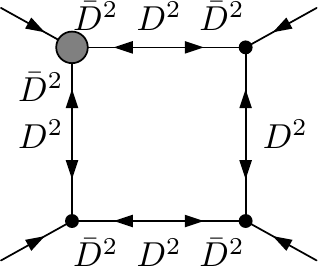} & ----- & ----- \\\hline
	\end{tabular} & 
	\begin{tabular}{| m{20mm} |c|c|}
		\hline
		Supergraph & D-algebra result & $\text{OP} \in \text{OP}_\nu$ \\\hline
		\vspace{2mm}\includegraphics[width=20mm]{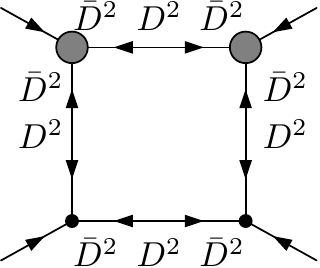} &  ----- & ----- \\\hline
		\vspace{2mm}\includegraphics[width=20mm]{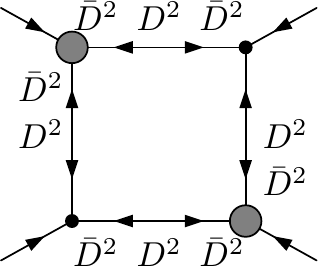} & ----- & ----- \\\hline
	\end{tabular} 
	\\	
	\begin{tabular}{| m{20mm} |c|c|}
		\hline
		Supergraph & D-algebra result & $\text{OP} \in \text{OP}_\nu$ \\\hline
		\multirow{4}{*}{\includegraphics[width=20mm]{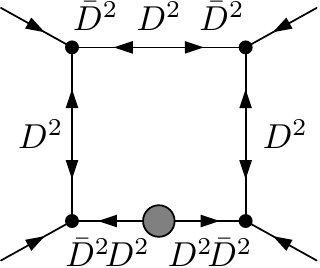}} & \multirow{4}{*}{$D^2 \hat K$} & \\
			& & $( B^* )$ \\
			& & $L L H_u H_u$ \\
			& & \\\hline
		\vspace{2mm}\includegraphics[width=20mm]{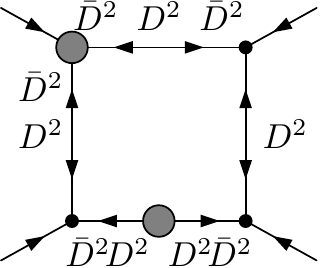} &  $D^2 \hat K \bar D^2 \hat K$ & ----- \\\hline
		\vspace{2mm}\includegraphics[width=20mm]{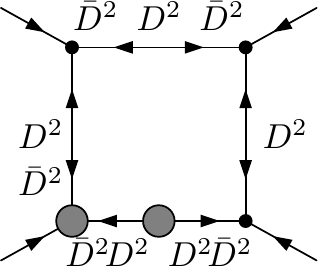} & $D^2 \hat K \bar D^2 \hat K$ & ----- \\\hline

		\vspace{2mm}\includegraphics[width=20mm]{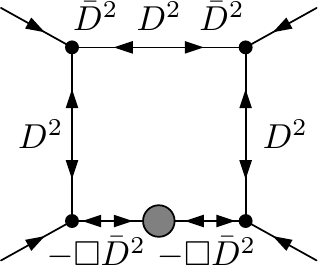} & ----- & ----- \\\hline
		\vspace{2mm}\includegraphics[width=20mm]{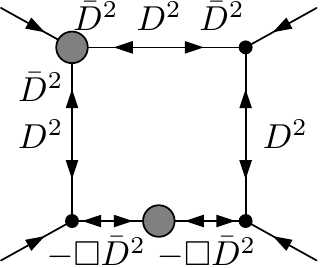} & ----- & ----- \\\hline
		\vspace{2mm}\includegraphics[width=20mm]{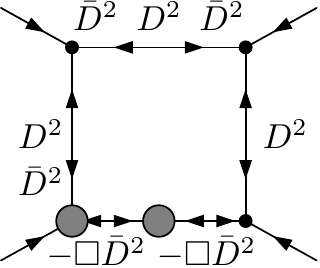} & ----- & ----- \\\hline
	\end{tabular} & 
	\begin{tabular}{| m{20mm} |c|c|}
		\hline
		Supergraph & D-algebra result & $\text{OP} \in \text{OP}_\nu$ \\\hline
		\vspace{2mm}\includegraphics[width=20mm]{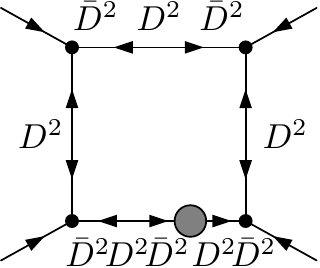} & $\hat K$ & ----- \\\hline
		\vspace{2mm}\includegraphics[width=20mm]{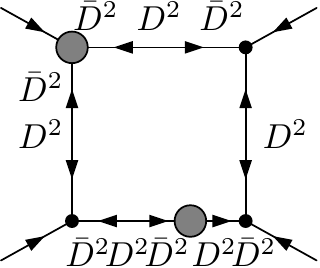} & $0$ & ----- \\\hline
		\vspace{2mm}\includegraphics[width=20mm]{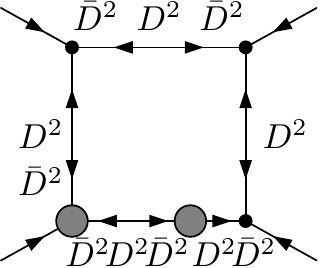} & $0$ & ----- \\\hline

		\vspace{2mm}\includegraphics[width=20mm]{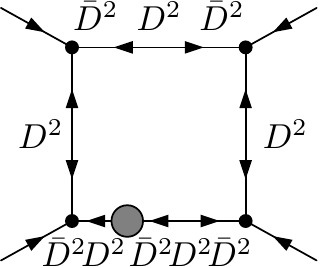} & $\hat K$ & ----- \\\hline
		\vspace{2mm}\includegraphics[width=20mm]{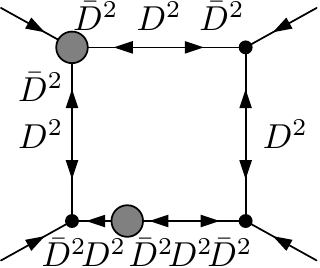} & $0$ & ----- \\\hline
		\vspace{2mm}\includegraphics[width=20mm]{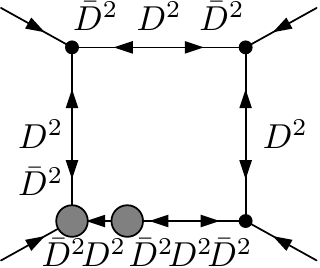} & $0$ & ----- \\\hline
	\end{tabular}
	\end{tabular}	
		\caption{Soft-$\nSUSY$ insertions up to order $3$ in the soft-$\nSUSY$ 
		scale for one-loop 1PI $\hat L \hat L \hat H_u \hat H_u$. 
		}\label{tb:SUSY_breaking_insertions_1PI_LLHuHu} 
	\end{table}

	\FloatBarrier

\section{One-loop topologies with self-energies} \label{app:susy_break_self}

	The topologies presented in this appendix are superficially divergent. Our assumption 
	is that in an actual model they are finite, so that $m_\nu$ is genuinely radiative. 
	One way by which such models can be constructed for any given topology is to postulate 
	a spontaneously broken symmetry that forbids the superficially divergent term at a more 
	fundamental level. For example, suppose that a given topology requires a 
	$\hat\Phi^\dagger_1 \hat\Phi_2$ self-energy, then the postulated symmetry should forbid 
	$\hat\Phi^\dagger_1 \hat\Phi_2$ but may allow, say, 
	$\hat\Phi^\dagger_1 \hat\Phi_2 \hat\rho$, where $\langle\rho\rangle \neq 0$ spontaneously 
	breaks the symmetry. Now, suppose that $\hat\Phi^\dagger_1 \hat\Phi_2 \hat\rho$ can arise only at loop
        level and is 1PI, then in the broken phase we have a radiative $\hat\Phi^\dagger_1 \hat\Phi_2$ which
        is necessarily convergent because $\int d^4\theta \, \hat\Phi^\dagger_1 \hat\Phi_2 \langle\hat\rho\rangle$ 
	has mass dimension 5. In order to construct genuine radiative models based on self-energy topologies 
	it may be necessary to consider more complicated models, as the simplest models in which a 
	symmetry forbids $\hat\Phi^\dagger_1\hat\Phi_2$ but allows $\hat\Phi^\dagger_1\hat\Phi_2\hat\rho$ 
	may also generate a tree-level contribution by allowing a superpotential term of the form 
	$\hat{\overline\Phi}_1\hat\Phi_2\hat\rho$. The simplest of these more complicated models 
	are those in which the self-energy topology is based on a dimension-4 superoperator that yields 
	a self-energy once $L$-number is broken. For example, $\hat\Phi^\dagger_1\hat\Phi_2\hat\rho_i$ 
	may be forbidden because $\hat\Phi^\dagger_1\hat\Phi_2$ carries $L$-number $+2$ while 
	the superfields whose scalar component break $L$-number, $\hat\rho_{1,2}$, carry an $L$-number 
	different from $-2$. Now, if $\hat\rho_1\hat\rho_2$ carries $L$-number $-2$, 
	$\hat\Phi^\dagger_1\hat\Phi_2\hat\rho_1\hat\rho_2$ is allowed. Then, if 
	$\hat\rho$'s can only interact with $\hat\Phi$'s by means of superfields charged under an 
	unbroken symmetry to which the actual leptons and $\hat\Phi$'s are blind (as the $U(1)_X$ 
	of the model example of~\secref{sec:model_example}), $\hat\Phi^\dagger_1\hat\Phi_2\hat\rho_1\hat\rho_2$ 
	is necessarily radiative and leads to a $\hat\Phi^\dagger_1\hat\Phi_2$ self-energy once $L$-number is broken. 

\medskip

	We start by considering tree-level 4-point supergraph topologies that are holomorphy compliant. 
	There are only two of such topologies, and which can be identified by the superoperators 
	$\hat A \hat B \hat C \hat D$ and $\hat A \hat B \hat C^\dagger \hat D^\dagger$. 
	Next, we consider self-energy insertions. These can be of four types: $\hat\Phi\hat{\overline\Phi}$, its $\Hc$, 
	$\hat\Phi\hat\Phi^\dagger$ and its $\Hc$. A self-energy can be inserted into the propagator 
	or into an external line. We will regard an insertion into $\hat C^\dagger$ as equivalent 
	to an insertion into $\hat D^\dagger$, since one can be obtained from the other by 
	relabelling the external lines. Similarly, an insertion into $\hat A$ is regarded 
	equivalent to an insertion into $\hat B$, $\hat C$ or $\hat D$. 
	Hence, there are $20$ one-loop 4-point topologies made with self-energies: $8$ 
	based on $\hat A \hat B \hat C \hat D$ and $12$ on $\hat A \hat B \hat C^\dagger \hat D^\dagger$. 

	Equipped with these topologies, we identify two external lines to be a pair 
	of $\hat L$'s, while the other two to be Higgses. In principle, the Higgses can be any of 
	the following configurations: $\hat H_u \hat H_u$, $\hat H_u \hat H^\dagger_d$ and 
	$\hat H^\dagger_d \hat H^\dagger_d$. We discard $3$ topologies 
	that cannot yield an $\OP \in \OPnu$: 
	\begin{itemize}
		\item Of the four topologies based on $\hat A \hat B \hat C^\dagger \hat D^\dagger$ in which 
		the self-energy insertion is into an external chiral line (say $\hat A$), only two have an external 
		pair of chiral lines. Since these chiral lines will be identified with a pair of $\hat L$'s, 
		we can label the two topologies according to the type of self-energy insertion 
		performed: $\hat L \hat A^\dagger$ and $\hat L \hat A$. Now, of these two topologies 
		only ``$\hat L \hat A$'' can yield an $\OP \in \OPnu$ because $\nSUSY$ 
		does not change the fact that the spinor projection of ``$\hat L \hat A^\dagger$'', i.e.\ 
		$L \tilde A^\dagger$, is proportional to external momenta. 
	\end{itemize}
	Of the $17$ surviving topologies we further discard the following $3$ 
	\beald
		\hat A \hat B \hat C^\dagger \hat D^\dagger & : && \hat L \hat L \hat H^\dagger_d \hat H^\dagger_d \,,\\
		\hat A \hat B \hat C^\dagger \hat D & : && \hat L \hat L \hat H^\dagger_d \hat H_u ~\text{and}~ \hat L \hat H_u \hat H^\dagger_d \hat L \,,\\
		\hat A \hat B D^2 \hat C \hat D^\dagger & : && \hat L \hat L D^2 \hat H_u \hat H^\dagger_d \,,
	\eeald
	since they yield a pure-$\nSUSYEWSB$ $\text{OP} \in \text{OP}_\nu$. The first 
	is based on $\hat A \hat B \hat C^\dagger \hat D^\dagger$ with a $\hat\Phi\hat\Phi^\dagger$ 
	self-energy insertion into the propagator by adjoining two chirality flips. 
	The second is based on $\hat A \hat B \hat \Phi \hat D$ with a $\hat{\overline \Phi}\hat C^\dagger$ 
	self-energy insertion into the external line $\hat \Phi - \hat C^\dagger$ by adjoining 
	the chirality flip $\hat\Phi\hat{\overline \Phi}$. The third is based on $\hat A \hat B \hat \Phi^\dagger \hat D^\dagger$ 
	with a $\hat{\overline \Phi}{}^\dagger \hat C$ self-energy insertion into the $\hat \Phi^\dagger - \hat C$ 
	line by adjoining the chirality flip $\hat\Phi^\dagger\hat{\overline \Phi}{}^\dagger$. 

	The surviving $7$ topologies in which the self-energy insertion is performed on the 
	propagator are depicted in the first column of~\tbref{tb:SGraph_1LoopTopSelfEnergies_LLHH}. 
	We note that the third row accounts for two topologies. The $7$ in which the insertion is 
	on the external line are listed in~\tbref{tb:SGraph_1LoopTopSelfEnergies_LLHH_cont}. 
	Notice that there are only two topologies with a self-energy insertion into an $\hat L$'s line: the 
	2nd and last rows of~\tbref{tb:SGraph_1LoopTopSelfEnergies_LLHH_cont}. 

	In the second column we show the corresponding superoperator(s), obtained by integrating by parts the $D$'s 
	in a way that avoids crossing the self-energy insertion. With this procedure, we are able to associate superoperators 
	to topologies made with self-energies that are identically zero in the SUSY limit (specifically, $\hat\Phi\hat{\overline \Phi}$ and 
	its $\Hc$). In the third column we identify the subset of $\SOPnu$ of each topology and in fourth column 
	we list the corresponding $L L H H$ operators and their schematic dependence on soft-$\nSUSY$, up 
	to order $3$ in $\Msoft$.  In order to obtain the fourth column, we considered soft-$\nSUSY$ insertions 
	as in~\secref{app:susy_break}. Particularly useful for this task was the catalogue of soft-$\nSUSY$ insertions 
	into the one-loop self-energies $\hat\Phi^\dagger\hat{\overline\Phi}{}^\dagger$ and $\hat\Phi^\dagger\hat\Phi$ 
	given in~\tbref{tb:SUSY_breaking_insertions_self_energy_PhiDPhiD} and~\tbref{tb:SUSY_breaking_insertions_self_energy_PhiDPhi}, 
	respectively.

	\begin{table}[h!t]
		\centering
		\renewcommand{\arraystretch}{1.2}
	\begin{tabular}{|c|c|l|c|}
		\hline
		Supergraph 	& \multicolumn{2}{c|}{Superoperator(s)} 	& $\OP \in \OPnu$	\\\hline 
		\multirow{3}{*}{\includegraphics[width=20mm]{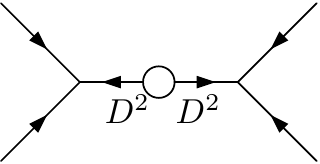}}
				& \multirow{3}{*}{$D^2 ( \hat A \hat B ) D^2 ( \hat C \hat D )$} & \multirow{3}{*}{$D^2 ( \hat L \hat L ) D^2 ( \hat H_u \hat H_u )$ {\tiny (type-II w/o)}}		& $( \Msoft^2 \mu^* ) L L H_u H^\dagger_d$  \\
				& & & $( A B^* \mu^* ) L L H_u H^\dagger_d$ \\
				& & & $( A \Msoft^2 ) L L H_u H_u$	\\\hline 
		\multirow{2}{*}{\includegraphics[width=20mm]{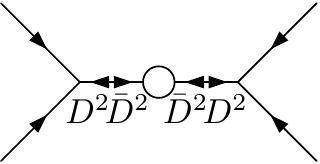}} & \multirow{2}{*}{$\hat A \hat B \hat C \hat D$} & $\hat L \hat L \hat H_u \hat H_u$ {\tiny (type-II w/)} & \multirow{2}{*}{$( B^* ) L L H_u H_u$} \\
				& & $\hat L \hat H_u \hat L \hat H_u$ {\tiny (type-I and -III)} & \\\hline
		\multirow{7}{*}{\includegraphics[width=20mm]{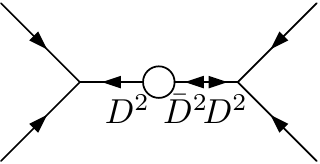}} & \multirow{7}{*}{$D^2 ( \hat A \hat B ) \hat C \hat D$} & $D^2 ( \hat L \hat L ) \hat H_u \hat H_u$ {\tiny (type-II w/o)} & $( A^* \mu^* ) L L H_u H^\dagger_d$ 		\\
				& & $D^2 ( \hat H_u \hat H_u ) \hat L \hat L$ {\tiny (type-II w/o)} & $( B^* \mu^* ) L L H_u H^\dagger_d$ 		\\
				& & & $( A^* \Msoft^2 \mu^* ) L L H_u H^\dagger_d$ \\
				& & & $( A^* A ) L L H_u H_u$ \\
				& & & $( A B^* ) L L H_u H_u$ \\
				& & & $( A^* B ) L L H_u H_u$ \\
				& & & $( \Msoft^2 ) L L H_u H_u$ \\\hline
		\multirow{3}{*}{\includegraphics[width=20mm]{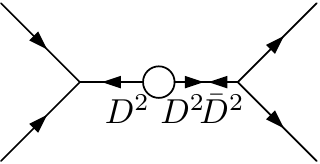}}
				& \multirow{3}{*}{$D^2 ( \hat A \hat B ) \hat C^\dagger \hat D^\dagger$} & \multirow{3}{*}{$D^2 ( \hat L \hat L ) \hat H^\dagger_d \hat H^\dagger_d$ {\tiny (type-II w/o)}} & $( B \mu ) L L H_u H^\dagger_d$		\\
				& & & $( A^* B ) L L H^\dagger_d H^\dagger_d$		\\
				& & & $( \Msoft^2 ) L L H^\dagger_d H^\dagger_d$ \\\hline 
		\multirow{2}{*}{\includegraphics[width=20mm]{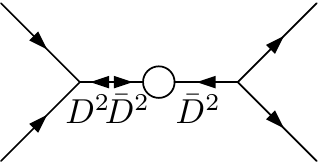}} & \multirow{2}{*}{$\hat A \hat B \bar D^2 ( \hat C^\dagger \hat D^\dagger )$} & \multirow{2}{*}{$\hat L \hat L \bar D^2 ( \hat H^\dagger_d \hat H^\dagger_d )$ {\tiny (type-II w/o)}} & $( B^* \mu ) L L H_u H^\dagger_d$ \\
				& & & $( A^* B^* ) L L H^\dagger_d H^\dagger_d$ \\\hline
		\multirow{8}{*}{\includegraphics[width=20mm]{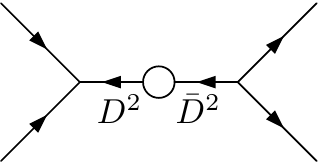}} & \multirow{8}{*}{$D^2 ( \hat A \hat B ) \bar D^2 ( \hat C^\dagger \hat D^\dagger )$} & \multirow{8}{*}{$D^2 ( \hat L \hat L ) \bar D^2 ( \hat H^\dagger_d \hat H^\dagger_d )$ {\tiny (type-II w/o)}} & $( A^* A \mu ) L L H_u H^\dagger_d$		\\
				& & & $( A B^* \mu ) L L H_u H^\dagger_d$ \\
				& & & $( A^* B \mu ) L L H_u H^\dagger_d$ \\
				& & & $( A^* B^* \mu ) L L H_u H^\dagger_d$ \\
				& & & $( A B \mu ) L L H_u H^\dagger_d$ \\
				& & & $( \Msoft^2 \mu ) L L H_u H^\dagger_d$ \\
				& & & $( A^* A^* A ) L L H^\dagger_d H^\dagger_d$ \\ 
				& & & $( A^* A B^* ) L L H^\dagger_d H^\dagger_d$ \\ 
				& & & $( A^* A^* B ) L L H^\dagger_d H^\dagger_d$ \\ 
				& & & $( A^* \Msoft^2 ) L L H^\dagger_d H^\dagger_d$ \\\hline 
	\end{tabular}
	\caption{One-loop supergraph topologies made with self-energy insertions into the propagator, and that only 
		yield $\nSUSYEWS$ contributions to $L L H H$. 
		The external superfields in first column's supergraphs are labelled as follows: $\hat A$ ($\hat C$) 
		and $\hat B$ ($\hat D$) are upper-left (-right) and lower-left (-right) external lines. 
		In the second column we display the corresponding superoperators, and where we have integrated by 
		parts the $D$'s in a way that avoids crossing over the self-energy insertions (see text). 
		In the third column we identify the corresponding $\SOP \in \SOPnu$ superoperators and their  
		underlying seesaw type, and where ``type-II w/'' and ``type-II w/o'' stand for a type-II seesaw with 
		and without a chirality flip, respectively. This column is identified with 
		the other columns by a map in which the left-to-right order by which superfields are written 
		corresponds to the alphabetic $\hat A$ to $\hat D$ sequence. In the fourth column we 
		list the $L L H H$ operators, and their dependence on soft-$\nSUSY$, that the third column's 
		superoperators generate up to order $3$ in $\Msoft$. 
	}\label{tb:SGraph_1LoopTopSelfEnergies_LLHH} 
	\end{table}

	\begin{table}[h!t]
		\centering
		\renewcommand{\arraystretch}{1.2}
	\begin{tabular}{|c|c|l|c|}
		\hline
		Supergraph 	& \multicolumn{2}{c|}{Superoperator(s)} 	& $\OP \in \OPnu$	\\\hline 
		\multirow{3}{*}{\includegraphics[width=20mm]{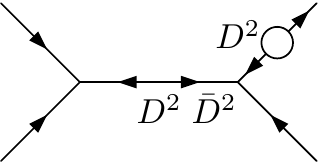}} & \multirow{3}{*}{$D^2 ( \hat A \hat B \hat D ) \hat C^\dagger$} & $D^2 ( \hat L \hat L \hat H_u ) \hat H^\dagger_d$ {\tiny (type-II w/)} & $( B \mu ) L L H_u H_u$ \\
				& & $D^2 ( \hat L \hat H_u \hat L ) \hat H^\dagger_d$ {\tiny (type-I and -III)} & $( A^* B ) L L H_u H^\dagger_d$ \\
				& & & $( \Msoft^2 ) L L H_u H^\dagger_d$ \\\hline 
		\multirow{2}{*}{\includegraphics[width=20mm]{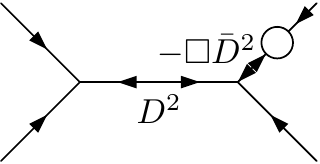}} & \multirow{2}{*}{$\hat A \hat B \hat C \hat D$} & $\hat L \hat L \hat H_u \hat H_u$, $\hat H_u \hat H_u \hat L \hat L$ {\tiny (type-II w/)} & \multirow{2}{*}{$( B^* ) L L H_u H_u$} 		\\
				& & $\hat L \hat H_u \hat L \hat H_u$, $\hat L \hat H_u \hat H_u \hat L$ {\tiny (type-I and -III)} & \\\hline 
		\multirow{9}{*}{\includegraphics[width=20mm]{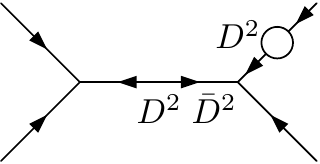}} & \multirow{9}{*}{$D^2 ( \hat A \hat B \hat D ) \hat C$} & $D^2 ( \hat L \hat L \hat H_u ) \hat H_u$ {\tiny (type-II w/)} & $( A^* \mu^* ) L L H_u H^\dagger_d$		\\
				& & $D^2 ( \hat L \hat H_u \hat L ) \hat H_u$ {\tiny (type-I and -III)} & $( B^* \mu^* ) L L H_u H^\dagger_d$ \\
				& & & $( A^* \Msoft^2 \mu^* ) L L H_u H^\dagger_d$ \\
				& & & $( A^* A ) L L H_u H_u$ \\
				& & & $( A B^* ) L L H_u H_u$ \\
				& & & $( A^* B ) L L H_u H_u$ \\
				& & & $( \Msoft^2 ) L L H_u H_u$ \\
				& & & \multicolumn{1}{l|}{only for type-II:} \\
				& & & ~~$( A^* B^* (\mu^*)^2 ) L L H^\dagger_d H^\dagger_d$ \\\hline
		\multirow{5}{*}{\includegraphics[width=20mm]{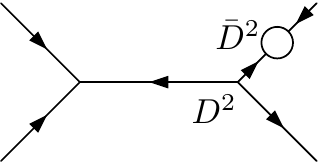}} & \multirow{5}{*}{$\bar D^2 ( D^2 ( \hat A \hat B ) \hat D^\dagger ) \hat C$} & \multirow{5}{*}{$D^2 ( \hat L \hat L ) \bar D^2 \hat H_d^\dagger \hat H_u$ {\tiny (type-II w/o)}} & $( A B^* \mu ) L L H_u H_u$		\\
				& & & $( \Msoft^2 \mu ) L L H_u H_u$ \\
				& & & $( B^* |\mu|^2 ) L L H_u H^\dagger_d$ \\
				& & & $( A^* \Msoft^2 ) L L H_u H^\dagger_d$ \\
				& & & $( A^* B^* \mu^*) L L H^\dagger_d H^\dagger_d$ \\\hline
		\multirow{3}{*}{\includegraphics[width=20mm]{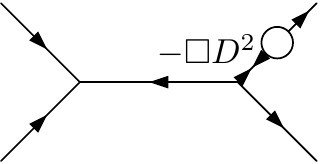}} & \multirow{3}{*}{$D^2 ( \hat A \hat B ) \hat C^\dagger \hat D^\dagger$} & \multirow{3}{*}{$D^2 ( \hat L \hat L ) \hat H^\dagger_d \hat H^\dagger_d$ {\tiny (type-II w/o)}} & $( B \mu ) L L H_u H^\dagger_d$		\\
				& & & $( A^* B ) L L H^\dagger_d H^\dagger_d$ \\
				& & & $( \Msoft^2 ) L L H^\dagger_d H^\dagger_d$ \\\hline
		\multirow{11}{*}{\includegraphics[width=20mm]{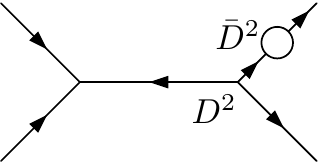}} & \multirow{11}{*}{$\bar D^2 ( D^2 ( \hat A \hat B ) \hat D^\dagger ) \hat C^\dagger$} & \multirow{11}{*}{$D^2 ( \hat L \hat L ) \bar D^2 \hat H^\dagger_d \hat H^\dagger_d$ {\tiny (type-II w/o)}} & $\mu ( A^* A ) L L H_u H^\dagger_d$		\\
				& & & $( A B^* \mu ) L L H_u H^\dagger_d$ \\ 
				& & & $( A^* B \mu ) L L H_u H^\dagger_d$ \\ 
				& & & $( \Msoft^2 \mu ) L L H_u H^\dagger_d$ \\
				& & & $( A \mu^2 ) L L H_u H_u$ \\
				& & & $( B \mu^2 ) L L H_u H_u$ \\
				& & & $( A \Msoft^2 \mu^2 ) L L H_u H_u$ \\
				& & & $( A^* A^* A ) L L H^\dagger_d H^\dagger_d$ \\
				& & & $( A^* \Msoft^2 ) L L H^\dagger_d H^\dagger_d$ \\
				& & & $( A^* A \mu ) L L H_u H^\dagger_d$ \\
				& & & $( A^* B \mu ) L L H_u H^\dagger_d$ \\\hline
		\multirow{2}{*}{\includegraphics[width=20mm]{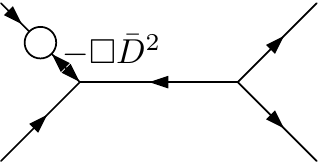}} & \multirow{2}{*}{$\hat A \hat B \bar D^2 ( \hat C^\dagger \hat D^\dagger )$} & \multirow{2}{*}{$\hat L \hat L \bar D^2 ( \hat H^\dagger_d \hat H^\dagger_d )$ {\tiny (type-II w/o)}} & $( B^* \mu ) L L H_u H^\dagger_d$	\\
				& & & $( A^* B^* ) L L H^\dagger_d H^\dagger_d$ \\\hline 
	\end{tabular}
	\caption{Same as in~\tbref{tb:SGraph_1LoopTopSelfEnergies_LLHH} but now for self-energy insertions 
		into external lines. 
	}\label{tb:SGraph_1LoopTopSelfEnergies_LLHH_cont} 
	\end{table}

	\begin{table}[h!t]
		\centering
		\renewcommand{\arraystretch}{1.2}
	\begin{tabular}{cc}
	\begin{tabular}{| m{20mm} |c|}
		\hline
		Supergraph & D-algebra result \\\hline
		\vspace{2mm}\includegraphics[width=20mm]{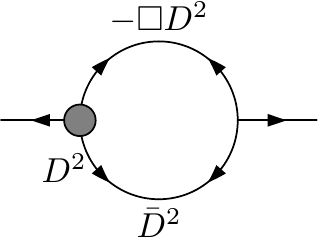} & ----- \\\hline
		\vspace{2mm}\includegraphics[width=20mm]{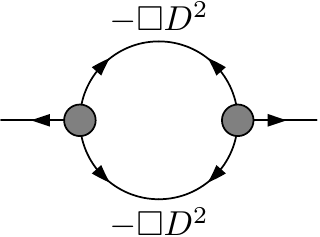} & ----- \\\hline
		\hline
		Supergraph & D-algebra result \\\hline
		\vspace{2mm}\includegraphics[width=20mm]{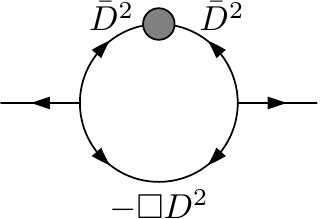} & $\bar D^2 \hat K$ \\\hline
		\vspace{2mm}\includegraphics[width=20mm]{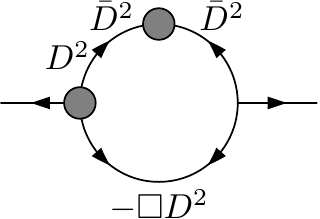} & $\bar D^2 \hat K D^2 \hat K$ \\\hline

		\vspace{2mm}\includegraphics[width=20mm]{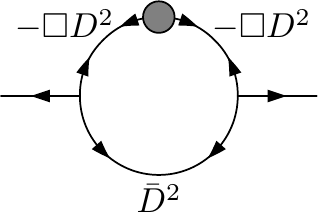} & ----- \\\hline
		\vspace{2mm}\includegraphics[width=20mm]{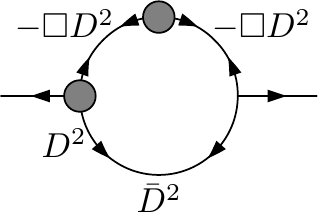} & ----- \\\hline
	\end{tabular} & 
	\begin{tabular}{| m{20mm} |c|}
		\hline
		Supergraph & D-algebra result \\\hline
		\vspace{2mm}\includegraphics[width=20mm]{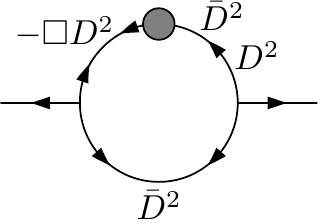} & $\hat K$ \\\hline
		\vspace{2mm}\includegraphics[width=20mm]{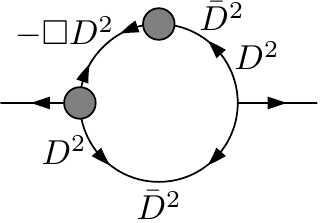} & $0$ \\\hline
		\vspace{2mm}\includegraphics[width=20mm]{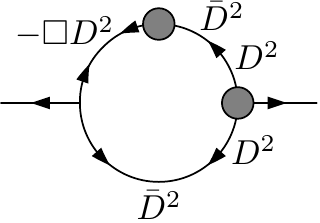} & $0$ \\\hline

		\vspace{2mm}\includegraphics[width=20mm]{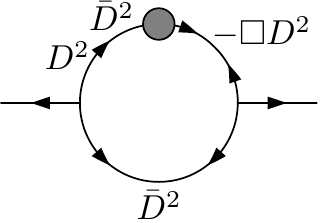} & $\hat K$ \\\hline
		\vspace{2mm}\includegraphics[width=20mm]{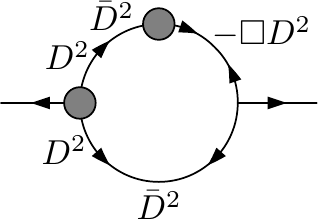} & $0$ \\\hline
		\vspace{2mm}\includegraphics[width=20mm]{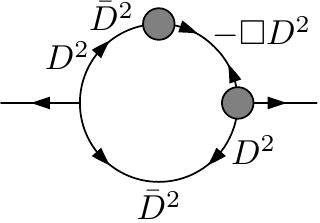} & $0$ \\\hline
	\end{tabular}
	\end{tabular}
	\caption{$\nSUSY$ insertions up to order $3$ in the soft-$\nSUSY$ scale for one-loop 
		$\hat\Phi^\dagger\hat{\overline \Phi}{}^\dagger$. 
		}\label{tb:SUSY_breaking_insertions_self_energy_PhiDPhiD} 
	\end{table}

	\begin{table}[h!t]
		\centering
		\renewcommand{\arraystretch}{1.2}
	\begin{tabular}{cc}
	\multicolumn{2}{c}{\begin{tabular}{| m{20mm} |c|}
		\hline
		Supergraph & D-algebra result \\\hline
		\vspace{2mm}\includegraphics[width=20mm]{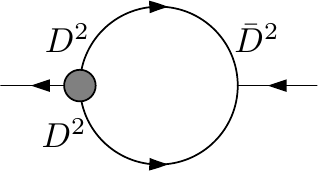} & $D^2 \hat K$ \\\hline
		\vspace{2mm}\includegraphics[width=20mm]{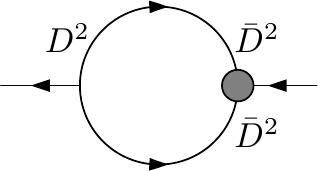} & $\bar D^2 \hat K$ \\\hline
		\vspace{2mm}\includegraphics[width=20mm]{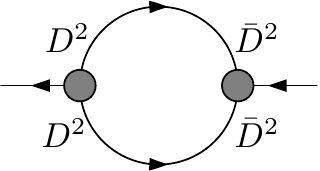} & $D^2 \hat K \bar D^2 \hat K$ \\\hline
	\end{tabular}} \\
	\begin{tabular}{| m{20mm} |c|}
		\hline
		Supergraph & D-algebra result \\\hline
		\vspace{2mm}\includegraphics[width=20mm]{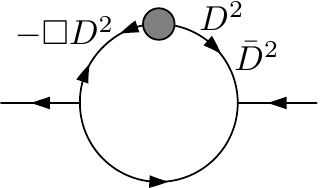} & $D^2 \hat K$ \\\hline
		\vspace{2mm}\includegraphics[width=20mm]{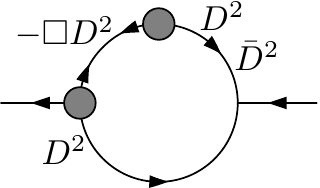} & $0$ \\\hline
		\vspace{2mm}\includegraphics[width=20mm]{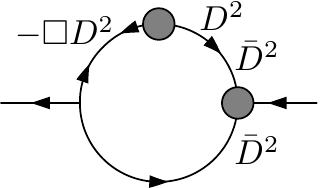} & $D^2 \hat K \bar D^2 \hat K$ \\\hline
		
		\vspace{2mm}\includegraphics[width=20mm]{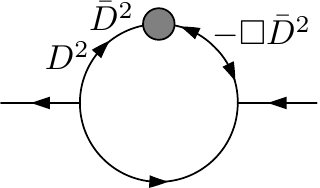} & $\bar D^2 \hat K$ \\\hline
		\vspace{2mm}\includegraphics[width=20mm]{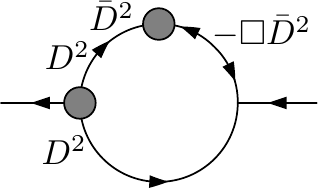} & $\bar D^2 \hat K D^2 \hat K$ \\\hline
		\vspace{2mm}\includegraphics[width=20mm]{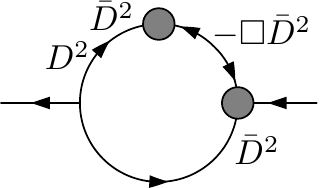} & $0$ \\\hline
	\end{tabular} & 
	\begin{tabular}{| m{20mm} |c|}
		\hline
		Supergraph & D-algebra result \\\hline
		\vspace{2mm}\includegraphics[width=20mm]{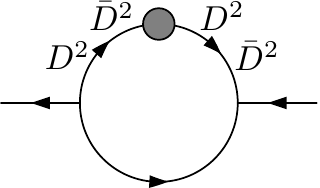} & $D^2 \bar D^2 \hat K , \hat K$ {\small (*)} \\\hline
		\vspace{2mm}\includegraphics[width=20mm]{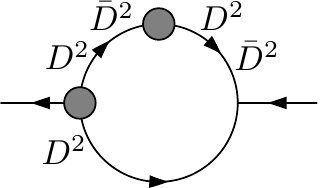} & $D^2 \hat K D^2 \bar D^2 \hat K$ \\\hline
		\vspace{2mm}\includegraphics[width=20mm]{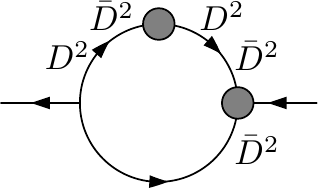} & $\bar D^2 \hat K D^2 \bar D^2 \hat K$ \\\hline

		\vspace{2mm}\includegraphics[width=20mm]{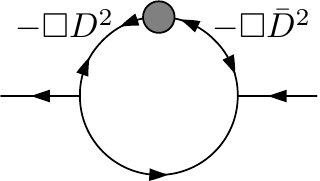} & $\hat K$ \\\hline
		\vspace{2mm}\includegraphics[width=20mm]{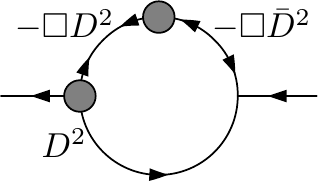} & $0$ \\\hline
		\vspace{2mm}\includegraphics[width=20mm]{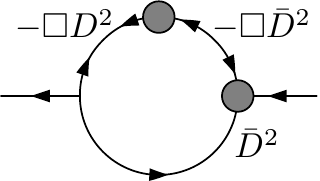} & $0$ \\\hline
	\end{tabular}
	\end{tabular}
	\caption{$\nSUSY$ insertions up to order $3$ in the soft-$\nSUSY$ scale for one-loop 
		$\hat\Phi^\dagger\hat\Phi$. (*) stands for an omitted term that vanishes as $\pext \to 0$. 
		}\label{tb:SUSY_breaking_insertions_self_energy_PhiDPhi} 
	\end{table}

	\FloatBarrier

\section{Soft SUSY breaking insertions in the model of~\cite{bmw}} \label{app:bmw} 

	The soft-$\nSUSY$ potential is parameterised according to the conventions set 
	at the beginning of~\appref{app:susy_break} and having~\eqref{eq:SPo_bmw} as 
	the superpotential of reference. 

\begin{figure}[h!t]
          \vspace{0.2in}
          \begin{center}
                \begin{tabular}{cc}
			\includegraphics[width=50mm]{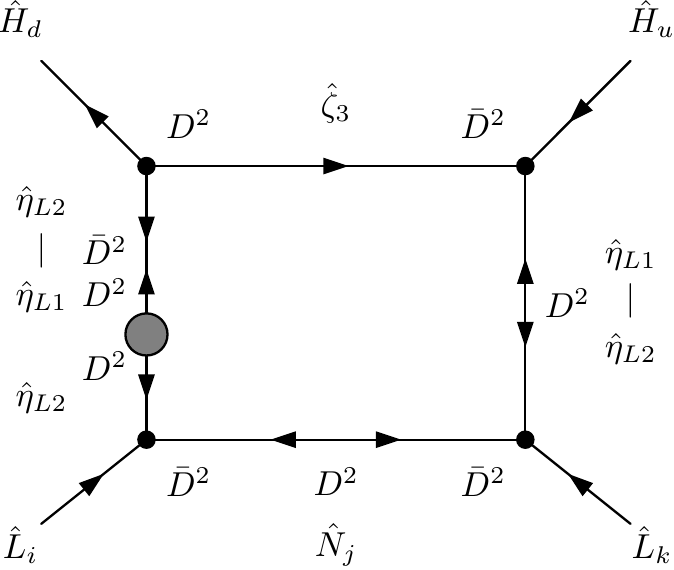} &
                	\includegraphics[width=50mm]{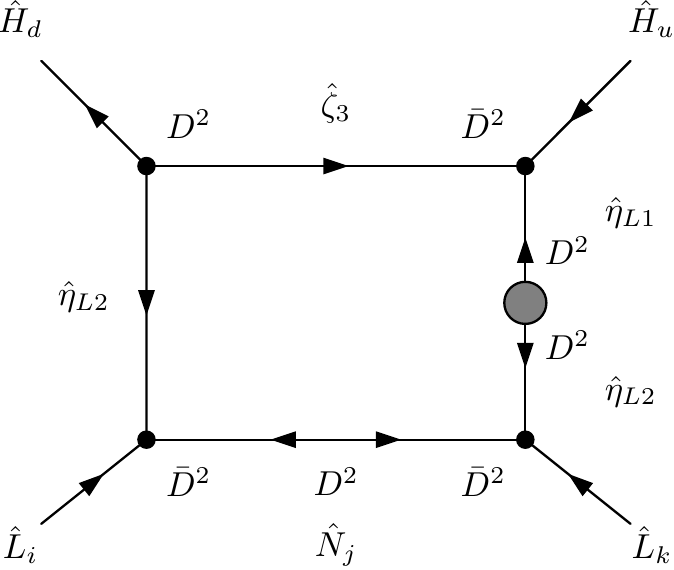}
		\end{tabular}
                \caption{Supergraphs for leading order $B_{\eta_L}$-term (grey blobs) contribution to $L L H_u H_d^\dagger$. 
                }\label{fig:LLHuHd_BetaL_ma}
          \end{center}
          \vspace{-0.2in}
        \end{figure}

	We have made a thorough calculation of soft-$\nSUSYEWS$ contributions to $L L H H$ up to order $2$ in the 
	soft-$\nSUSY$ scale. This allowed us to confirm that the only type of soft-$\nSUSY$ insertions into 
	$\hat L \hat L \hat H_u \hat H_u$ -- which can be identified 
	by their dependence on $f_{10}^2$ in the expression given below -- that yielded an $L L H H$ were 
	$B$-terms, in agreement with~\tbref{tb:SUSY_breaking_insertions_1PI_LLHuHu}. 
	In the simplifying limit of $M_{N_i} = \mu_{s3} = \mu_{L2} = M_N$ we find that the 
	effective Lagrangian contains 
	\beal
		\frac{1}{384 \pi^2 M_N^3} \Bigg[ & -f_{10}^2 \Bigg(  
			B_N 
			+ \left( 1 + \left( \frac{f^*_9}{f_{10}} \frac{\mu_L}{M_N} \right)^2 \right) B_{\zeta_3} 
			+ 2 B_{\eta_L}
			+ \frac{4 f^*_9}{f_{10}} \frac{\mu_L}{M_N} \bigg( (\Msoft^2)_{\zeta_3} + (\Msoft^2)_{\eta_{L2}} \bigg) 
		\Bigg) H_u H_u \Bigg. \nonumber\\
		& +2 f^*_9 f_{10} \Bigg( 
			2 A^*_9 M_N 
			+ B_N - B_{\zeta_3} 
			- \frac{|\mu_L|^2}{M_N^2} B_{\eta_L} 
			-\frac{2 f^*_9}{f_{10}} \frac{\mu_L}{M_N} \bigg( (\Msoft^2)_{\zeta_3} + (\Msoft^2)_{\eta_{L2}} \bigg) 
		\Bigg) H_u H^\dagger_d \nonumber\\
		& +(f^*_9)^2 \Bigg( 
			4 A^*_9 M_N 
			+ B_N - B_{\zeta_3} - 2 B_{\eta_L} 
		\Bigg) H^\dagger_d H^\dagger_d 
		\Bigg. \Bigg] \, L \,\mathbf{f_{16}} \mathbf{f^T_{16}} \, L \,.
	\eeal
	A fortuitous cancellation in the all masses equal limit prevents a $\mu_L$-independent $B_{\eta_L}$-term 
	contribution to $L L H_u H_d^\dagger$ from appearing. This cancellation happens between the 
	diagram with a $B_{\eta_L}$ inserted into the $L \to H^\dagger_d$ line and the diagram with a 
	$B_{\eta_L}$ inserted into the $L \to H_u$ line, as shown in~\figref{fig:LLHuHd_BetaL_ma}. 
	To be precise, the $\mu_L$-independent $B_{\eta_L}$-term contributions coming from the first and second 
	supergraphs add up to 
	\bea
		\lim_{M_N = \mu_{L2} = \mu_{s3}} \Big( M_N \mu_{L2}^2 E_0 + M_N \Big[ D_0 + \mu_{L2}^2 E_0 \Big] \Big) 
		\int d^4\theta \, D^2 \left( \frac{\hat X^\dagger \hat X}{M^2_X} \right)_{B_{\eta_L}} 
		\hat L \hat L \hat H_u \hat H_d^\dagger = 0 \,,
	\eea
	respectively, and where $D_0$ and $E_0$ are the following scalar one-loop integrals evaluated 
	at $\pext = 0$: $D_0(...,M_N^2,\mu_{L2}^2,\mu_{s3}^2,\mu_{L2}^2)$ and 
	$E_0(...,M_N^2,\mu_{L2}^2,\mu_{L2}^2,\mu_{s3}^2,\mu_{L2}^2)$, respectively. $\hat X^\dagger \hat X$ is a 
	$\nSUSY$ spurion insertion (cf.~\eqref{eq:SoftSUSYbreak_vertices}). 
	The remainder of the second supergraph generates the $\mu_L$-dependent 
	$B_{\eta_L}$-term contribution: 
	\be
		M_N E_0 
		\int d^4\theta \, \bar D^2 D^2 \left( \frac{\hat X^\dagger \hat X}{M^2_X} \right)_{B_{\eta_L}} 
			\hat L \hat L D^2 \hat H_u \hat H^\dagger_d 
		\supset |\mu_L|^2 B_{\eta_L} M_N E_0 \, L L H_u H^\dagger_d \,.
	\ee

\section{Radiative renormalisable couplings in SUSY} \label{app:radiative_couplings} 

	In this appendix we show that, by relying just on the renormalisability of the 
	superpotential, some four-scalar couplings can be genuinely radiative. 

	Let $\hat X_i$ be a chiral scalar superfield of components $\phi_i$, $\chi_i$ and $F_i$. 
	In each statement we increase $i$ whenever a field/superfield is introduced that does 
	not need to have the identity of a previously introduced field/superfield. 
	For instance, when an $F_i$ is used, we say that it contains 
	some $\phi$'s labelled by increasing the counter $i$. In this way no 
	{\it a priori} assumption is made regarding the form of the superpotential. 

\begin{figure}[h!t]
          \vspace{-0.0in}
          \begin{center}
                \includegraphics[width=160mm]{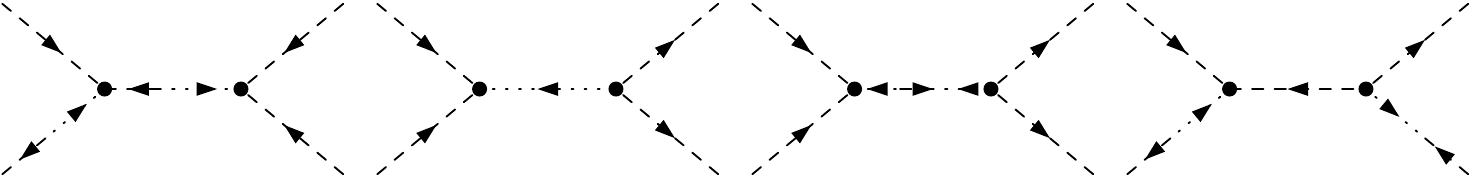}
                \caption{All possible tree-level topologies of 
		$\phi^\dagger \phi^3$ and $(\phi^\dagger\phi)^2$ under the 
		assumption of a renormalisable superpotential.
		Exchanges of $F_Y$, $\bar\phi_Y$ ($F_Y$-induced) and $\phi_Y$, 
		as mentioned in the text, correspond to the last three diagrams. 
                }\label{fig:4Scalar_Tree_Topologies}
          \end{center}
          \vspace{-0.25in}
        \end{figure}

	The only radiative possibility for renormalisable spinor-scalar interactions is (schematically) 
	\be
		\frac{1}{M} \int d^4\theta \, \hat X^\dagger_1 \hat X_2 \hat X_3 \supset 
			\frac{1}{M} F^\dagger_1 \chi_2 \chi_3 \supset \phi_4 \chi_2 \chi_3 \,.
	\ee
	This means that $\hat X_2 \hat X_3 \hat X_4$ is symmetric and, since it is renormalisable, 
	allowed in the superpotential. Thus, there is a tree-level contribution to $\phi_4 \chi_2 \chi_3$. 
	Regarding three-scalar interactions, the possibilities are 
	\beald
		& \int d^4\theta \, \hat X^\dagger_1 \hat X_2 \supset F^\dagger_1 F_2 \supset M \phi_3 \phi_4 \phi^\dagger_5 \,,\\
		& \frac{1}{M} \int d^4\theta \, \hat X^\dagger_1 \hat X_2 \hat X_3 \supset \frac{1}{M} F^\dagger_1 F_2 \phi_3 \supset M \phi_4 \phi^\dagger_5 \phi_3 \,,
	\eeald
	where both say that $\hat X_2 \hat X_3 \hat X_4$ is symmetric, and thus allowed in the superpotential. In addition, 
	the first necessitates $\hat X_1 \hat X_3 \hat X_4, \hat X_2 \hat X_5 \subset \mathcal{W}$, while the second necessitates 
	$\hat X_2 \hat X_5, \hat X_1 \hat X_4 \subset \mathcal{W}$. Hence, in both cases there is a tree-level contribution to 
	$\phi_3 \phi_4 \phi^\dagger_5$ once $F_2$ is integrated out. Regarding four-scalar interactions, we have 
	\beald
		\text{(a)} &&& \int d^4\theta \, \hat X^\dagger_1 \hat X_2 \supset \phi_3 \phi_4 \phi^\dagger_5 \phi^\dagger_6 \,,\\
		\text{(b)} &&& \frac{1}{M} \int d^4\theta \, \hat X^\dagger_1 \hat X_2 \hat X_3 \supset \phi_4 \phi^\dagger_5 \phi^\dagger_6 \phi_3 \,,\\
		\text{(c)} &&& \frac{1}{M} \int d^4\theta \, \hat X^\dagger_1 \hat X_2 \hat X_3 \supset \phi_4 \phi_5 \phi^\dagger_6 \phi_3 \,,\\
		\text{(d)} &&& \frac{1}{M^2} \int d^4\theta \, \hat X^\dagger_1 \hat X_2 \hat X_3 \hat X_4 \supset \phi_5 \phi^\dagger_6 \phi_3 \phi_4 \,,\\
		\text{(e)} &&& \frac{1}{M^2} \int d^4\theta \, \hat X^\dagger_1 \hat X_2 \hat X^\dagger_3 \hat X_4 \supset \phi_5 \phi^\dagger_6 \phi^\dagger_3 \phi_4 \,.
	\eeald
	(a) and (b) entail a tree-level contribution. (c) and (d) have tree-level contributions if and only if there exists 
	a representation $\hat Y \sim \hat X_2$ such that $\hat Y \hat X_3 \hat X_4 \hat X_5$ is generated 
	at tree-level; in the case of (c), this happens if $\hat X_1$ is massive. 
	(e) has a tree-level contribution if and only if the model contains 
	a representation $\hat Y$ such that 
	\be
		\hat Y \hat X_4 \hat X_5 , \hat Y \hat X_3 \hat X_6 \subset \mathcal{W} ~~
		\text{or} ~~ 
		\hat Y \hat X_1 \hat X_3 , \hat Y \hat X_2 \hat X_4 \subset \mathcal{W} \,,
	\ee
	corresponding to the tree-level exchange of $F_Y$ or $\phi_Y$, respectively. 
	If former's case $\hat Y$ has a mass term, there is also a contribution due to an 
	exchange of $\bar\phi_Y$ and the sum of the two gives 
	\be
		\phi^\dagger_3 \phi^\dagger_6 \frac{-\Box}{-\Box - M^2_Y} ( \phi_4 \phi_5 ) \,.
	\ee 
	For an easier understanding of the ``only if'' part of these assertions, we show in~\figref{fig:4Scalar_Tree_Topologies} all 
	possible realisations of tree-level $\phi^\dagger \phi^3$ and $(\phi^\dagger\phi)^2$ under the assumption of a 
	renormalisable superpotential. We use auxiliary fields, shown as dotted lines with an arrowhead, to make 
	clear the holomorphy of the superpotential. 

	To conclude, four-scalar couplings coming from (c), (d) or (e) are possible radiative 
	couplings in a supersymmetric setting.

\section{Model example} \label{app:model_example}

	Our conventions regarding $SU(2)_L$ contractions in the superpotential of~\eqref{eq:SPot_model_example} 
	are fully specified by the following. 
	Reading each term from left to right, let $\hat A$ be the first doublet superfield and $\hat B$ the second, 
	and let $\epsilon_{12} = 1$ be the totally anti-symmetric tensor. 
	Then, 
	\beald
		\mathbf{3} \otimes \mathbf{2} \otimes \mathbf{2} \supset \mathbf{1} : &~~&& \hat \Delta \hat A \hat B := \hat\Delta^\alpha \underline{\hat A}{}^a T^\alpha_{ab} \hat B^b 
			= \hat\Delta^\alpha \hat A^a ( \epsilon T^\alpha )_{ab} \hat B^b \,, \\
		 &&& ( \hat \Delta \hat A \hat B )^\dagger := \hat\Delta^{\dagger\alpha} \hat B^{\dagger a} T^\alpha_{ab} \underline{\hat A^\dagger}{}^b 
			= \hat\Delta^{\dagger\alpha} \hat B^{\dagger a} ( -T^\alpha \epsilon )_{ab} \hat A^{\dagger b} \,,\\
		\mathbf{3} \otimes \mathbf{3} \supset \mathbf{1} : &&& \hat\Delta\hat{\overline \Delta} := 2 \hat\Delta^\alpha \hat{\overline \Delta}{}^\beta \text{Tr}\left[ T^\alpha T^\beta \right] \,,\\
		\mathbf{2} \otimes \mathbf{2} \supset \mathbf{1} : &&& \hat A \hat B := ( \hat A \hat B ) := \hat A^a \epsilon_{ab} \hat B^b \,, \label{eq:SU2L_contraction_DEF}
	\eeald
	where $\underline{\hat A}{}^a := -\epsilon_{ab} \hat A^b$. 
	
	Useful identities are 
	\beald
		\left[ \hat L^a_i ( \epsilon T^\alpha )_{ab} \hat L^b_j \right] \left[ \hat H_u^c ( \epsilon T^\alpha )_{cd} \hat H_u^d \right]
			 & = -\frac{1}{2} \left( \hat L_i \hat H_u \right) \left( \hat L_j \hat H_u \right) \,,\\
		\left[ \hat L^a_i ( \epsilon T^\alpha )_{ab} \hat L^b_j \right] \left[ H_u^c ( \epsilon T^\beta T^\alpha )_{cd} H_u^d \right] 
			 & = \frac{1}{2} \left( \hat L_{(i} \hat H_u \right) \left[ \hat H_u^a ( \epsilon T^\beta )_{ab} \hat L_{j)}^b \right] \,, \\
		\left[ \hat L^a_i ( \epsilon T^\alpha )_{ab} \hat L^b_j \right] \left[ \hat H_u^c ( \epsilon T^\beta T^\alpha )_{cd} \hat H_u^d \right] \left[ \hat H^{\dagger e} T^\beta_{ef} \hat H^f \right] 
			 & = \bead[t]
				& \frac{1}{4} \left( \hat L_{(i} \hat H_u \right) \left( \hat H_u \hat H \right) \hat H^\dagger \hat L_{j)} \\
			 	& + \frac{1}{8} \left( \hat L_i \hat H_u \right) \left( \hat L_j \hat H_u \right) \hat H^\dagger \hat H \,,
			\eead
	\eeald
	where indices within $(~)$ are symmetrised in a normalised way.

\medskip

	We define the following abbreviations for scalar one-loop integrals~\cite{Passarino:1978jh} 
	evaluated at $\pext = 0$: 
	\beald
		& C_0 := C_0(0,0,0,M^2_{X_1},M^2_{X_2},M^2_{X_3}) \,, \\
		& D_{0,0} := D_0(0,0,0,0,0,0,M^2_{X_1},M^2_{X_2},M^2_{X_3},0) \,, \\
		& D_{0,i} := D_0(0,0,0,0,0,0,M^2_{X_1},M^2_{X_2},M^2_{X_3},M^2_{X_i}) \,, \\
		& E_{0,i} := E_0(0,0,0,0,0,0,0,0,0,0,M^2_{X_1},M^2_{X_2},M^2_{X_3},M^2_{X_3},M^2_{X_i}) \,,
	\eeald
	where $i = 1,2,3$.

	\subsection{Dimension-7 $\text{OP} \in \text{OP}_\nu$} \label{app:model_example_dim7}

		The supergraphs of~\figref{fig:SuperGraph_Gauge_dim7_model_example} involving each of 
		the $SU(2)_L \otimes U(1)_Y$ gauge vector superfields turn out to add up to an overall 
		dependence in which all $M_{X_i}^2$ are equally weighed. This is due to a partial cancellation 
		between upper and lower diagrams. Hence, the possibility of attaching 
		$\hat V_{SU(2)_L}$ to either $\hat X_1 / \hat{\overline X}_1$ or $\hat X_2 / \hat{\overline X}_2$ 
		simplifies to a multiplicative factor of $2$. We thus find that the effective K\"ahler potential contains 
		\be
			\frac{\mathbf{a} M_{X_3} C_0}{32 \pi^2 M_\Delta^2} D^2 ( \hat L^a \hat L^b ) \hat H_u^c \hat H_u^d \left[ 
				g'^2 \hat V_{U(1)_Y} \epsilon_{ac} \epsilon_{bd} 
				+ 2 g^2 \hat V^\alpha_{SU(2)_L} \epsilon_{ac} ( \epsilon T^\alpha )_{db} 
			\right] \,. \label{eq:SOP_LLHHV_app}
		\ee
		Hence, the effective Lagrangian contains 
		\be
			\frac{g^2 \mathbf{a} M_{X_3} C_0}{32 \pi^2 M_\Delta^2} \left( \frac{1}{2 c_w^2} \left( L H_u \right) \left( L H_u \right) 
				\left[ H^\dagger_u H_u + c_{2 w} H^\dagger_d H_d \right] + \left( L H_u \right) \left( H_u H_d \right) H^\dagger_d L \right) \,. \label{eq:gLLHHHH_app}
		\ee
		The other supergraphs give 
		\beal
		& \bead
			& -\frac{\mathbf{a} M_{X_3}}{32 \pi^2 M^2_\Delta} \sum_{i = 1}^2 |\lambda_i|^2 ( D_{0,3} + M^2_{X_i} E_{0,i} ) \int d^4\theta \, D^2 ( \hat L^a_i \hat L^b_j ) \hat H_u^c \hat H_u^d \hat H_u^{\dagger e} \hat H_u^f \epsilon_{ac} \epsilon_{bd} \delta_{ef} \\
			& +\frac{\mathbf{b} M_{X_3} M_{X_1} M_{X_2}}{32 \pi^2 M_\Delta} \sum_{i = 1}^2 |\lambda_i|^2 E_{0,i} \int d^4\theta \left( \hat L_i \hat H_u \right) \left( \hat L_j \hat H_u \right) \hat H^\dagger_u \hat H_u 
		\eead \label{eq:SOP_LLHHHH_app} \\
		& \bead
			& & \supset & -\frac{\mathbf{a} M_{X_3} |\mu|^2}{16 \pi^2 M^2_\Delta} \sum_{i = 1}^2 |\lambda_i|^2 ( D_{0,3} + M^2_{X_i} E_{0,i} )\left[ 
				\frac{1}{2} \left( L H_u \right) \left( L H_u \right) H^\dagger_d H_d 
				- \left( L H_u \right) \left( H_u H_d \right) H^\dagger_d L 
			\right] \\ 
			&&& + \frac{\mathbf{b} M_{X_3} M_{X_1} M_{X_2} \mu}{32 \pi^2 M_\Delta} \sum_{i = 1}^2 |\lambda_i|^2 E_{0,i} \left( L H_u \right) \left( L H_u \right) \left( H_u H_d \right) \,.
		\eead \label{eq:muLLHHHH_app}
		\eeal

	\subsection{Dimension-5 $\text{OP} \in \text{OP}_\nu$} \label{app:model_example_dim5}

		Inspection of~\tbref{tb:SUSY_breaking_insertions_A_D2LLHuHu_LLD2HuHu},~\tbref{tb:SUSY_breaking_insertions_B_D2LLHuHu_LLD2HuHu},~\tbref{tb:SUSY_breaking_insertions_M2_D2LLHuHu_LLD2HuHu},~\tbref{tb:SUSY_breaking_insertions_BDEL_MDEL2_D2LLHuHu_LLD2HuHu} and~\tbref{tb:SUSY_breaking_insertions_1PR_LLHuHu} reveals that, 
		up to order $3$ in soft-$\nSUSY$, there are $22$ terms\footnote{Recall 
		that in those tables we suppressed insertions that were redundant 
		due to some symmetry of the supergraph topology. In here, we are counting 
		them provided they involve a distinct set of superfields.} contributing to 
		$L L H_u H_u$ ($3$ of them proportional to $\bar\lambda_X$), 
		$10$ terms to $L L H_u H^\dagger_d$ and $1$ to 
		$L L H_d^\dagger H_d^\dagger$. To be specific, their 
		contribution to the effective Lagrangian reads  
		\beal
			-\frac{\lambda_1\lambda_2}{32 \pi^2 M_\Delta^2} \Bigg( 
				\Bigg[ 
				& \lambda^*_X M_{X_3} \Bigg( \bead[t]
					& \sum_{i = 1}^2 \left[ (\Msoft^2)_{X_i} ( C_0 + M_{X_i}^2 D_{0,i} ) 
					+ (\Msoft^2)_{\overline X_i} M_{X_i}^2 D_{0,i} \right] \Bigg. \Bigg. \Bigg. \\
					& + \left[ (\Msoft^2)_{X_3} + (\Msoft^2)_{\overline X_3} \right] ( C_0 + M_{X_3}^2 D_{0,3} ) 
					+ A^*_X ( A_1 + A_2 ) C_0 \Bigg)
				\eead \nonumber\\
			& - \bar\lambda_X M_\Delta \Bigg( 
					M_{X_3} ( B_{X_1} M_{X_2} D_{0,1} + B_{X_2} M_{X_1} D_{0,2} ) 
					+ B_{X_3} M_{X_1} M_{X_2} D_{0,3} 
				\Bigg) \nonumber\\
			& + \lambda^*_X ( A_1 + A_2 ) \Bigg( M_{X_3} \sum_{i = 1}^2 B_{X_i} M_{X_i} D_{0,i} 
					+ B_{X_3} ( C_0 + M_{X_3}^2 D_{0,3} ) - \frac{M_{X_3} B_\Delta C_0}{M_\Delta} \Bigg) \nonumber\\
			& + \lambda^*_X A^*_X \Bigg( M_{X_3} \sum_{i = 1}^2 B_{X_i} M_{X_i} D_{0,i} 
					+ B_{X_3} M_{X_3}^2 D_{0,3} \Bigg)
				\Bigg] H_u H_u \nonumber
		\eeal
		\beal
			\phantom{-\frac{\lambda_1\lambda_2}{32 \pi^2 M_\Delta^2} \Bigg(} & - 2 \mu^* \lambda^*_X \Bigg[ \bead[t] & M_{X_3} \left( 
							A^*_X C_0 + \sum_{i = 1}^2 M_{X_i} B_{X_i} D_{0,i} 
							-\frac{B_\Delta C_0}{M_\Delta} \right) 
						+ B_{X_3} \left( C_0 + M_{X_3}^2 D_{0,3} \right ) \nonumber\\
							& + A^*_X M_{X_3} \Bigg( \bead[t] & \sum_{i = 1}^2 (\Msoft^2)_{X_i} D_{0,i} 
								+ \frac{1}{2} \left[ (\Msoft^2)_{X_3} + (\Msoft^2)_{\overline X_3} \right] D_{0,3} \\
										& -\frac{(\Msoft^2)_{\Delta} C_0}{M_\Delta^2} \Bigg)  
			\Bigg] H_u H^\dagger_d \eead \eead \nonumber\\
			& + (\mu^*)^2 \lambda^*_X A^*_X B_{X_3} D_{0,3} \, H^\dagger_d H^\dagger_d 
			\Bigg) L \, \boldsymbol{\lambda_L} L \,. \label{eq:model_example_full3SUSYbreak}
		\eeal
		These results have been confirmed by standard means of calculation, and further 
		checked against algorithmic evaluations with FeynArts/FormCalc~\cite{feynarts}. To 
		generate the necessary model files we have used to our advantage 
		FeynRule's~\cite{feynrules} support for superfields.


\begin{thebibliography}{99}


\bibitem{seesaw:I}
%%\cite{Minkowski:1977sc}
%\bibitem{Minkowski:1977sc}
  P.~Minkowski,
  %``mu --> e gamma at a Rate of One Out of 1-Billion Muon Decays?,''
  Phys.\ Lett.\ B {\bf 67} (1977) 421;
  %%CITATION = PHLTA,B67,421;%%
  %
%%\cite{Ramond:1979py}
%\bibitem{Ramond:1979py}
  M.~Gell-Mann, P.~Ramond and R.~Slansky,
  ``The Family Group in Grand Unified Theories,''
  in {\it Sanibel Conference} (1979), CALT-68-700, 
  reprinted in hep-ph/9809459; 
  %%CITATION = HEP-PH/9809459;%%
%%\cite{GellMann:1980vs}
%\bibitem{GellMann:1980vs}
  M.~Gell-Mann, P.~Ramond and R.~Slansky, in {\it Complex Spinors and
  Unified Theories} eds. P.~Van.~Nieuwenhuizen and D.~Z.~Freedman,
  {\it Supergravity} (North-Holland, Amsterdam, 1979), 
  p.315 [Print-80-0576 (CERN)], reprinted in arXiv:1306.4669;
  %
  T.~Yanagida, in {\it Proceedings of the Workshop on the Unified Theory
  and the Baryon Number in the Universe}, eds. O.~Sawada and
  A.~Sugamoto (KEK, Tsukuba, 1979), p.95;
  %
  S.~L.~Glashow, in {\it Quarks and Leptons}, eds. M.~L\'evy {\it et
  al.} (Plenum Press, New York, 1980), p.687;
  %
%%\cite{Mohapatra:1979ia}
%\bibitem{Mohapatra:1979ia}
  R.~N.~Mohapatra and G.~Senjanovi\'c,
  %``Neutrino Mass And Spontaneous Parity Nonconservation,''
  Phys.\ Rev.\ Lett.\  {\bf 44} (1980) 912. 
  %%CITATION = PRLTA,44,912;%%



\bibitem{seesaw:IandII}
%%\cite{Schechter:1980gr}
%\bibitem{Schechter:1980gr}
  J.~Schechter and J.~W.~F.~Valle,
  %``Neutrino Masses in SU(2) x U(1) Theories,''
  Phys.\ Rev.\ D {\bf 22} (1980) 2227;
  %%CITATION = PHRVA,D22,2227;%%
%%\cite{Mohapatra:1980yp}
%\bibitem{Mohapatra:1980yp}
  R.~N.~Mohapatra and G.~Senjanovic,
  %``Neutrino Masses and Mixings in Gauge Models with Spontaneous Parity Violation,''
  Phys.\ Rev.\ D {\bf 23} (1981) 165;
  %%CITATION = PHRVA,D23,165;%%
%%\cite{Lazarides:1980nt}
%\bibitem{Lazarides:1980nt}
  G.~Lazarides, Q.~Shafi and C.~Wetterich,
  %``Proton Lifetime and Fermion Masses in an SO(10) Model,''
  Nucl.\ Phys.\ B {\bf 181} (1981) 287;
  %%CITATION = NUPHA,B181,287;%%
%%\cite{Wetterich:1981bx}
%\bibitem{Wetterich:1981bx}
  C.~Wetterich,
  %``Neutrino Masses and the Scale of B-L Violation,''
  Nucl.\ Phys.\ B {\bf 187} (1981) 343;
  %%CITATION = NUPHA,B187,343;%%
%%\cite{Schechter:1981cv}
%\bibitem{Schechter:1981cv}
  J.~Schechter and J.~W.~F.~Valle,
  %``Neutrino Decay and Spontaneous Violation of Lepton Number,''
  Phys.\ Rev.\ D {\bf 25} (1982) 774.
  %%CITATION = PHRVA,D25,774;%%



%\cite{Weinberg:1979sa}
\bibitem{Weinberg:1979sa}
  S.~Weinberg,
  %``Baryon and Lepton Nonconserving Processes,''
  Phys.\ Rev.\ Lett.\  {\bf 43} (1979) 1566.
  %%CITATION = PRLTA,43,1566;%%



\bibitem{direct}
%%\cite{Aseev:2011dq}
%\bibitem{Aseev:2011dq}
  V.~N.~Aseev {\it et al.}  [Troitsk Collaboration],
  %``An upper limit on electron antineutrino mass from Troitsk experiment,''
  Phys.\ Rev.\ D {\bf 84} (2011) 112003
  [arXiv:1108.5034 [hep-ex]].
  %%CITATION = ARXIV:1108.5034;%%

\bibitem{indirect}
%%\cite{Moresco:2012by}
%\bibitem{Moresco:2012by}
  M.~Moresco, L.~Verde, L.~Pozzetti, R.~Jimenez and A.~Cimatti,
  %``New constraints on cosmological parameters and neutrino properties using the expansion rate of the Universe to z~1.75,''
  JCAP {\bf 1207} (2012) 053
  [arXiv:1201.6658 [astro-ph.CO]];
  %%CITATION = ARXIV:1201.6658;%%
%%\cite{RiemerSorensen:2011fe}
%\bibitem{RiemerSorensen:2011fe}
  S.~Riemer-Sorensen, C.~Blake, D.~Parkinson, T.~M.~Davis, S.~Brough, M.~Colless, C.~Contreras and W.~Couch {\it et al.},
  %``The WiggleZ Dark Energy Survey: Cosmological neutrino mass constraint from blue high-redshift galaxies,''
  Phys.\ Rev.\ D {\bf 85} (2012) 081101
  [arXiv:1112.4940 [astro-ph.CO]];
  %%CITATION = ARXIV:1112.4940;%%
%%\cite{Xia:2012na}
%\bibitem{Xia:2012na}
  J.~-Q.~Xia, B.~R.~Granett, M.~Viel, S.~Bird, L.~Guzzo, M.~G.~Haehnelt, J.~Coupon and H.~J.~McCracken {\it et al.},
  %``Constraints on Massive Neutrinos from the CFHTLS Angular Power Spectrum,''
  JCAP {\bf 1206} (2012) 010
  [arXiv:1203.5105 [astro-ph.CO]].
  %%CITATION = ARXIV:1203.5105;%%



%\cite{Dev:2013oxa}
\bibitem{Dev:2013oxa}
  C.~H.~Lee, P.~S.~Bhupal Dev and R.~N.~Mohapatra,
  %``Natural TeV-scale left-right seesaw mechanism for neutrinos and experimental tests,''
  Phys.\ Rev.\ D {\bf 88} (2013) 9,  093010
  [arXiv:1309.0774 [hep-ph]].
  %%CITATION = ARXIV:1309.0774;%%



%% <START> HIGHER DIMENSIONAL OPnu

%\cite{Babu:2009aq}
\bibitem{Babu:2009aq}
  K.~S.~Babu, S.~Nandi and Z.~Tavartkiladze,
  %``New Mechanism for Neutrino Mass Generation and Triply Charged Higgs Bosons at the LHC,''
  Phys.\ Rev.\ D {\bf 80} (2009) 071702
  [arXiv:0905.2710 [hep-ph]].
  %%CITATION = ARXIV:0905.2710;%%

%\cite{Bonnet:2009ej}
\bibitem{Bonnet:2009ej}
  F.~Bonnet, D.~Hernandez, T.~Ota and W.~Winter,
  %``Neutrino masses from higher than d=5 effective operators,''
  JHEP {\bf 0910} (2009) 076
  [arXiv:0907.3143 [hep-ph]].
  %%CITATION = ARXIV:0907.3143;%%

%% </END>



% inverse (double) seesaw 
\bibitem{inverse}
%%\cite{Mohapatra:1986aw}
%\bibitem{Mohapatra:1986aw}
  R.~N.~Mohapatra,
  %``Mechanism for Understanding Small Neutrino Mass in Superstring Theories,''
  Phys.\ Rev.\ Lett.\  {\bf 56} (1986) 561;
  %%CITATION = PRLTA,56,561;%%
%%\cite{Mohapatra:1986bd}
%\bibitem{Mohapatra:1986bd}
  R.~N.~Mohapatra and J.~W.~F.~Valle,
  %``Neutrino Mass and Baryon Number Nonconservation in Superstring Models,''
  Phys.\ Rev.\ D {\bf 34} (1986) 1642;
  %%CITATION = PHRVA,D34,1642;%%
%%\cite{GonzalezGarcia:1988rw}
%\bibitem{GonzalezGarcia:1988rw}
  M.~C.~Gonzalez-Garcia and J.~W.~F.~Valle,
  %``Fast Decaying Neutrinos and Observable Flavor Violation in a New Class of Majoron Models,''
  Phys.\ Lett.\ B {\bf 216} (1989) 360.
  %%CITATION = PHLTA,B216,360;%%



%% <START> RADIATIVE SEESAW 

% 1-loop
\bibitem{seesaw:oneloop}
%%\cite{Zee:1980ai}
%\bibitem{Zee:1980ai}
  A.~Zee,
  %``A Theory of Lepton Number Violation, Neutrino Majorana Mass, and Oscillation,''
  Phys.\ Lett.\ B {\bf 93} (1980) 389
   [Erratum-ibid.\ B {\bf 95} (1980) 461];
  %%CITATION = PHLTA,B93,389;%%
%%\cite{Pilaftsis:1991ug}
%\bibitem{Pilaftsis:1991ug}
  A.~Pilaftsis,
  %``Radiatively induced neutrino masses and large Higgs neutrino couplings in the standard model with Majorana fields,''
  Z.\ Phys.\ C {\bf 55} (1992) 275
  [hep-ph/9901206];
  %%CITATION = HEP-PH/9901206;%%
%%\cite{Dev:2012sg}
%\bibitem{Dev:2012sg}
  P.~S.~B.~Dev and A.~Pilaftsis,
  %``Minimal Radiative Neutrino Mass Mechanism for Inverse Seesaw Models,''
  Phys.\ Rev.\ D {\bf 86} (2012) 113001
  [arXiv:1209.4051 [hep-ph]].
  %%CITATION = ARXIV:1209.4051;%%

% 2-loop
%\cite{Babu:1988ki}
\bibitem{Babu:1988ki}
  K.~S.~Babu,
  %``Model of 'Calculable' Majorana Neutrino Masses,''
  Phys.\ Lett.\ B {\bf 203} (1988) 132.
  %%CITATION = PHLTA,B203,132;%%

% 3-loop
%\cite{Krauss:2002px}
\bibitem{Krauss:2002px}
  L.~M.~Krauss, S.~Nasri and M.~Trodden,
  %``A Model for neutrino masses and dark matter,''
  Phys.\ Rev.\ D {\bf 67} (2003) 085002
  [hep-ph/0210389];
  %%CITATION = HEP-PH/0210389;%%
%%\cite{Cheung:2004xm}
%\bibitem{Cheung:2004xm}
  K.~Cheung and O.~Seto,
  %``Phenomenology of TeV right-handed neutrino and the dark matter model,''
  Phys.\ Rev.\ D {\bf 69} (2004) 113009
  [hep-ph/0403003].
  %%CITATION = HEP-PH/0403003;%%

%\cite{Ma:2006km}
\bibitem{Ma:2006km}
  E.~Ma,
  %``Verifiable radiative seesaw mechanism of neutrino mass and dark matter,''
  Phys.\ Rev.\ D {\bf 73} (2006) 077301
  [hep-ph/0601225].
  %%CITATION = HEP-PH/0601225;%%

% 3-loop
%\cite{Aoki:2008av}
\bibitem{Aoki:2008av}
  M.~Aoki, S.~Kanemura and O.~Seto,
  %``Neutrino mass, Dark Matter and Baryon Asymmetry via TeV-Scale Physics without Fine-Tuning,''
  Phys.\ Rev.\ Lett.\  {\bf 102} (2009) 051805
  [arXiv:0807.0361 [hep-ph]];
  %%CITATION = ARXIV:0807.0361;%%
%%\cite{Aoki:2009vf}
%\bibitem{Aoki:2009vf}
  M.~Aoki, S.~Kanemura and O.~Seto,
  %``A Model of TeV Scale Physics for Neutrino Mass, Dark Matter and Baryon Asymmetry and its Phenomenology,''
  Phys.\ Rev.\ D {\bf 80} (2009) 033007
  [arXiv:0904.3829 [hep-ph]].
  %%CITATION = ARXIV:0904.3829;%%

% 3-loop
%\cite{Gustafsson:2012vj}
\bibitem{Gustafsson:2012vj}
  M.~Gustafsson, J.~M.~No and M.~A.~Rivera,
  %``Predictive Model for Radiatively Induced Neutrino Masses and Mixings with Dark Matter,''
  Phys.\ Rev.\ Lett.\  {\bf 110} (2013) 21,  211802
  [arXiv:1212.4806 [hep-ph]].
  %%CITATION = ARXIV:1212.4806;%%

%\cite{Ma:1998dn}
\bibitem{Ma:1998dn}
  E.~Ma,
  %``Pathways to naturally small neutrino masses,''
  Phys.\ Rev.\ Lett.\  {\bf 81} (1998) 1171
  [hep-ph/9805219].
  %%CITATION = HEP-PH/9805219;%%

%\cite{FileviezPerez:2009ud}
\bibitem{FileviezPerez:2009ud}
  P.~Fileviez Perez and M.~B.~Wise,
  %``On the Origin of Neutrino Masses,''
  Phys.\ Rev.\ D {\bf 80} (2009) 053006
  [arXiv:0906.2950 [hep-ph]].
  %%CITATION = ARXIV:0906.2950;%%

%\cite{Bonnet:2012kz}
\bibitem{Bonnet:2012kz}
  F.~Bonnet, M.~Hirsch, T.~Ota and W.~Winter,
  %``Systematic study of the d=5 Weinberg operator at one-loop order,''
  JHEP {\bf 1207} (2012) 153 
  [arXiv:1204.5862 [hep-ph]].
  %%CITATION = ARXIV:1204.5862;%%

%% </END>



%\cite{Martin:1999hc}
\bibitem{Martin:1999hc}
  S.~P.~Martin,
  %``Dimensionless supersymmetry breaking couplings, flat directions, and the origin of intermediate mass scales,''
  Phys.\ Rev.\ D {\bf 61} (2000) 035004
  [hep-ph/9907550].
  %%CITATION = HEP-PH/9907550;%%



%% <START> HARD SUSY BREAKING m_nu 

%\cite{Frere:1999uv}
\bibitem{Frere:1999uv}
  J.~M.~Frere, M.~V.~Libanov and S.~V.~Troitsky,
  %``Neutrino masses from nonstandard supersymmetry breaking terms,''
  Phys.\ Lett.\ B {\bf 479} (2000) 343 
  [hep-ph/9912204].
  %%CITATION = HEP-PH/9912204;%%

%\cite{ArkaniHamed:2000bq}
\bibitem{ArkaniHamed:2000bq}
  N.~Arkani-Hamed, L.~J.~Hall, H.~Murayama, D.~Tucker-Smith and N.~Weiner,
  %``Small neutrino masses from supersymmetry breaking,''
  Phys.\ Rev.\ D {\bf 64} (2001) 115011
  [hep-ph/0006312].
  %%CITATION = HEP-PH/0006312;%%

%\cite{Frere:2003ys}
\bibitem{Frere:2003ys}
  J.~M.~Frere and E.~Ma,
  %``Gaugino induced quark and lepton masses in a truly minimal left-right model,''
  Phys.\ Rev.\ D {\bf 68} (2003) 051701
  [hep-ph/0305155].
  %%CITATION = HEP-PH/0305155;%%

%\cite{Demir:2007dt}
\bibitem{Demir:2007dt}
  D.~A.~Demir, L.~L.~Everett and P.~Langacker,
  %``Dirac Neutrino Masses from Generalized Supersymmetry Breaking,''
  Phys.\ Rev.\ Lett.\  {\bf 100} (2008) 091804
  [arXiv:0712.1341 [hep-ph]].
  %%CITATION = ARXIV:0712.1341;%%

%% </END>



%% <START> SUSY breaking mediators as seesaw mediators 

%\cite{Joaquim:2006uz}
\bibitem{Joaquim:2006uz}
  F.~R.~Joaquim and A.~Rossi,
  %``Gauge and Yukawa mediated supersymmetry breaking in the triplet seesaw scenario,''
  Phys.\ Rev.\ Lett.\  {\bf 97} (2006) 181801
  [hep-ph/0604083];
  %%CITATION = HEP-PH/0604083;%%
%%\cite{Joaquim:2006mn}
%\bibitem{Joaquim:2006mn}
  F.~R.~Joaquim and A.~Rossi,
  %``Phenomenology of the triplet seesaw mechanism with Gauge and Yukawa mediation of SUSY breaking,''
  Nucl.\ Phys.\ B {\bf 765} (2007) 71
  [hep-ph/0607298].
  %%CITATION = HEP-PH/0607298;%%

%\cite{Mohapatra:2008wx}
\bibitem{Mohapatra:2008wx}
  R.~N.~Mohapatra, N.~Okada and H.~-B.~Yu,
  %``nu-GMSB with Type III Seesaw and Phenomenology,''
  Phys.\ Rev.\ D {\bf 78} (2008) 075011
  [arXiv:0807.4524 [hep-ph]].
  %%CITATION = ARXIV:0807.4524;%%
 
%\cite{FileviezPerez:2009im}
\bibitem{FileviezPerez:2009im}
  P.~Fileviez Perez, H.~Iminniyaz, G.~Rodrigo and S.~Spinner,
  %``Gauge Mediated SUSY Breaking via Seesaw,''
  Phys.\ Rev.\ D {\bf 81} (2010) 095013
  [arXiv:0911.1360 [hep-ph]].
  %%CITATION = ARXIV:0911.1360;%%

%% </END> 



\bibitem{seesaw:III}
%%\cite{Foot:1988aq}
%\bibitem{Foot:1988aq}
  R.~Foot, H.~Lew, X.~G.~He and G.~C.~Joshi,
  %``Seesaw Neutrino Masses Induced by a Triplet of Leptons,''
  Z.\ Phys.\ C {\bf 44} (1989) 441.
  %%CITATION = ZEPYA,C44,441;%%



%\cite{Rossi:2002zb}
\bibitem{Rossi:2002zb}
  A.~Rossi,
  %``Supersymmetric seesaw without singlet neutrinos: Neutrino masses and lepton flavor violation,''
  Phys.\ Rev.\ D {\bf 66} (2002) 075003
  [hep-ph/0207006]. 
  %%CITATION = HEP-PH/0207006;%%
 


%% <START> KAHLER OPnu

% Both L L Hu Hd* and L L Hd* Hd* Kahler operators 
%\cite{Casas:2002sn}
\bibitem{Casas:2002sn}
  J.~A.~Casas, J.~R.~Espinosa and I.~Navarro,
  %``New supersymmetric source of neutrino masses and mixings,''
  Phys.\ Rev.\ Lett.\  {\bf 89} (2002) 161801
  [hep-ph/0206276].
  %%CITATION = HEP-PH/0206276;%%

% Kahler operator L L Hd* Hd*
%\cite{Brignole:2010nh}
\bibitem{Brignole:2010nh}
  A.~Brignole, F.~R.~Joaquim and A.~Rossi,
  %``Beyond the standard seesaw: Neutrino masses from Kahler operators and broken supersymmetry,''
  JHEP {\bf 1008} (2010) 133
  [arXiv:1007.1942 [hep-ph]].
  %%CITATION = ARXIV:1007.1942;%%

%% </END>



\bibitem{SUSYradiativeHIERARCHY}
%%\cite{Ibanez:1982xg}
%\bibitem{Ibanez:1982xg}
  L.~E.~Ibanez,
  %``Hierarchical Suppression of Radiative Quark and Lepton Masses in Supersymmetric {GUTs},''
  Phys.\ Lett.\ B {\bf 117} (1982) 403;
  %%CITATION = PHLTA,B117,403;%%
%%\cite{Lahanas:1982et}
%\bibitem{Lahanas:1982et}
  A.~B.~Lahanas and D.~Wyler,
  %``Radiative Fermion Masses and Supersymmetry,''
  Phys.\ Lett.\ B {\bf 122} (1983) 258.
  %%CITATION = PHLTA,B122,258;%%



%\cite{Borzumati:1999sp}
\bibitem{Borzumati:1999sp}
  F.~Borzumati, G.~R.~Farrar, N.~Polonsky and S.~D.~Thomas,
  %``Soft Yukawa couplings in supersymmetric theories,''
  Nucl.\ Phys.\ B {\bf 555} (1999) 53
  [hep-ph/9902443].
  %%CITATION = HEP-PH/9902443;%%



%\cite{Batra:2008rc}
\bibitem{Batra:2008rc}
  P.~Batra and E.~Ponton,
  %``Supersymmetric electroweak symmetry breaking,''
  Phys.\ Rev.\ D {\bf 79} (2009) 035001
  [arXiv:0809.3453 [hep-ph]].
  %%CITATION = ARXIV:0809.3453;%%



%\cite{Gates:1983nr}
\bibitem{Gates:1983nr}
  S.~J.~Gates, M.~T.~Grisaru, M.~Rocek and W.~Siegel,
  %``Superspace Or One Thousand and One Lessons in Supersymmetry,''
  Front.\ Phys.\  {\bf 58} (1983) 1
  [hep-th/0108200].
  %%CITATION = HEP-TH/0108200;%%



%\cite{Girardello:1981wz}
\bibitem{Girardello:1981wz}
  L.~Girardello and M.~T.~Grisaru,
  %``Soft Breaking of Supersymmetry,''
  Nucl.\ Phys.\ B {\bf 194} (1982) 65.
  %%CITATION = NUPHA,B194,65;%%



%% <STARTS> MODELS IN THE LITERATURE

\bibitem{fm}
%%\cite{Franceschini:2013aha}
%\bibitem{Franceschini:2013aha}
  R.~Franceschini and R.~N.~Mohapatra,
  %``Radiatively Induced Type II seesaw and Vector-like 5/3 Charge Quarks,''
  Phys.\ Rev.\ D {\bf 89} (2014) 055013
  [arXiv:1306.6108 [hep-ph]].
  %%CITATION = ARXIV:1306.6108;%%

\bibitem{bmw}
%%\cite{Bhattacharya:2013nya}
%\bibitem{Bhattacharya:2013nya}
  S.~Bhattacharya, E.~Ma and D.~Wegman,
  %``Supersymmetric left-right model with radiative neutrino mass and multipartite dark matter,''
  Eur.\ Phys.\ J.\ C {\bf 74} (2014) 2902
  [arXiv:1308.4177 [hep-ph]].
  %%CITATION = ARXIV:1308.4177;%%

\bibitem{kmsy}
%%\cite{Kanemura:2013uva}
%\bibitem{Kanemura:2013uva}
  S.~Kanemura, N.~Machida, T.~Shindou and T.~Yamada,
  %``A UV complete model for radiative seesaw scenarios and electroweak baryogenesis based on the supersymmetric gauge theory,''
  Phys.\ Rev.\ D {\bf 89} (2014) 1,  013005
  [arXiv:1309.3207 [hep-ph]].
  %%CITATION = ARXIV:1309.3207;%%

%% ------------------------------

%\cite{Passarino:1978jh}
\bibitem{Passarino:1978jh}
  G.~Passarino and M.~J.~G.~Veltman,
  %``One Loop Corrections for e+ e- Annihilation Into mu+ mu- in the Weinberg Model,''
  Nucl.\ Phys.\ B {\bf 160} (1979) 151;
  %%CITATION = NUPHA,B160,151;%%
%%\cite{Denner:1991kt}
%\bibitem{Denner:1991kt}
  A.~Denner,
  %``Techniques for calculation of electroweak radiative corrections at the one loop level and results for W physics at LEP-200,''
  Fortsch.\ Phys.\  {\bf 41} (1993) 307
  [arXiv:0709.1075 [hep-ph]];
  %%CITATION = ARXIV:0709.1075;%%
%%\cite{Mertig:1990an}
%\bibitem{Mertig:1990an}
  R.~Mertig, M.~Bohm and A.~Denner,
  %``FEYN CALC: Computer algebraic calculation of Feynman amplitudes,''
  Comput.\ Phys.\ Commun.\  {\bf 64} (1991) 345;
  %%CITATION = CPHCB,64,345;%%
We follow the conventions of: 
%%\cite{Hahn:1998yk}
%\bibitem{Hahn:1998yk}
  T.~Hahn and M.~Perez-Victoria,
  %``Automatized one loop calculations in four-dimensions and D-dimensions,''
  Comput.\ Phys.\ Commun.\  {\bf 118} (1999) 153
  [hep-ph/9807565].
  %%CITATION = HEP-PH/9807565;%%

%% ------------------------------

\bibitem{Kanemura}
%%\cite{Kanemura:2012uy}
%\bibitem{Kanemura:2012uy}
  S.~Kanemura, T.~Shindou and T.~Yamada,
  %``A light Higgs scenario based on the TeV-scale supersymmetric strong dynamics,''
  Phys.\ Rev.\ D {\bf 86} (2012) 055023
  [arXiv:1206.1002 [hep-ph]];
  %%CITATION = ARXIV:1206.1002;%%
%%\cite{Kanemura:2012hr}
%\bibitem{Kanemura:2012hr}
  S.~Kanemura, E.~Senaha, T.~Shindou and T.~Yamada,
  %``Electroweak phase transition and Higgs boson couplings in the model based on supersymmetric strong dynamics,''
  JHEP {\bf 1305} (2013) 066
  [arXiv:1211.5883 [hep-ph]].
  %%CITATION = ARXIV:1211.5883;%%

%% </END>



%% <START> TOOLS 

\bibitem{feynarts}
%%\cite{Hahn:2000kx}
%\bibitem{Hahn:2000kx}
  T.~Hahn,
  %``Generating Feynman diagrams and amplitudes with FeynArts 3,''
  Comput.\ Phys.\ Commun.\  {\bf 140} (2001) 418
  [hep-ph/0012260].
  %%CITATION = HEP-PH/0012260;%%

\bibitem{feynrules}
%%\cite{Alloul:2013bka}
%\bibitem{Alloul:2013bka}
  A.~Alloul, N.~D.~Christensen, C.~Degrande, C.~Duhr and B.~Fuks,
  ``FeynRules 2.0 - A complete toolbox for tree-level phenomenology,''
  arXiv:1310.1921 [hep-ph].
  %%CITATION = ARXIV:1310.1921;%%

%% </END>


\end{thebibliography}
\end{document}